\documentclass[letter,bibyear]{aa} % for the letters

\usepackage{natbib}
\usepackage{graphicx}
\usepackage{txfonts}
\newcommand{\jms}{J.~Mol.~Spectrosc.}   % Journal of Molecular Spectroscopy
\newcommand{\jmst}{J.~Mol.~Struct.}   % Journal of Molecular Structure

\newcommand{\kms}{km s$^{-1}$}

\newcommand{\once}{10$^{11}$\,cm$^{-2}$}
\newcommand{\doce}{10$^{12}$\,cm$^{-2}$}
\newcommand{\trece}{10$^{13}$\,cm$^{-2}$}

\begin{document}

\title{Discovery of two isomers of ethynyl cyclopentadiene in TMC-1:
Abundances of CCH and CN derivatives of hydrocarbon cycles\thanks{Based on observations carried out
with the Yebes 40m telescope (projects 19A003,
20A014, 20D023, and 21A011). The 40m
radio telescope at Yebes Observatory is operated by the Spanish Geographic 
Institute
(IGN, Ministerio de Transportes, Movilidad y Agenda Urbana).}}

\author{
J.~Cernicharo\inst{1},
M.~Ag\'undez\inst{1},
R.~I.~Kaiser\inst{2},
C.~Cabezas\inst{1},
B.~Tercero\inst{3,4},
N.~Marcelino\inst{4},
J.~R.~Pardo\inst{1}, and
P.~de~Vicente\inst{3}
}

\institute{Grupo de Astrof\'isica Molecular, Instituto de F\'isica Fundamental (IFF-CSIC),
C/ Serrano 121, 28006 Madrid, Spain\\ \email jose.cernicharo@csic.es
\and Department of Chemistry, University of Hawaii at Manoa, Honolulu, HI 96822, USA
\and Centro de Desarrollos Tecnol\'ogicos, Observatorio de Yebes (IGN), 19141 Yebes, Guadalajara, Spain
\and Observatorio Astron\'omico Nacional (OAN, IGN), Madrid, Spain
}

\date{Received; accepted}

\abstract{
We report the detection of two isomers of ethynyl cyclopentadiene ($c$-C$_5$H$_5$CCH), 
namely 1- and 2-ethynyl-1,3-cyclopentadiene, in the direction of TMC-1. We derive column densities 
of (1.4$\pm$0.2)$\times$\doce\, and
(2.0$\pm$0.4)$\times$\doce, respectively, for these two cyclopentadiene derivatives, which imply that they 
are about ten times less abundant than
cyclopentadiene. We also report the tentative detection of ethynyl benzene (C$_6$H$_5$CCH), for which
we estimate a column density of (2.5$\pm$0.4)$\times$\doce.
We derived abundances for the corresponding cyano derivatives
of cyclopentadiene and benzene and found values significantly lower than previously reported. 
The rotational temperature of the ethynyl and cyano
derivatives of these cycles is about 9\, K, that is, very close to the gas kinetic temperature of the cloud.
The abundance ratio of the 1- and 2- isomers of ethynyl cyclopentadiene is 
1.4$\pm$0.5, while for the two isomers of cyano cyclopentadiene it is 
2.4$\pm$0.6. The relative abundances of CCH over CN derivatives
is 7.7$\pm$2.2 for cyclopentadiene, which probably reflects the abundance ratio of the radicals 
CCH and CN; this ratio is only 2.1$\pm$0.5 for benzene, which suggests that additional reactions 
besides cyano radicals with benzene are involved in 
the formation of benzonitrile. The formation of these cycles is reasonably well 
accounted for through a chemical scheme based on neutral-neutral reactions. It 
is predicted that benzene should be as abundant as cyclopentadiene in TMC-1.
}

\titlerunning{c-C$_5$H$_5$CCH in TMC-1}
\authorrunning{Cernicharo et al.}

\maketitle

\section{Introduction}

The QUIJOTE\footnote{\textbf{Q}-band \textbf{U}ltrasensitive \textbf{I}nspection \textbf{J}ourney 
to the \textbf{O}bscure \textbf{T}MC-1 \textbf{E}nvironment} \citep{Cernicharo2021a}
and the GOTHAM\footnote{\textbf{G}BT \textbf{O}bservations of \textbf{T}MC-1: 
\textbf{H}unting \textbf{A}romatic \textbf{M}olecules} \citep{McGuire2018}
line surveys are providing exciting results of the abundance of 
hydrocarbons and their ethynyl and cyano derivatives 
in the cold pre-stellar core Taurus Molecular Cloud 1 (TMC-1). Species such as
the propargyl radical (CH$_2$CCH), vinylacetylene 
(CH$_2$CHCCH), ethynylallene (H$_2$CCCHCCH), and cyclic 
hydrocarbons such as cyclopentadiene ($c$-C$_5$H$_6$), $o$-benzyne (C$_6$H$_4$), 
and indene (c-C$_9$H$_8$) \citep{Agundez2021a,Cernicharo2021a,Cernicharo2021b,Cernicharo2021c,
Cernicharo2021d} have been detected using the QUIJOTE line survey
through a line-by-line identification process.
Using spectral stacking techniques, the GOTHAM line survey has provided
the detections of the cyano derivatives of cyclopentadiene, benzene, and naphthalene
\citep{McGuire2021,Lee2021,McGuire2018}.
These results reveal a new and unexpected
chemistry that requires a profound revision of the chemical processes at work in cold dark clouds such as TMC-1.
In order to provide an adequate reference for chemical models, reliable molecular abundances
need to be obtained. Moreover, observing distinct ethynyl and cyano derivatives of hydrocarbons
can provide important constraints on the reactivity of CCH and CN radicals with unsaturated 
acyclic and cyclic hydrocarbons.

In this letter we report the discovery of two isomers of ethynyl cyclopentadiene ($c$-C$_5$H$_5$CCH) and the tentative
detection of ethynyl benzene (C$_6$H$_5$CCH, hereinafter referred to as $EBZ$) towards TMC-1. From our data we also derive column densities 
for the corresponding cyano derivatives of benzene and cyclopentadiene, previously detected by \cite{McGuire2018}, \cite{McCarthy2021}, and 
\citet{Lee2021}, providing a rigorous confirmation of the presence of these species in TMC-1 based on a line-by-line detection procedure and a coherent and homogeneous set of abundances for the CCH and CN
derivatives of cyclopentadiene and benzene.

\section{Observations} \label{observations}

New receivers, built within the Nanocosmos project\footnote{\texttt{https://nanocosmos.iff.csic.es/}}
and installed at the Yebes 40m radio telescope, were used
for the observations of TMC-1
($\alpha_{J2000}=4^{\rm h} 41^{\rm  m} 41.9^{\rm s}$ and $\delta_{J2000}=
+25^\circ 41' 27.0''$). A detailed description of the system is 
given by \citet{Tercero2021}.
The receiver consists of two cold high electron mobility transistor amplifiers that cover the
31.0-50.3 GHz band with horizontal and vertical             
polarizations. Receiver temperatures in the runs conducted in 2020 vary from 22 K at 32 GHz
to 42 K at 50 GHz. Some power adaptation in the down-conversion chains have reduced
the receiver temperatures in 2021 to 16\,K at 32 GHz and 30\,K at 50 GHz.
The backends are $2\times8\times2.5$ GHz fast Fourier transform spectrometers
with a spectral resolution of 38.15 kHz,
providing the whole coverage of the Q band in both polarizations. 
All observations were performed in the frequency 
switching mode with frequency throws of 8 and 10 MHz. The main beam efficiency varies from 0.6 at
32 GHz to 0.43 at 50 GHz. 
Pointing corrections were derived from nearby quasars and SiO masers,
and errors were always within 2-3$''$. The telescope 
beam size is 56$''$ and 31$''$ at 31 and 50 GHz, respectively.
The intensity scale used in this work, antenna temperature
($T_A^*$), was calibrated using two absorbers at different temperatures and the
atmospheric transmission model ATM \citep{Cernicharo1985, Pardo2001}.
Calibration uncertainties were adopted to be 10~\%.
All data were analysed using the GILDAS package\footnote{\texttt{http://www.iram.fr/IRAMFR/GILDAS}}.
Details of the QUIJOTE line survey are provided by \citet{Cernicharo2021d}.
The 1$\sigma$ sensitivity of the survey varies between 0.17 and 0.30 mK between 31 and 50 GHz.

\begin{figure}[]
\centering
\includegraphics[scale=0.14,angle=0]{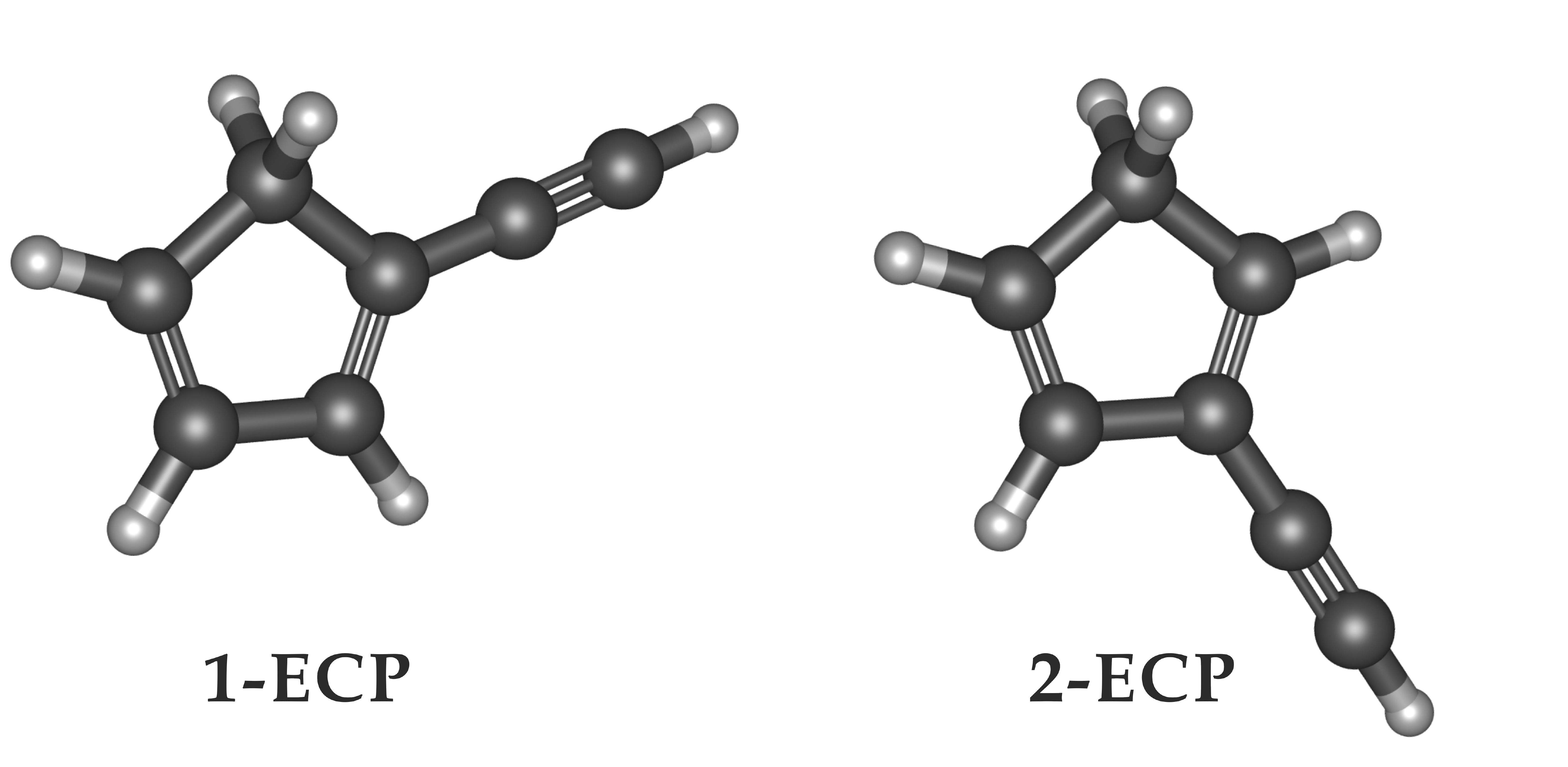}
\caption{Scheme of the two lowest energy isomers of ethynyl cyclopentadiene. }
\label{fig_structure}
\end{figure}

\section{Detection of cycles in TMC-1} 
\label{results}

Line identification in this work was done using the catalogues 
MADEX \citep{Cernicharo2012}, CDMS \citep{Muller2005}, and JPL \citep{Pickett1998}. 
By September 2021, the MADEX code contained 6377 spectral
entries corresponding to the ground and vibrationally excited states, together
with the corresponding isotopologues, of 1696 molecules. 

The recent detection
of cyclopentadiene \citep{Cernicharo2021a} and of its cyano derivatives
\citep{McCarthy2021,Lee2021} suggests that other derivatives of
cyclopentadiene could be present in this source, in particular the ethynyl ones, 1- and 
2-ethynyl-1,3-cyclopentadiene (hereinafter referred to as 1-$ECP$ and 2-$ECP$, respectively; see Fig. \ref{fig_structure}). 
These two isomers of ethynyl cyclopentadiene were observed in the laboratory by \citet{McCarthy2020}. We used these
data to fit the rotational and distortion constants in order to predict
the frequencies of their rotational transitions within the Q band (see Sects.
3.1.1 and 3.1.2). The calculated uncertainties for these transitions are 10-25 kHz (0.1-0.3 \kms).

\subsection{The isomers of ethynyl cyclopentadiene} \label{ab_initio}

The substitution of a hydrogen atom by an ethynyl (CCH) 
group in cyclopentadiene ($c$-C$_5$H$_6$)  yields three possible isomers of ethynyl cyclopentadiene. Quantum chemical calculations at the  
MP2/6-311++G(d,p) level of theory \citep{Moller1934, Frisch1984} predict 
1-$ECP$ as the most stable isomer, 
followed closely 
by 2-$ECP$ (6 kJ\,mol$^{-1}$) and then 
5-ethynyl-1,3-cyclopentadiene, whose energy is far higher
(by 27 kJ\,mol$^{-1}$). The structures 
of the two isomers 1-$ECP$ and 2-$ECP$ are shown in Fig. \ref{fig_structure}. 
The two isomers are moderately polar, with dipole moments along their $a$- 
and $b$-inertial axes. The predicted $\mu_a$ values for 1-$ECP$ and 2-$ECP$ 
are 0.81 and 1.11 D, respectively, while the $\mu_b$ values are 0.32 and 0.37 D, respectively,
in good agreement with previous calculations \citep{Lee2019}. Both isomers
were observed in the laboratory by \citet{McCarthy2020} (we note that
2-$ECP$ was named 5-ethynyl-1,3-cyclopentadiene in their work). The low
value of $\mu_b$ for the two isomers results in $b$-type transitions that are around ten times
weaker than the $a$-type ones. In the following, we start our search with the isomer with
the largest dipole moment, namely 2-$ECP$.

\begin{figure*}
\centering
\includegraphics[scale=0.54]{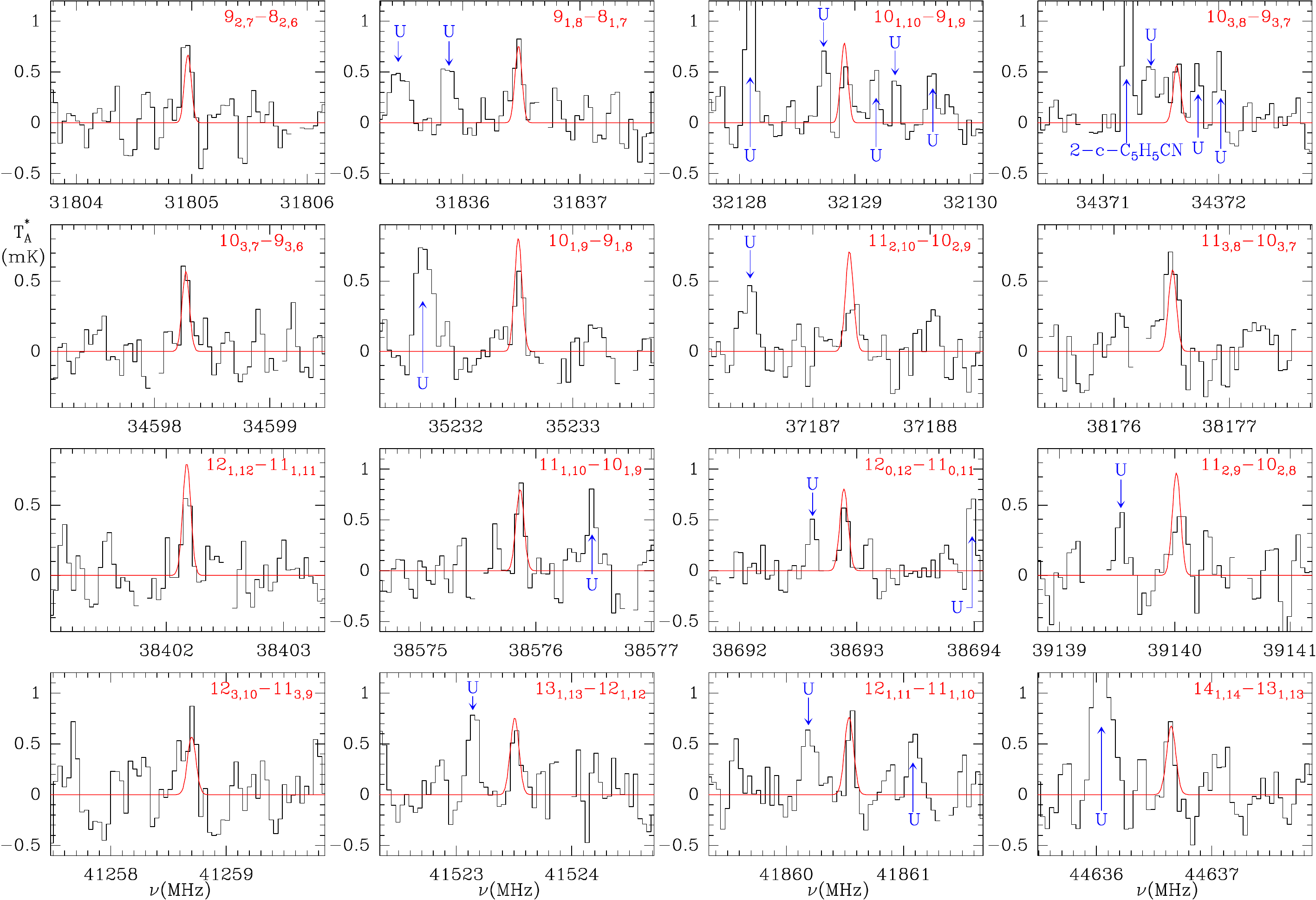}
\caption{Subset of the observed lines of 2-$ECP$ in the 31-50 GHz 
frequency range towards TMC-1.
Line parameters for the complete list of detected lines of 2-$ECP$ are given in Table \ref{line_parameters_2ecp}.
The abscissa corresponds to the rest frequency assuming a local standard of rest velocity of 5.83
km s$^{-1}$. 
The ordinate is the antenna temperature corrected for atmospheric and telescope losses in mK.
The red line shows the synthetic spectrum obtained from a fit to the observed line profiles,
which provides T$_r$=9$\pm$1\,K and N(2-$ECP$)=(1.4$\pm$0.2)$\times$\doce. 
The rotational quantum
numbers are indicated in each panel. Blanked channels correspond to negative features produced
in the folding of the frequency switching data. 
}
\label{fig_2-ecp}
\end{figure*}

\subsubsection{Detection of 2-ethynyl-1,3-cyclopentadiene (2-$ECP$)} \label{detection_2ecp}

The laboratory data for 2-$ECP$, the isomer with the largest dipole moment along the $a$ axis, cover the frequency range
6.5-25.8 GHz with $J_{max}$=8 \citep{McCarthy2020}. \citet{McCarthy2020} quote an uncertainty on their 
frequencies of 2 kHz.
However, their own fit provides a standard deviation of 4.5 kHz. We thus assigned an uncertainty of 4 kHz to all
their measured frequencies, except for the 7$_{1,6}$-6$_{1,5}$ line, for which
they quote an uncertainty of 10 kHz. The resulting rotational and distortion constants
with the new uncertainties do not show a significant variation with respect to their
constants. However, the standard deviation of the fit improves to 3 kHz if
the distortion constant $\delta_J$ is also included in the fit 
(see Table \ref{table_new_constants_ecp}).
This new set of constants was used to
predict the frequencies of the rotational lines of 2-$ECP$ in the 31-50 GHz domain.

A total of 24 $a$-type lines of 2-$ECP$ were detected in TMC-1 above the 3$\sigma$ level with the QUIJOTE line survey. 
Some of them are shown in
Fig. \ref{fig_2-ecp}. The derived line parameters
are given in Appendix \ref{observed_lines_ecp} (see Table 
\ref{line_parameters_2ecp}). A fit to the observed line profiles assuming
a source diameter of 80$''$ \citep{Fosse2001}
provides a rotational temperature of 9.0$\pm$1.0\,K and
a column density of (1.4$\pm$0.2)$\times$\doce. The synthetic spectra are compared
with observations in Fig. \ref{fig_2-ecp} (red line). 
With the adopted source diameter, the molecular emission fills the
main beam of the telescope at all observed frequencies.

All observed lines of 2-$ECP$ correspond
to values of $J$ between 9-14 and $K_a\le$3; hence,
the associated upper level energies cover the range 8-20\,K. Rotational temperatures below
8\,K underestimate the emission of the transitions arising from the higher energy levels.
Nevertheless, transitions involving energy levels between 8 and 12 K are not very sensitive
to the rotational temperature and can be fitted with rotational temperatures as low as 
6\,K with a modest variation in the column density. This effect is discussed in
detail in Appendix A of \citet{Cernicharo2021e}.
A similar situation has been found for the cyano derivatives of cyclopentadiene and
benzene (see our Appendices \ref{cyano_cyclopentadiene} and \ref{cyano_benzene}), which
have considerably larger dipole moments. The near thermalization of the rotational levels
of the derivatives of cyclopentadiene and benzene is most likely
due the large collisional
rates we could expect for these molecules, which exhibit a much larger geometrical cross-section
than linear molecules such as HC$_5$N and HC$_7$N, for which rotational temperatures
around 8-9\,K have also been found \citep{Cernicharo2020}. A similar rotational temperature of $\sim$9\,K 
has been derived for indene and cyclopentadiene \citep{Cernicharo2021a}.

\subsubsection{Detection of 1-ethynyl-1,3-cyclopentadiene (1-$ECP$)} \label{detection_1ecp}

The laboratory data for 1-$ECP$ \citep{McCarthy2020} cover the frequency range
6.5-24.9 GHz with $J_{max}$=7 and $K_a\le$3. \citet{McCarthy2020} quote a standard deviation for their fit
of 2.7 kHz. However, it is possible to reduce this value to 1.1 kHz by fitting the
distortion constant $\delta_K$. The new rotational and distortion constants,
which are given in Table \ref{table_new_constants_ecp}, were used to
predict the frequencies of the rotational lines of 1-$ECP$ in the 31-50 GHz domain.

If the abundances of the two isomers were identical, the $a$-type lines of 1-$ECP$ would be
1.9 times weaker that those of 2-$ECP$ (the squared ratio of the $a$ component
of the dipole moment). Hence, we expect to detect only the strongest transitions of
1-$ECP$. A total of 14 lines of 1-$ECP$ were detected in TMC-1 above the 3$\sigma$ level with the current 
sensitivity of the QUIJOTE line survey. 
Some of them are shown in
Fig. \ref{fig_1-ecp}. The derived line parameters
are given in Appendix \ref{observed_lines_ecp} (see Table 
\ref{line_parameters_1ecp}). A fit to the observed line profiles
assuming the same source size (80$''$) and rotational temperature (9\,K) as for 2-$ECP$ 
provides a column density of (2.0$\pm$0.4)$\times$\doce. Therefore, 
the isomer 2-$ECP$ is a factor of 1.4$\pm$0.5 more abundant than 1-$ECP$. 

The detection of the two isomers of ethynyl cyclopentadiene is robust since it is based on the
detection of a significant number of individual transitions. Taking into account the
column density of (1.2$\pm$0.3)$\times$\trece\, derived for cyclopentadiene \citep{Cernicharo2021a}, 
1-$ECP$ and 2-$ECP$
are less abundant than cyclopentadiene by factors of 6.0$\pm$2.5 and 8.6$\pm$3.3, respectively.

Finally, improved molecular constants for 1-$ECP$ and 2-$ECP$ resulting from a 
merged fit to the laboratory
data and the observed frequencies in TMC-1 are provided in Appendix A.1 
(see Table \ref{table_new_constants_ecp}).

\subsection{Ethynyl benzene} \label{ethynyl_benzene}

Laboratory spectroscopy for $EBZ$, 
C$_6$H$_5$CCH,
has been provided by different authors covering frequencies up to 
340 GHz, $J$ up
to 140, and $K_a$ up to 48 \citep{Cox1975,Dreizler2004,Kisiel2010}. Hence, 
the frequency predictions in the
range of our line survey are rather accurate, with calculated uncertainties below 1 kHz. 
The dipole moment of the molecule is low, 0.66\,D \citep{Cox1975}.
Hence, we could expect weak emission lines in our data. The molecule has two pairs of identical
hydrogen nuclei, which introduces an $ortho$ and $para$ spin statistic. $Ortho$ and $para$
levels correspond to $K_a$ even and odd, respectively. The ratio of
statistical weights is 5/3. The lowest energy $para$ level (1$_{1,1}$) is 0.33 K above
the ground $ortho$ level (0$_{0,0}$). Due to the low dipole moment and the
$\sim$1.7 factor in the statistical weights, we do not expect to
have enough sensitivity in our data to detect the $para$ transitions. A quick examination
of all $K_a$=0 and 2 lines reveals only four lines at the 3$\sigma$ limit of the survey.
All explored lines are summarized in Appendix \ref{appendix_ebz} 
(see Table \ref{line_parameters_ebz}).
The four detected lines
are shown in Fig. \ref{fig_ebz}. With this limited number of lines, it is not possible to claim a detection.
Tentatively, we derive a column density of (2.5$\pm$0.4)$\times$\doce\, for an assumed rotational 
temperature of 9\,K and a source size of 80$''$. Taking into account the density of unknown features, we consider
that a stacking of our data is hazardous and that a definitive detection has to wait
for the improvement of the QUIJOTE line survey. Assuming that the derived column density is a 
3$\sigma$ limit,
the relative abundance of ethynyl cyclopentadiene (the two isomers) and $EBZ$ is $\ge$1.4, which
suggests 
that their potential cyclopentadiene and benzene precursors have abundances of the same order in TMC-1.

\subsection{Cyano derivatives of cyclopentadiene and benzene} \label{cyano_derivatives}

Two cyano derivatives of cyclopentadiene (hereinafter referred to as 1-$CCP$ and 2-$CCP$; see Appendix \ref{cyano_cyclopentadiene})
were detected in TMC-1 using stacking techniques by \citet{McCarthy2021} and \citet{Lee2021}. A few individual lines of 1-$CCP$ were reported by \citet{Lee2021}.
There are some discrepancies between
the column densities reported by these authors for 1-$CCP$. \citet{McCarthy2021} derive
$N$(1-$CCP$)=(1.44$\pm$0.17)$\times$\doce\, and $T_{rot}$=6.0$\pm$0.3\,K, 
while \citet{Lee2021} find a column density of (8.3$\pm$0.1)$\times$\once\, and a
rotational temperature of 6.00$\pm$0.03\,K. For 2-$CCP$, \citet{Lee2021} derive a column density
of 1.9$\times$\once.
In order to provide a coherent and homogeneous set of column densities, we analyse the lines of the two isomers in Appendix \ref{cyano_cyclopentadiene}.
The observed lines of 1-$CCP$ are shown in Fig. \ref{fig_1-CCP} and those of 
2-$CCP$ in Fig. \ref{fig_2-CCP}. Line parameters for the two species are
given in Tables \ref{line_parameters_1-ccp} and \ref{line_parameters_2-ccp}, respectively.
We derive a rotational
temperature of 9.0$\pm$1.0\,K for both species and column densities of (3.1$\pm$0.3)$\times$\once\,
and (1.3$\pm$0.2)$\times$\once\, for 1-$CCP$ and 2-$CCP$, respectively. 
The isomer 1-$CCP$ is 2.4 times more abundant than 2-$CCP$, which is twice lower than the abundance ratio found by \citet{Lee2021}.
The discrepancies with
previous works are discussed in Appendix \ref{cyano_cyclopentadiene}. 

Benzonitrile, C$_6$H$_5$CN, was previously detected towards TMC-1 by \citet{McGuire2018} through stacking
techniques and some well-detected individual lines. They
obtain a rotational temperature of 7\,K and a column density of 4$\times$\once. 
In a more recent work, \citet{Burkhardt2021} derive a column density of 1.6$\times$\doce\ (i.e.
a factor of four higher than previously reported) and a rotational temperature of 6.1$\pm$0.3\,K.
In Appendix \ref{cyano_benzene} we discuss the 100 individual lines of this species detected
with high sensitivity with QUIJOTE's data. They are shown in Figs. \ref{fig_bn1}, \ref{fig_bn2}, \ref{fig_bn3},
and \ref{fig_bn4}. We obtain a rotational temperature of 9.0$\pm$0.5\,K and a total
column density of (1.2$\pm$0.1)$\times$\doce. A rotational temperature of 6\,K cannot explain
the observed emission of lines with $K_a\ge$4 (see the caption of Fig. \ref{fig_bn1} and Appendix
\ref{cyano_benzene}).

All column densities derived in this work are summarized in Table \ref{column_densities}.

\tiny
\begin{table}
\centering
\small
\caption{Abundances of ethynyl and cyano species in TMC-1.}
\label{column_densities}
\begin{tabular}{lccl}
\hline
Molecule                  & N (cm$^{-2}$)       & Abundance$^a$           & Comments\\
\hline              
$c$-C$_5$H$_6$            & 1.3$\times10^{13}$     & 1.3$\times$10$^{-09}$&     1    \\
1-$c$-C$_5$H$_5$CCH       & 1.4$\times10^{12}$     & 1.4$\times$10$^{-10}$&     2    \\
2-$c$-C$_5$H$_5$CCH       & 2.0$\times10^{12}$     & 2.0$\times$10$^{-10}$&     2    \\
1-$c$-C$_5$H$_5$CN        & 3.1$\times10^{11}$     & 3.1$\times$10$^{-11}$&     2,A  \\
2-$c$-C$_5$H$_5$CN        & 1.3$\times10^{11}$     & 1.3$\times$10$^{-11}$&     2,B  \\                                                               
C$_6$H$_5$CCH             &$\sim$2.5$\times10^{12}$& 2.5$\times$10$^{-10}$&     2,C  \\
C$_6$H$_5$CN              & 1.2$\times10^{12}$     & 1.2$\times$10$^{-10}$&     2,D  \\
$c$-C$_9$H$_8$            & 1.6$\times10^{13}$     & 1.6$\times$10$^{-09}$&     1,E  \\
\hline                                                                                             
\end{tabular}
\tablefoot{\\
\tablefoottext{a}{Assuming a column density of molecular hydrogen of
10$^{22}$ cm$^{-2}$ \citep{Cernicharo1987}. We note that there is a significant difference 
in the beam size for the QUIJOTE and GOTHAM line surveys. This could explain, at least partially, the differences 
in the derived parameters.
} 
\tablefoottext{1}{\citet{Cernicharo2021a}.} 
\tablefoottext{2}{This work.}
\tablefoottext{A}{A value of 1.44$\times$\doce\  was reported by 
\citet{McCarthy2021} and of 8.3$\times$\once\, by \citet{Lee2021}.}
\tablefoottext{B}{A value of 1.9$\times$\once\, was derived by \citet{Lee2021}.}
\tablefoottext{C}{Tentative detection.}
\tablefoottext{D}{A value of 4.0$\times$\once\, was derived by \citet{McGuire2018}. This
value was revised to 1.6$\times$\once\, by \citet{Burkhardt2021}.}
\tablefoottext{E}{A value of 9.6$\times$\doce\, was reported by \citet{Burkhardt2021}.}
%\tablefoottext{F}{\citet{McGuire2021} have
%derived a column density of 7.4$\times$\once, and of 7.1$\times$\once, 
%for the isomers 1- and 2- of cyanonaphthalene.}\\
}
\end{table}
\normalsize

\section{Chemistry of cycles in TMC-1}

It is remarkable that, given their chemical complexity, the CCH and CN derivatives of cyclopentadiene 
and benzene are observed with relatively large abundances in TMC-1. To understand how these species can
be formed, we built a chemical model similar to that used in previous recent studies of TMC-1 
(e.g. \citealt{Cernicharo2021d}).  Briefly, we adopted typical conditions of cold dark clouds -- namely, a gas temperature of 10 K, a volume density of H nuclei of 2\,$\times$\,10$^4$ cm$^{-3}$, a cosmic-ray 
ionization rate of H$_2$ of 1.3\,$\times$\,10$^{-17}$ s$^{-1}$, a visual extinction of 30 mag, and the 
so-called low-metal elemental abundances (e.g. \citealt{Agundez2013}) -- with the exception of oxygen, 
for which we decreased the abundance (see below). The core of the chemical network is the RATE12 network of the UMIST 
database \citep{McElroy2013}, 
with updates from \cite{Loison2015}, \cite{Marcelino2021}, \cite{Agundez2021a,Agundez2021b}, 
and \cite{Cernicharo2021d}. We also included a specific chemistry for the CCH and CN derivatives of 
$c$-C$_5$H$_6$ and C$_6$H$_6$. We assumed that they are destroyed through reactions with C, C$^+$, and H$^+$. The chemical scheme of formation of these species is based on neutral-neutral reactions and is 
discussed in detail in Appendix~\ref{app:chem_scheme}. Briefly, reactions 
of CCH and CN with $c$-C$_5$H$_6$ and C$_6$H$_6$ lead to the CCH/CN 
derivatives of each cycle \citep{Balucani1999,Jones2010}. 
The formation of the precursor hydrocarbon cycles 
$c$-C$_5$H$_6$ and C$_6$H$_6$ relies on reactions between small hydrocarbon 
radicals and butadiene (CH$_2$CHCHCH$_2$), which acts as a key species that 
opens the chemistry to hydrocarbon cycles at 10\,K
\citep{He2020a,Jones2011}. Additional routes to the CCH/CN 
derivatives of $c$-C$_5$H$_6$ involve reactions of C$_3$H/C$_2$N 
with butadiene. The chemical 
scheme thus involves essentially neutral-neutral reactions, with the 
exception of benzene, which has been hypothesized to also be formed by a route involving 
ion-neutral reactions that result in the precursor ion C$_6$H$_7^+$ 
(see e.g. \citealt{Agundez2021a}).

\begin{figure*}
\centering
\includegraphics[width=0.9\textwidth]{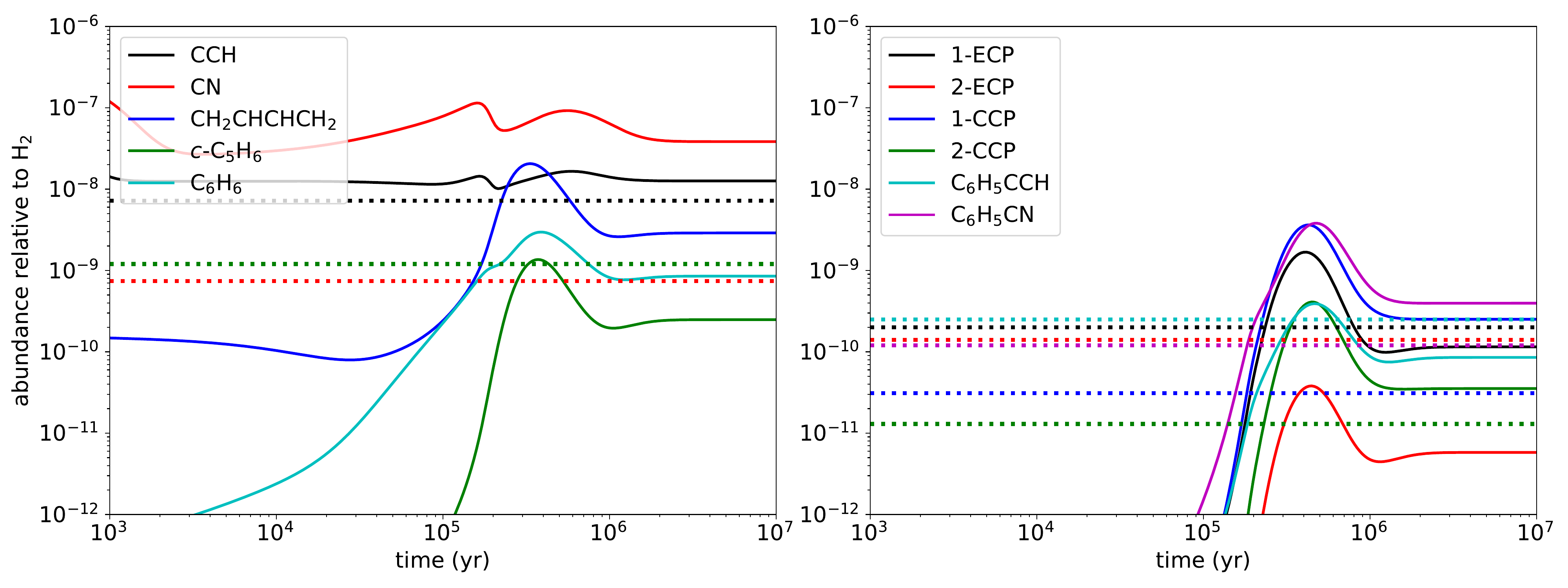}
\caption{Calculated abundances of CCH and CN derivatives of $c$-C$_5$H$_6$ 
and C$_6$H$_6$ (right panel) and of their precursors (left panel). The 
horizontal dotted lines correspond to the abundances observed in TMC-1.}
\label{fig:abun}
\end{figure*}

If we set the gas-phase elemental abundance of oxygen to O/H = 
3.3\,$\times$\,10$^{-4}$ (C/O = 0.55), the calculated peak abundances of 
all CCH/CN derivatives of $c$-C$_5$H$_6$ and C$_6$H$_6$ remain two to 
three orders of magnitude below the observed values. If the gas-phase 
elemental abundance of oxygen is decreased so that C/O = 1 
(O/H = 1.8\,$\times$\,10$^{-4}$), then the agreement between the chemical 
model and the observations is much better (see Fig.~\ref{fig:abun}). Producing 
$c$-C$_5$H$_6$ with an abundance as high as $\sim$\,10$^{-9}$ relative to H$_2$ 
and CCH/CN derivatives of $c$-C$_5$H$_6$ and C$_6$H$_6$ with abundances in 
the range 10$^{-11}$-10$^{-10}$ seems to require a C/O elemental gas-phase ratio 
close to that in TMC-1. This is in agreement with a recent study on the elemental 
abundances in TMC-1 \citep{Fuente2019}.

A significant failure of the chemical model concerns the abundance ratios between 
the CCH and the CN derivatives of both $c$-C$_5$H$_6$ and C$_6$H$_6$, which 
are below one, while the observed ones are above unity. This is ultimately caused 
by the overabundance of CN in the chemical model, a feature also found in previous 
chemical models of cold dark clouds (e.g. \citealt{Agundez2013,Daranlot2013}), 
which deserves a dedicated study. As a consequence, the calculated CCH/CN ratio 
is $<$\,1 at any time (see the left panel in Fig.~\ref{fig:abun}), while the observed 
CCH/CN ratio in TMC-1 is $\sim$\,10 \citep{Pratap1997}. This fact translates to 
the CCH and CN derivatives of $c$-C$_5$H$_6$ and C$_6$H$_6$, for which 
calculated CCH/CN ratios are $<$\,1, while the observed ones are $>$\,1 (see  the
right panel in Fig.~\ref{fig:abun}). It is worth noting that the CCH/CN ratio 
observed for cyclopentadiene,  (1-ECP + 2-ECP)/(1-CCP + 2-CCP) = 7.7, is close 
to the CCH/CN ratio itself, $\sim$\,10 \citep{Pratap1997}, while the CCH/CN 
ratio for benzene, C$_6$H$_5$CCH/C$_6$H$_5$CN $\leq$ 2.1, is significantly 
lower. This fact suggests that additional routes to the reaction CN + 
C$_6$H$_6$ could form C$_6$H$_5$CN in TMC-1. In the case of the 
CCH and CN derivatives of C$_3$H$_4$, the CCH/CN ratio is 3.5 
\citep{Marcelino2021,Cernicharo2021b}, which is in between the values found 
for the derivatives of $c$-C$_5$H$_6$ and C$_6$H$_6$.

Concerning the abundance ratios between different isomers, for $c$-C$_5$H$_5$CCH 
the two isomers 1-$ECP$ and 2-$ECP$ are observed with similar abundances. However, 
the chemical model calculates 1-$ECP$ to be much more abundant than 2-$ECP$ (see 
the right panel in Fig.~\ref{fig:abun}) because, in addition to the reaction C$_2$H 
+ $c$-C$_5$H$_6$, which produces the two isomers, 
the reaction between C$_3$H and butadiene yields 1-$ECP$ but 
not 2-$ECP$ (see Appendix~\ref{app:chem_scheme}). Similarly, 
for $c$-C$_5$H$_5$CN the chemical 
model calculates a 1-$CCP$/2-$CCP$ ratio higher than observed because 1-$CCP$ has more formation routes. We note, however, 
that the branching ratios adopted in the chemical model for the production of 
the different isomers are uncertain.

A success of the chemical model is that it correctly reproduces the abundance 
of cyclopentadiene. If the chemical scheme discussed in 
Appendix~\ref{app:chem_scheme} is complete, the abundances of the non-polar 
molecules benzene and butadiene in TMC-1 should be of the order of the 
calculated ones. That is, benzene should be present with an abundance of 
some 10$^{-9}$ relative to H$_2$, similar to that of cyclopentadiene, while 
butadiene should be around ten times more abundant than benzene (see the left 
panel in Fig.~\ref{fig:abun}). If butadiene is that abundant, it could be 
detected indirectly, for example through its protonated form or a polar derivative 
such as C$_4$H$_5$CN.

\section{Conclusions}

We have reported the detection in TMC-1 of two isomers of ethynyl cyclopentadiene 
($c$-C$_5$H$_5$CCH), namely 1-$ECP$ and 2-$ECP$, and the 
tentative detection of $EBZ$ (C$_6$H$_5$CCH). In addition, 
we report an exhaustive line-by-line detection of the cyano derivatives of 
cyclopentadiene and benzene ($c$-C$_5$H$_5$CN and C$_6$H$_5$CN). This 
allowed us to provide a coherent set of column densities for the various 
CCH and CN derivatives of cyclopentadiene and benzene in TMC-1. A chemical 
model that includes chemical routes to these cycles based on neutral-neutral 
reactions is reasonably successful in explaining the order of magnitude 
of the observed abundances. It is predicted that benzene should have an 
abundance similar to that of cyclopentadiene in TMC-1.

\begin{acknowledgements}

We thank ERC for funding through grant ERC-2013-Syg-610256-NANOCOSMOS 
and Ministerio de Ciencia e Innovaci\'on of Spain (MICIU) for 
funding support through projects PID2019-106110GB-I00, 
PID2019-107115GB-C21, and PID2019-106235GB-I00. M.A. 
thanks MICIU for grant RyC-2014-16277.

\end{acknowledgements}

\onecolumn
\begin{appendix}

\section{Observed lines of 1-$ECP$, 2-$ECP$, and $EBZ$}\label{observed_lines_ecp}

\begin{figure*}
\centering
\includegraphics[scale=0.60]{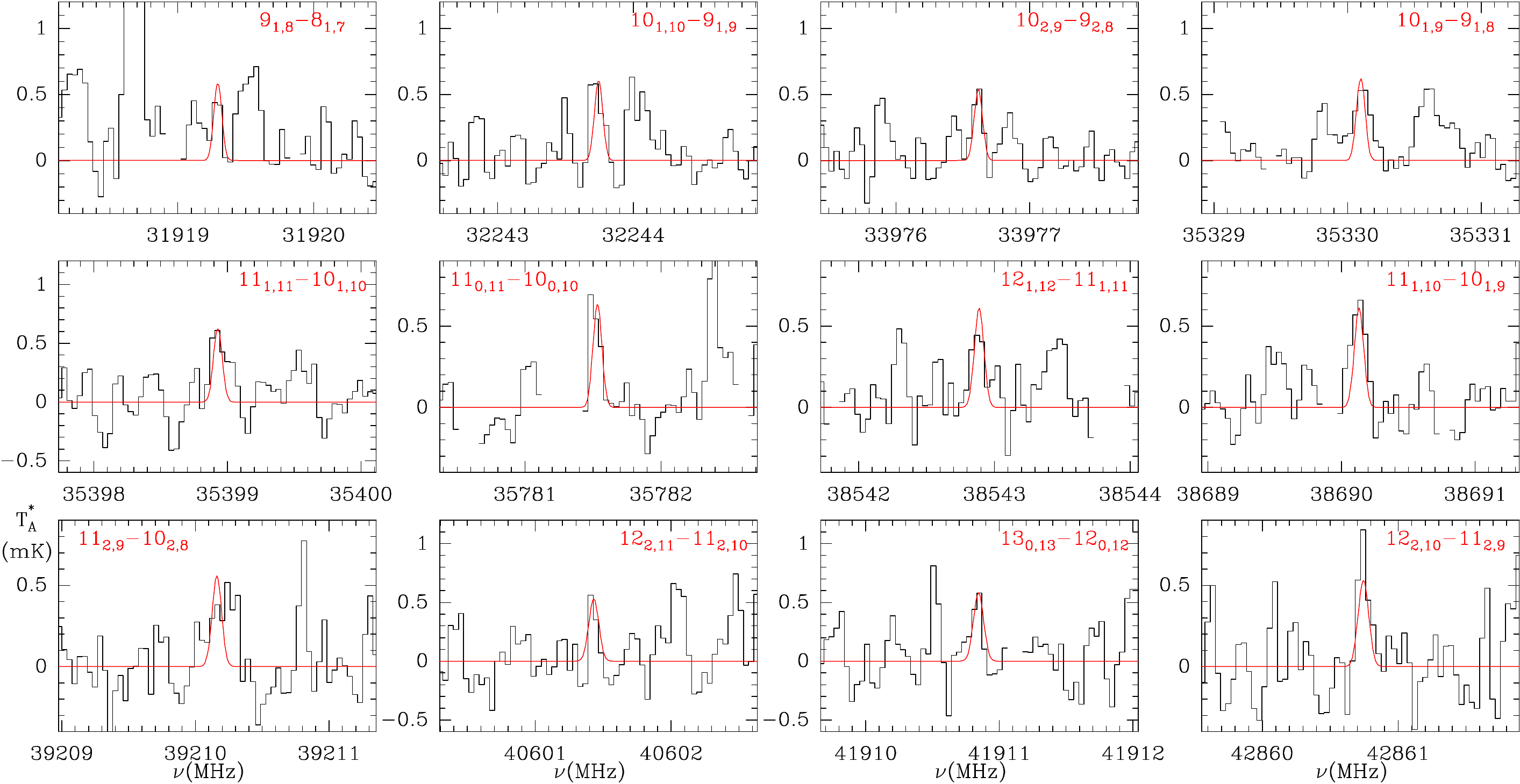}
\caption{
Observed lines of 1-$ECP$ in the 31-50 GHz frequency range towards TMC-1. 
Line parameters for the complete list of detected lines of 1-$ECP$ are given in Table \ref{line_parameters_1ecp}.
The abscissa corresponds to the rest frequency assuming a local standard of rest velocity of 5.83
km s$^{-1}$. 
The ordinate is the antenna temperature corrected for atmospheric and telescope losses in mK.
The red line shows the synthetic spectrum obtained for an assumed
T$_r$ of 9\,K and N(1-$ECP$)=(2.0$\pm$0.4)$\times$\doce. 
The rotational quantum
numbers are indicated in each panel. Blanked channels correspond to negative features produced
in the folding of the frequency switching data. 
}
\label{fig_1-ecp}
\end{figure*}

Line parameters were derived from a Gaussian fit to the observed lines using
the GILDAS package. A velocity
coverage of $\pm$15\,\kms\, was selected for each line. Observed frequencies were derived assuming a local standard of rest velocity of 5.83 \kms \citep{Cernicharo2020}. 
The predicted frequencies in MADEX \citep{Cernicharo2012}, which arise from a fit to the laboratory data of
\citet{McCarthy2020}, agree within 5-30 kHz with the observed ones. These differences are always
below 2$\times\sigma\times\Delta\nu$, where $\Delta\nu$ is 
the estimated frequency uncertainty of the observed lines. The derived line parameters for
2-$ECP$ and 1-$ECP$ are given in Tables \ref{line_parameters_2ecp} and 
\ref{line_parameters_1ecp}, respectively. 
Selected lines of 1-$ECP$ and 2-$ECP$ are shown in Figs. \ref{fig_1-ecp} and \ref{fig_2-ecp}, respectively.
A merged fit to the laboratory and TMC-1 frequencies
is discussed in Appendix \ref{new_constants_ecp}. 

\begin{table*}
\centering
\small
\caption{Observed line parameters of 2-$ECP$.}
\label{line_parameters_2ecp}
\begin{tabular}{lcccccr}
\hline
Transition       & $\nu_{pred}$\,$^a$ & $\nu_{obs}\,^b$ &  $\int$ $T_A^*$ dv $^c$ & $\Delta$v\,$^d$ & $T_A^*$\,$^e$ & N  \\
$(J_{K_{\rm a},K_{\rm c}})_{\rm u}-(J_{K_{\rm a},K_{\rm c}})_{\rm l}$                 & (MHz)         & (MHz)        & (mK\,km\,s$^{-1}$)   & (km\,s$^{-1}$)& (mK)   &  \\
\hline
$ 9_{2,7} -8_{2,6}$  &31804.974$\pm$0.010&31804.956$\pm$0.010&0.81$\pm$0.14&0.88$\pm$0.16&0.87$\pm$0.18& \\ %plot
$ 9_{1,8} -8_{1,7}$  &31836.466$\pm$0.009&31836.469$\pm$0.010&0.80$\pm$0.21&1.01$\pm$0.25&0.74$\pm$0.18& \\ %plot
$10_{1,10}-9_{1,9}$  &32128.912$\pm$0.006&32128.929$\pm$0.020&0.50$\pm$0.12&0.82$\pm$0.23&0.57$\pm$0.16& \\ %plot
$10_{0,10}-9_{0,9}$  &32584.193$\pm$0.006&                   &             &             &             &A\\
$10_{2,9}-9_{2,8}$   &33877.354$\pm$0.008&                   &             &             & $\le$0.54   &B\\
$10_{3,8}-9_{3,7}$   &34371.631$\pm$0.010&34371.645$\pm$0.020&0.35$\pm$0.13&0.53$\pm$0.22&0.63$\pm$0.16& \\
$10_{3,7}-9_{3,6}$   &34598.278$\pm$0.012&34598.267$\pm$0.010&0.47$\pm$0.10&0.68$\pm$0.20&0.65$\pm$0.16& \\ %plot
$10_{1,9}-9_{1,8}$   &35232.520$\pm$0.012&35232.541$\pm$0.010&0.44$\pm$0.08&0.70$\pm$0.15&0.59$\pm$0.16& \\ %plot
$11_{1,11}-10_{1,10}$&35271.182$\pm$0.009&35271.206$\pm$0.020&0.25$\pm$0.06&0.33$\pm$0.20&0.65$\pm$0.17&C\\
$10_{2,8}-9_{2,7}$   &35477.328$\pm$0.015&                   &             &             &$\le$0.55    & \\ %possible feature 35477.395$\pm$30
$11_{0,11}-10_{0,10}$&35640.891$\pm$0.008&35640.883$\pm$0.010&0.30$\pm$0.09&0.58$\pm$0.17&0.50$\pm$0.17& \\
$11_{2,10}-10_{2,9}$ &37187.287$\pm$0.010&37187.341$\pm$0.020&0.40$\pm$0.12&1.12$\pm$0.30&0.34$\pm$0.16& \\
$11_{3,9}-10_{3,8}$  &37818.228$\pm$0.014&                   &             &             &$\le$0.52    & \\ %feature at exact frequency 0.3mK
$11_{3,8}-10_{3,7}$  &38176.533$\pm$0.017&38176.490$\pm$0.010&0.77$\pm$0.14&1.02$\pm$0.21&0.70$\pm$0.16&D\\
$12_{1,12}-11_{1,11}$&38402.186$\pm$0.012&38402.176$\pm$0.010&0.40$\pm$0.09&0.61$\pm$0.16&0.62$\pm$0.16& \\ %plot
$11_{1,10}-10_{1,9}$ &38575.843$\pm$0.015&38575.866$\pm$0.010&0.53$\pm$0.12&0.54$\pm$0.13&0.54$\pm$0.16& \\ %plot
$12_{0,12}-11_{0,11}$&38692.902$\pm$0.012&38692.899$\pm$0.010&0.41$\pm$0.08&0.57$\pm$0.15&0.67$\pm$0.17& \\ %plot
$11_{2,9}-10_{2,8}$  &39139.995$\pm$0.021&39140.044$\pm$0.030&0.46$\pm$0.14&0.97$\pm$0.30&0.45$\pm$0.17& \\
$12_{2,11}-11_{2,10}$&40476.885$\pm$0.014&                   &             &             &             &E\\
$12_{3,10}-11_{3,9}$ &41258.658$\pm$0.019&41258.707$\pm$0.020&0.45$\pm$0.18&0.50$\pm$0.18&0.90$\pm$0.26&E\\
$13_{1,13}-12_{1,12}$&41523.491$\pm$0.016&41523.514$\pm$0.020&0.42$\pm$0.15&0.62$\pm$0.25&0.63$\pm$0.24& \\ %plot
$13_{0,13}-12_{0,12}$&41746.171$\pm$0.016&41746.185$\pm$0.020&0.53$\pm$0.17&0.89$\pm$0.27&0.56$\pm$0.25& \\ 
$12_{3,9}-11_{3,8}$  &41795.647$\pm$0.024&41795.617$\pm$0.020&0.30$\pm$0.11&0.54$\pm$0.20&0.53$\pm$0.20& \\ %plot??
$12_{1,11}-11_{1,10}$&41860.507$\pm$0.018&41860.552$\pm$0.010&0.39$\pm$0.09&0.40$\pm$0.15&0.99$\pm$0.20& \\ %plot
$12_{2,10}-11_{2,9}$ &42779.782$\pm$0.028&42779.776$\pm$0.020&0.37$\pm$0.10&0.66$\pm$0.25&0.52$\pm$0.20&F\\
$13_{2,12}-12_{2,11}$&43745.506$\pm$0.017&                   &             &             & $\le$0.62   & \\
$14_{1,14}-13_{1,13}$&44636.685$\pm$0.021&44636.635$\pm$0.010&0.51$\pm$0.13&0.55$\pm$0.15&0.88$\pm$0.24& \\ %plot
$14_{0,14}-13_{0,13}$&44803.670$\pm$0.021&44803.652$\pm$0.010&1.05$\pm$0.15&0.59$\pm$0.10&1.68$\pm$0.26& D\\ %plot ??
$13_{1,12}-12_{1,11}$&45083.511$\pm$0.021&45083.564$\pm$0.020&0.62$\pm$0.15&0.48$\pm$0.13&1.21$\pm$0.30& \\
$13_{3,10}-12_{3,9}$ &45456.801$\pm$0.033&45456.835$\pm$0.020&0.38$\pm$0.10&0.25$\pm$0.20&1.43$\pm$0.30& \\
$13_{2,11}-12_{2,10}$&46386.401$\pm$0.036&                   &             &             &$\le$0.9     & \\
$14_{2,13}-13_{2,12}$&46992.950$\pm$0.021&46992.978$\pm$0.010&0.60$\pm$0.16&0.44$\pm$0.14&1.26$\pm$0.30& \\
$15_{1,15}-14_{1,14}$&47743.272$\pm$0.027&                   &             &             & $\le$0.9    & \\
$15_{0,15}-14_{0,14}$&47866.359$\pm$0.027&                   &             &             & $\le$1.0    & \\
\hline
\end{tabular}
\tablefoot{\\
For the observational parameters we adopted the uncertainty of the Gaussian fit provided by \texttt{GILDAS}.
Upper limits correspond to 3$\sigma$ values.\\
\tablefoottext{a}{Predicted frequency from the rotational and distortion constants
derived from a fit to the lines observed by \citet{McCarthy2020} 
(see Sect. \ref{detection_2ecp}).}\\
\tablefoottext{b}{Observed frequency assuming a v$_{LSR}$ of 5.83 \kms.}\\
\tablefoottext{c}{Integrated line intensity in mK\,km\,s$^{-1}$.}\\
\tablefoottext{d}{Line width at half intensity derived by fitting a Gaussian function to
the observed line profile (in km\,s$^{-1}$).}\\
%\tablefoottext{e}{v$_{LSR}$ assuming that the predicted frequencies as the rest frequencies of the lines (in km\,s$^{-1}$).}\\
\tablefoottext{e}{Antenna temperature in millikelvin.}\\
\tablefoottext{A}{Blended with a negative feature resulting from the folding of the frequency switching data.}\\
\tablefoottext{B}{Possible blend with a negative feature (see note A). Fit unreliable.}\\
\tablefoottext{C}{Partially blended with a negative feature (see note A). Fit still possible.}\\
\tablefoottext{D}{Possible blend with an unidentified line. The line appears too strong compared with other transitions.}\\
\tablefoottext{E}{Fully blended with one of the hyperfine components of H$_2$CCN.}\\
\tablefoottext{F}{Possible blend with an unidentified line. The derived intensity is in line with that of other transitions.}\\
}
\end{table*}
\clearpage

\begin{table*}
\centering
\small
\caption{Observed line parameters of 1-$ECP$.}
\label{line_parameters_1ecp}
\begin{tabular}{lcccccr}
\hline
Transition       & $\nu_{pred}$\,$^a$ & $\nu_{obs}\,^b$ &  $\int$ $T_A^*$ dv $^c$ & $\Delta$v\,$^d$ & $T_A^*$\,$^e$ & N  \\
$(J_{K_{\rm a},K_{\rm c}})_{\rm u}-(J_{K_{\rm a},K_{\rm c}})_{\rm l}$                 & (MHz)         & (MHz)        & (mK\,km\,s$^{-1}$)   & (km\,s$^{-1}$)& (mK)   &  \\
\hline
$ 9_{3, 6}- 8_{3, 5}$&31125.481$\pm$0.003&                   &             &             & $\le$0.53       \\
$ 9_{2, 7}- 8_{2, 6}$&31859.742$\pm$0.004&                   &             &             & $\le$0.55       \\
$ 9_{1, 8}- 8_{1, 7}$&31919.294$\pm$0.003&31919.294$\pm$0.020&0.35$\pm$0.09&0.55$\pm$0.25&0.44$\pm$0.15& A\\
$10_{1,10}- 9_{1, 9}$&32243.744$\pm$0.006&32243.721$\pm$0.015&0.61$\pm$0.11&0.91$\pm$0.18&0.64$\pm$0.14&  \\
$10_{0,10}- 9_{0, 9}$&32711.504$\pm$0.005&                   &             &             &             & B\\
$10_{2, 9}- 9_{2, 8}$&33976.618$\pm$0.005&33976.611$\pm$0.010&0.44$\pm$0.10&0.73$\pm$0.18&0.57$\pm$0.15&  \\ 
$10_{3, 8}- 9_{3, 7}$&34456.710$\pm$0.004&                   &             &             & $\le$0.55    &  \\
$10_{3, 7}- 9_{3, 6}$&34670.863$\pm$0.005&                   &             &             & $\le$0.51    &  \\
$10_{1, 9}- 9_{1, 8}$&35330.100$\pm$0.004&35330.107$\pm$0.010&0.80$\pm$0.14&1.20$\pm$0.30&0.52$\pm$0.16&  \\
$11_{1,11}-10_{1,10}$&35398.929$\pm$0.008&35398.938$\pm$0.020&0.57$\pm$0.15&0.82$\pm$0.29&0.65$\pm$0.18&  \\
$10_{2, 8}- 9_{2, 7}$&35538.796$\pm$0.005&                   &             &             &$\le$0.50    & C\\
$11_{0,11}-10_{0,10}$&35781.534$\pm$0.008&35781.525$\pm$0.010&0.54$\pm$0.10&0.73$\pm$0.15&0.70$\pm$0.18&  \\
$11_{2,10}-10_{2, 9}$&37298.932$\pm$0.006&                   &             &             &$\le$0.50    & D\\
$11_{3, 9}-10_{3, 8}$&37912.444$\pm$0.005&                   &             &             &$\le$0.55    &  \\
$11_{3, 8}-10_{3, 7}$&38251.486$\pm$0.007&                   &             &             &$\le$0.59    &  \\
$12_{1,12}-11_{1,11}$&38542.888$\pm$0.011&38542.877$\pm$0.015&0.28$\pm$0.08&0.71$\pm$0.22&0.37$\pm$0.13&  \\
$11_{1,10}-10_{1, 9}$&38690.146$\pm$0.005&38690.125$\pm$0.015&0.71$\pm$0.12&0.98$\pm$0.18&0.65$\pm$0.14&  \\
$12_{0,12}-11_{0,11}$&38845.920$\pm$0.011&                   &             &             &$\le$0.51    &  \\
$11_{2, 9}-10_{2, 8}$&39210.161$\pm$0.007&39210.148$\pm$0.020&0.35$\pm$0.10&0.77$\pm$0.30&0.41$\pm$0.16& E\\
$12_{2,11}-11_{2,10}$&40601.469$\pm$0.009&40601.428$\pm$0.020&0.27$\pm$0.09&0.28$\pm$0.15&0.55$\pm$0.20& F\\
$12_{3,10}-11_{3, 9}$&41362.551$\pm$0.007&                   &             &             &$\le$<0.60   &  \\
$13_{1,13}-12_{1,12}$&41677.124$\pm$0.015&                   &             &             &$\le$<0.66   & G\\
$12_{3, 9}-11_{3, 8}$&41871.672$\pm$0.009&                   &             &             &$\le$<0.70   &  \\
$13_{0,13}-12_{0,12}$&41910.880$\pm$0.015&41910.870$\pm$0.015&0.37$\pm$0.12&0.60$\pm$0.25&0.58$\pm$0.18&  \\
$12_{1,11}-11_{1,10}$&41993.447$\pm$0.007&41993.453$\pm$0.025&0.64$\pm$0.15&0.70$\pm$0.32&0.86$\pm$0.20& H\\
$12_{2,10}-11_{2, 9}$&42860.722$\pm$0.010&42860.741$\pm$0.015&0.54$\pm$0.14&0.61$\pm$0.19&0.83$\pm$0.22&  \\
$13_{2,12}-12_{2,11}$&43883.551$\pm$0.011&                   &             &             &$\le$0.60    &  \\
$14_{1,14}-13_{1,13}$&44803.174$\pm$0.019&                   &             &             &$\le$0.76    &  \\
$13_{3,11}-12_{3,10}$&44804.061$\pm$0.009&                   &             &             &$\le$0.69    &  \\
$14_{0,14}-13_{0,13}$&44979.670$\pm$0.019&                   &             &             &$\le$0.87    &  \\ 
$13_{1,12}-12_{1,11}$&45236.599$\pm$0.010&                   &             &             &$\le$0.92    &  \\ 
$13_{3,10}-12_{3, 9}$&45533.079$\pm$0.013&                   &             &             &$\le$1.05    &  \\ 
$13_{2,11}-12_{2,10}$&46480.158$\pm$0.012&                   &             &             &$\le$0.91    &  \\ 
$14_{2,13}-13_{2,12}$&47144.916$\pm$0.015&                   &             &             &$\le$1.15    &  \\ 
$15_{1,15}-14_{1,14}$&47922.512$\pm$0.025&                   &             &             &             & I\\
$15_{0,15}-14_{0,14}$&48053.480$\pm$0.025&                   &             &             &$\le$1.15    &  \\
$14_{3,12}-13_{3,11}$&48234.023$\pm$0.012&                   &             &             &$\le$1.25    &  \\
$14_{1,13}-13_{1,12}$&48420.229$\pm$0.013&48420.188$\pm$0.015&0.55$\pm$0.13&0.48$\pm$0.12&1.0$\pm$0.29 &  \\
$14_{3,11}-13_{3,10}$&49233.211$\pm$0.017&                   &             &             &$\le$1.18    &  \\
\hline
\end{tabular}
\tablefoot{\\
For the observational parameters we adopted the uncertainty of the Gaussian fit provided by \texttt{GILDAS}.
Upper limits correspond to 3$\sigma$ values.\\
\tablefoottext{a}{Predicted frequency from the rotational and distortion constants
derived from a fit to the lines observed by \citet{McCarthy2020} 
(see Sect. \ref{detection_1ecp}).}\\
\tablefoottext{b}{Observed frequency assuming a v$_{LSR}$ of 5.83 \kms.}\\
\tablefoottext{c}{Integrated line intensity in mK\,km\,s$^{-1}$.}\\
\tablefoottext{d}{Line width at half intensity derived by fitting a Gaussian function to
the observed line profile (in km\,s$^{-1}$).}\\
%\tablefoottext{e}{v$_{LSR}$ assuming that the predicted frequencies as the rest frequencies of the lines (in km\,s$^{-1}$).}\\
\tablefoottext{e}{Antenna temperature in millikelvin.}\\
\tablefoottext{A}{Marginal detection.}\\
\tablefoottext{B}{Blended with the $J$=29-28 transition of HC$_7$N at 32711.672 MHz.}\\
\tablefoottext{C}{A 2.5$\sigma$ feature appears at the predicted frequency.}\\
\tablefoottext{D}{A 0.4 mK feature appears at -83 kHz of the predicted frequency. The line profile
is affected in its blue frequency wing by a negative feature produced in the
folding of the frequency switching data. Fit unreliable.}\\
\tablefoottext{E}{The line is blended with two unidentified features. Doubtful fit.}\\
\tablefoottext{F}{Possible blend with a negative feature in the red wing of the line profile.
Doubtful fit.}\\
\tablefoottext{G}{Feature of 0.5 mK at -130 kHz. It cannot be assigned to 1-$ECP$.}\\
\tablefoottext{H}{The line is blended in its red wing with a unknown feature.
Doubtful fit.}\\
\tablefoottext{I}{The line is fully blended with a negative feature produced in the folding
of the frequency switching data.}\\
}
\end{table*}
\clearpage

\subsection{Improved rotational constants for 1-$ECP$ and 2-$ECP$} \label{new_constants_ecp}

The observed lines of 2-$ECP$ (see Table \ref{line_parameters_2ecp}) and 1-$ECP$ 
(see Table \ref{line_parameters_1ecp}) were merged with the
laboratory data \citep{McCarthy2020} to provide a new set of rotational 
and distortion constants. They are given in Table \ref{table_new_constants_ecp} and can be used to 
predict the frequencies of the rotational transitions for the two observed isomers
of cyclopentadiene up to 50 GHz with an 
accuracy better than 15 kHz ($K_a\le3$). The calculated and the observed-minus-calculated frequencies
for 2-$ECP$ and 1-$ECP$ are given in Tables \ref{fitted_lines_2ecp} and \ref{fitted_lines_1ecp}, 
respectively.

\begin{table*}
\centering
\small
\caption{Rotational and distortion constants for 1-$ECP$ and 2-$ECP$.}
\label{table_new_constants_ecp}
\begin{tabular}{|lcc|cc|}
\hline
                       & 1-$ECP^a$         & 1-$ECP^b$       & 2-$ECP^c$         & 2-$ECP^b$        \\
Constant               &   (MHz)           &   (MHz)         &   (MHz)           &   (MHz)          \\
\hline                                                                                             
$A$                    &8374.116845(672)   &8374.116982(777) &   8262.3702(10)   &  8262.3708(15)   \\   
$B$                    &1879.207441(354)   &1879.206888(286) &   1876.250982(284)&  1876.252295(342)\\
$C$                    &1549.116825(323)   &1549.117494(236) &   1543.195532(187)&  1543.194285(305)\\
$\Delta_J$ 10$^{-5}$   &   6.862(218)      &   6.643(162)    &      4.879(292)   &     5.646(179)   \\
$\Delta_{JK}$ 10$^{-3}$&   2.4044(163)     &   2.4271(154)   &      2.5504(205)  &     2.5219(288)  \\
$\delta_J$ 10$^{-5}$   &   1.815(113)      &   1.6858(645)   &      1.082(216)   &     1.0217(707)  \\
$\delta_K$ 10$^{-4}$   &   15.59(145)      &   13.04(106)    &                   &     6.98(113)    \\
\hline                                                                                             
\hline                                                                                             
N$_{lines}$            & 30                &   44            & 39                & 63               \\
$\nu_{max}$ (GHz)      &24.974             &   48.420        & 25.805            & 46.992           \\
$(J,K_a)_{max}$        & 7,3               &   14,3          & 8,3               & 14,3             \\
$\sigma$ (kHz)         & 1.3               &    8.3          & 3.0               &   16.0           \\ 
\hline             
\end{tabular}
\tablefoot{\\
Values between parentheses correspond to the uncertainties of the parameters 
in units of the last significant digits.\\
\tablefoottext{a}{Rotational and distortion constants from a fit to the 
lines of 1-$ECP$ measured in the laboratory by \citet{McCarthy2020}, 
including the distortion constant $\delta_K$.}\\
\tablefoottext{b}{Merged fit to the laboratory and space frequencies.}\\
\tablefoottext{c}{Rotational and distortion constants from a fit to 
lines of 2-$ECP$ measured in the laboratory by \citet{McCarthy2020}
with the uncertainties of 2 kHz changed to 4 kHz.}\\
}
\end{table*}

\begin{table*}
\centering
\small
\caption{Observed and calculated line frequencies for 2-$ECP$.}
\label{fitted_lines_2ecp}
\begin{tabular}{crrrr}
\hline
Transition       & $\nu_{obs}$\,$^a$ & $\nu_{cal}\,^b$ & $\nu_{obs}$-$\nu_{cal}$$^c$& Notes  \\
                 & (MHz)             & (MHz)           & (MHz)                  & \\
\hline
$ 2_{1, 2}- 1_{1, 1}$& 6505.8227$\pm$0.004& 6505.8264$\pm$0.0005&-0.0037& 1\\
$ 1_{1, 0}- 1_{0, 1}$& 6719.1716$\pm$0.004& 6719.1700$\pm$0.0014& 0.0016& 1\\
$ 2_{0, 2}- 1_{0, 1}$& 6826.2008$\pm$0.004& 6826.2014$\pm$0.0005&-0.0006& 1\\
$ 2_{1, 1}- 2_{0, 2}$& 7064.9046$\pm$0.004& 7064.9047$\pm$0.0015&-0.0001& 1\\
$ 2_{1, 1}- 1_{1, 0}$& 7171.9370$\pm$0.004& 7171.9361$\pm$0.0006& 0.0010& 1\\
$ 3_{1, 2}- 3_{0, 3}$& 7607.0222$\pm$0.004& 7607.0213$\pm$0.0016& 0.0009& 1\\
$ 4_{0, 4}- 3_{1, 3}$& 7942.9745$\pm$0.004& 7942.9736$\pm$0.0018& 0.0010& 1\\
$ 4_{1, 3}- 4_{0, 4}$& 8372.6059$\pm$0.004& 8372.6036$\pm$0.0021& 0.0023& 1\\
$ 5_{1, 4}- 5_{0, 5}$& 9394.8545$\pm$0.004& 9394.8561$\pm$0.0028&-0.0016& 1\\
$ 1_{1, 1}- 0_{0, 0}$& 9805.5638$\pm$0.004& 9805.5612$\pm$0.0015& 0.0026& 1\\
$ 3_{1, 3}- 2_{1, 2}$& 9750.9497$\pm$0.004& 9750.9517$\pm$0.0007&-0.0020& 1\\
$ 3_{0, 3}- 2_{0, 2}$&10207.6962$\pm$0.004&10207.6961$\pm$0.0007& 0.0002& 1\\
$ 3_{2, 2}- 2_{2, 1}$&10258.2745$\pm$0.004&10258.2731$\pm$0.0007& 0.0014& 1\\
$ 3_{2, 1}- 2_{2, 0}$&10308.9078$\pm$0.004&10308.9107$\pm$0.0007&-0.0029& 1\\
$ 3_{1, 2}- 2_{1, 1}$&10749.8126$\pm$0.004&10749.8126$\pm$0.0008&-0.0000& 1\\
$ 5_{0, 5}- 4_{1, 4}$&11804.6273$\pm$0.004&11804.6243$\pm$0.0020& 0.0030& 1\\
$ 2_{1, 2}- 1_{0, 1}$&12891.9402$\pm$0.004&12891.9412$\pm$0.0016&-0.0010& 1\\
$ 4_{1, 4}- 3_{1, 3}$&12987.1482$\pm$0.004&12987.1487$\pm$0.0009&-0.0005& 1\\
$ 4_{0, 4}- 3_{0, 3}$&13551.9681$\pm$0.004&13551.9690$\pm$0.0009&-0.0009& 1\\
$ 4_{2, 3}- 3_{2, 2}$&13667.8217$\pm$0.004&13667.8189$\pm$0.0009& 0.0028& 1\\
$ 4_{3, 2}- 3_{3, 1}$&13702.0907$\pm$0.004&13702.0855$\pm$0.0017& 0.0052& 1\\
$ 4_{3, 1}- 3_{3, 0}$&13703.8882$\pm$0.004&13703.8985$\pm$0.0017&-0.0103& 1\\
$ 4_{2, 2}- 3_{2, 1}$&13793.6166$\pm$0.004&13793.6158$\pm$0.0009& 0.0008& 1\\
$ 4_{1, 3}- 3_{1, 2}$&14317.5513$\pm$0.004&14317.5513$\pm$0.0010&-0.0000& 1\\
$ 6_{0, 6}- 5_{1, 5}$&15684.0022$\pm$0.004&15684.0029$\pm$0.0022&-0.0007& 1\\
$ 5_{1, 5}- 4_{1, 4}$&16212.1187$\pm$0.004&16212.1190$\pm$0.0011&-0.0003& 1\\
$ 5_{0, 5}- 4_{0, 4}$&16848.7979$\pm$0.004&16848.7994$\pm$0.0011&-0.0015& 1\\
$ 5_{2, 4}- 4_{2, 3}$&17068.9149$\pm$0.004&17068.9111$\pm$0.0011& 0.0038& 1\\
$ 5_{3, 3}- 4_{3, 2}$&17137.5081$\pm$0.004&17137.5036$\pm$0.0020& 0.0045& 1\\
$ 5_{3, 2}- 4_{3, 1}$&17143.8417$\pm$0.004&17143.8366$\pm$0.0020& 0.0051& 1\\
$ 5_{2, 3}- 4_{2, 2}$&17317.2679$\pm$0.004&17317.2686$\pm$0.0011&-0.0007& 1\\
$ 5_{1, 4}- 4_{1, 3}$&17871.0508$\pm$0.004&17871.0519$\pm$0.0012&-0.0011& 1\\
$ 4_{1, 4}- 3_{0, 3}$&18596.1415$\pm$0.004&18596.1441$\pm$0.0022&-0.0026& 1\\
$ 6_{1, 6}- 5_{1, 5}$&19424.1049$\pm$0.004&19424.1046$\pm$0.0012& 0.0003& 1\\
$ 6_{0, 6}- 5_{0, 5}$&20091.4959$\pm$0.004&20091.4977$\pm$0.0013&-0.0018& 1\\
$ 6_{1, 5}- 5_{1, 4}$&21405.5877$\pm$0.004&21405.5901$\pm$0.0013&-0.0024& 1\\
$ 7_{1, 7}- 6_{1, 6}$&22621.9636$\pm$0.004&22621.9644$\pm$0.0014&-0.0008& 1\\
$ 7_{1, 6}- 6_{1, 5}$&24915.6787$\pm$0.010&24915.6815$\pm$0.0016&-0.0028& 1\\
$ 8_{1, 8}- 7_{1, 7}$&25805.2041$\pm$0.004&25805.2049$\pm$0.0015&-0.0008& 1\\
$ 9_{2, 7}- 8_{2, 6}$&31804.9560$\pm$0.010&31804.9681$\pm$0.0036&-0.0121& 2\\
$ 9_{1, 8}- 8_{1, 7}$&31836.4690$\pm$0.010&31836.4750$\pm$0.0024&-0.0060& 2\\
$10_{1,10}- 9_{1, 9}$&32128.9290$\pm$0.020&32128.9081$\pm$0.0021& 0.0209& 2\\
$10_{3, 8}- 9_{3, 7}$&34371.6450$\pm$0.020&34371.6387$\pm$0.0036& 0.0064& 2\\
$10_{3, 7}- 9_{3, 6}$&34598.2670$\pm$0.010&34598.2718$\pm$0.0042&-0.0048& 2\\
$10_{1, 9}- 9_{1, 8}$&35232.5410$\pm$0.010&35232.5337$\pm$0.0031& 0.0073& 2\\
$11_{1,11}-10_{1,10}$&35271.2060$\pm$0.020&35271.1752$\pm$0.0026& 0.0308& 2\\
$11_{0,11}-10_{0,10}$&35640.8830$\pm$0.010&35640.8855$\pm$0.0026&-0.0025& 2\\
$11_{2,10}-10_{2, 9}$&37187.3410$\pm$0.020&37187.3126$\pm$0.0043& 0.0284& 2\\
$11_{3, 8}-10_{3, 7}$&38176.4900$\pm$0.010&38176.5234$\pm$0.0057&-0.0334& 2\\
$12_{1,12}-11_{1,11}$&38402.1760$\pm$0.010&38402.1750$\pm$0.0035& 0.0010& 2\\
$11_{1,10}-10_{1, 9}$&38575.8660$\pm$0.010&38575.8633$\pm$0.0038& 0.0027& 2\\
$12_{0,12}-11_{0,11}$&38692.9020$\pm$0.010&38692.8915$\pm$0.0035& 0.0105& 2\\
$11_{2, 9}-10_{2, 8}$&39140.0440$\pm$0.030&39139.9920$\pm$0.0067& 0.0520& 2\\
$12_{3,10}-11_{3, 9}$&41258.7070$\pm$0.020&41258.6781$\pm$0.0052& 0.0289& 2\\
$13_{1,13}-12_{1,12}$&41523.5140$\pm$0.020&41523.4740$\pm$0.0047& 0.0400& 2\\
$13_{0,13}-12_{0,12}$&41746.1850$\pm$0.020&41746.1544$\pm$0.0047& 0.0306& 2\\
$12_{3,9}-11_{3,8}$  &41795.6170$\pm$0.020&41795.6321$\pm$0.0075&-0.0151& 2\\
$12_{1,11}-11_{1,10}$&41860.5520$\pm$0.010&41860.5363$\pm$0.0049& 0.0157& 2\\
$12_{2,10}-11_{2, 9}$&42779.7760$\pm$0.020&42779.7823$\pm$0.0085&-0.0063& 2\\
$14_{1,14}-13_{1,13}$&44636.6350$\pm$0.010&44636.6596$\pm$0.0063&-0.0246& 2\\
$14_{0,14}-13_{0,13}$&44803.6520$\pm$0.010&44803.6444$\pm$0.0063& 0.0076& 2\\
$13_{1,12}-12_{1,11}$&45083.5640$\pm$0.020&45083.5500$\pm$0.0062& 0.0140& 2\\
$13_{3,10}-12_{3, 9}$&45456.8350$\pm$0.020&45456.7837$\pm$0.0112& 0.0513& 2\\
$14_{2,13}-13_{2,12}$&46992.9780$\pm$0.010&46992.9999$\pm$0.0088&-0.0219& 2\\
\hline
\end{tabular}
\tablefoot{\\
\tablefoottext{a}{Observed line frequencies and assigned uncertainties.}\\
\tablefoottext{b}{Calculated frequencies and uncertainties.}\\
\tablefoottext{c}{Observed-minus-calculated frequencies.}\\
\tablefoottext{1}{Laboratory data from \citet{McCarthy2020}.}\\
\tablefoottext{2}{Frequencies observed in TMC-1 assuming a v$_{LSR}$ of 5.83 km\,s$^{-1}$ (see Table
\ref{line_parameters_2ecp}).}\\
}
\end{table*}

\clearpage
\begin{table*}
\centering
\small
\caption{Observed and calculated line frequencies for 1-$ECP$.}
\label{fitted_lines_1ecp}
\begin{tabular}{crrrr}
\hline
Transition       & $\nu_{obs}$\,$^a$ & $\nu_{cal}\,^b$ & $\nu_{obs}$-$\nu_{cal}$$^c$& Notes  \\
                 & (MHz)             & (MHz)           & (MHz)                  & \\
\hline
$ 2_{1, 2}- 1_{1, 1}$& 6526.5517$\pm$0.002& 6526.5533$\pm$0.0002&-0.0016&1\\
$ 1_{1, 0}- 1_{0, 1}$& 6824.9928$\pm$0.002& 6824.9920$\pm$0.0008& 0.0008&1\\
$ 2_{0, 2}- 1_{0, 1}$& 6844.3837$\pm$0.002& 6844.3828$\pm$0.0002& 0.0009&1\\
$ 2_{1, 1}- 1_{1, 0}$& 7186.7197$\pm$0.002& 7186.7206$\pm$0.0003&-0.0009&1\\
$ 5_{1, 4}- 5_{0, 5}$& 9468.7286$\pm$0.002& 9468.7268$\pm$0.0012& 0.0018&1\\
$ 3_{1, 3}- 2_{1, 2}$& 9782.3011$\pm$0.002& 9782.3002$\pm$0.0003& 0.0009&1\\
$ 1_{1, 1}- 0_{0, 0}$& 9923.2290$\pm$0.002& 9923.2320$\pm$0.0008&-0.0030&1\\
$ 3_{0, 3}- 2_{0, 2}$&10236.0247$\pm$0.002&10236.0231$\pm$0.0003& 0.0016&1\\
$ 3_{2, 1}- 2_{2, 0}$&10333.8478$\pm$0.002&10333.8506$\pm$0.0003&-0.0028&1\\
$ 3_{1, 2}- 2_{1, 1}$&10772.2630$\pm$0.002&10772.2641$\pm$0.0004&-0.0011&1\\
$ 2_{1, 2}- 1_{0, 1}$&13021.4616$\pm$0.002&13021.4612$\pm$0.0009& 0.0004&1\\
$ 4_{1, 4}- 3_{1, 3}$&13029.4062$\pm$0.002&13029.4049$\pm$0.0004& 0.0013&1\\
$ 4_{0, 4}- 3_{0, 3}$&13591.6516$\pm$0.002&13591.6512$\pm$0.0004& 0.0004&1\\
$ 4_{2, 3}- 3_{2, 2}$&13703.6639$\pm$0.002&13703.6635$\pm$0.0004& 0.0004&1\\
$ 4_{3, 2}- 3_{3, 1}$&13736.7971$\pm$0.002&13736.7980$\pm$0.0008&-0.0009&1\\
$ 4_{3, 1}- 3_{3, 0}$&13738.5067$\pm$0.002&13738.5064$\pm$0.0008& 0.0003&1\\
$ 4_{2, 2}- 3_{2, 1}$&13825.2872$\pm$0.002&13825.2873$\pm$0.0004&-0.0001&1\\
$ 4_{1, 3}- 3_{1, 2}$&14348.0230$\pm$0.002&14348.0245$\pm$0.0005&-0.0015&1\\
$ 5_{1, 5}- 4_{1, 4}$&16265.6263$\pm$0.002&16265.6259$\pm$0.0004& 0.0005&1\\
$ 5_{0, 5}- 4_{0, 4}$&16901.2787$\pm$0.002&16901.2786$\pm$0.0004& 0.0001&1\\
$ 5_{2, 4}- 4_{2, 3}$&17114.2495$\pm$0.002&17114.2489$\pm$0.0005& 0.0007&1\\
$ 5_{2, 3}- 4_{2, 2}$&17354.5142$\pm$0.002&17354.5150$\pm$0.0005&-0.0008&1\\
$ 5_{1, 4}- 4_{1, 3}$&17910.0601$\pm$0.002&17910.0614$\pm$0.0005&-0.0013&1\\
$ 6_{1, 6}- 5_{1, 5}$&19489.2327$\pm$0.002&19489.2329$\pm$0.0005&-0.0002&1\\
$ 6_{0, 6}- 5_{0, 5}$&20158.1916$\pm$0.002&20158.1912$\pm$0.0005& 0.0004&1\\
$ 6_{2, 5}- 5_{2, 4}$&20514.6554$\pm$0.002&20514.6529$\pm$0.0007& 0.0025&1\\
$ 6_{2, 4}- 5_{2, 3}$&20926.1530$\pm$0.002&20926.1532$\pm$0.0008&-0.0002&1\\
$ 6_{1, 5}- 5_{1, 4}$&21453.8466$\pm$0.002&21453.8466$\pm$0.0006& 0.0000&1\\
$ 7_{0, 7}- 6_{0, 6}$&23360.7767$\pm$0.002&23360.7759$\pm$0.0006& 0.0008&1\\
$ 7_{1, 6}- 6_{1, 5}$&24974.1324$\pm$0.002&24974.1330$\pm$0.0008&-0.0006&1\\
$ 9_{1, 8}- 8_{1, 7}$&31919.2940$\pm$0.020&31919.2920$\pm$0.0016& 0.0020&2\\
$10_{1,10}- 9_{1, 9}$&32243.7210$\pm$0.015&32243.7383$\pm$0.0024&-0.0173&2\\
$10_{2, 9}- 9_{2, 8}$&33976.6110$\pm$0.010&33976.6075$\pm$0.0027& 0.0035&2\\
$10_{1, 9}- 9_{1, 8}$&35330.1070$\pm$0.010&35330.0961$\pm$0.0022& 0.0109&2\\
$11_{1,11}-10_{1,10}$&35398.9380$\pm$0.020&35398.9204$\pm$0.0034& 0.0176&2\\
$11_{0,11}-10_{0,10}$&35781.5250$\pm$0.010&35781.5254$\pm$0.0033&-0.0004&2\\
$12_{1,12}-11_{1,11}$&38542.8770$\pm$0.015&38542.8772$\pm$0.0048&-0.0002&2\\
$11_{1,10}-10_{1, 9}$&38690.1250$\pm$0.015&38690.1393$\pm$0.0029&-0.0143&2\\
$11_{2, 9}-10_{2, 8}$&39210.1480$\pm$0.020&39210.1666$\pm$0.0054&-0.0186&2\\
$12_{2,11}-11_{2,10}$&40601.4280$\pm$0.020&40601.4505$\pm$0.0047&-0.0225&2\\
$13_{0,13}-12_{0,12}$&41910.8700$\pm$0.015&41910.8654$\pm$0.0064& 0.0046&2\\
$12_{1,11}-11_{1,10}$&41993.4530$\pm$0.025&41993.4358$\pm$0.0039& 0.0172&2\\
$12_{2,10}-11_{2, 9}$&42860.7410$\pm$0.015&42860.7280$\pm$0.0069& 0.0130&2\\
$14_{1,13}-13_{1,12}$&48420.1880$\pm$0.015&48420.2025$\pm$0.0068&-0.0145&2\\
\hline
\end{tabular}
\tablefoot{\\
\tablefoottext{a}{Observed line frequencies and assigned uncertainties.}\\
\tablefoottext{b}{Calculated frequencies and uncertainties.}\\
\tablefoottext{c}{Observed-minus-calculated frequencies.}\\
\tablefoottext{1}{Laboratory data from \citet{McCarthy2020}.}\\
\tablefoottext{2}{Frequencies observed in TMC-1 assuming a v$_{LSR}$ of 5.83 km\,s$^{-1}$.}\\
}
\end{table*}

\clearpage

\subsection{Ethynyl benzene}\label{appendix_ebz}
\begin{figure*}
\centering
\includegraphics[scale=0.60,angle=0]{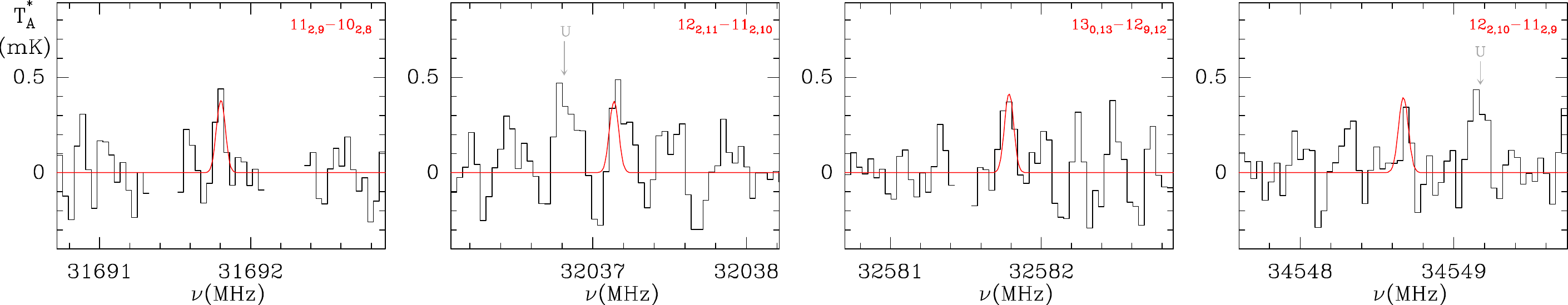}
\caption{Four lines of $EBZ$ (C$_6$H$_5$CCH) observed in the 31-50 GHz 
frequency range towards TMC-1. 
The abscissa corresponds to the rest frequency assuming a local standard of rest velocity of 5.83
km s$^{-1}$. 
The ordinate is the antenna temperature corrected for atmospheric and telescope losses in mK.
The red line shows the synthetic spectrum obtained from a fit to the observed line profiles
adopting a rotational temperature of 9\,K, 
which provides N($EBZ$)=(2.5$\pm$0.4)$\times$\doce\,
(see Sect. \ref{ethynyl_benzene}).
The rotational quantum
numbers are indicated in each panel. Blanked channels correspond to negative features produced
in the folding of the frequency switching data.
}
\label{fig_ebz}
\end{figure*}

For $EBZ$, the searched lines are given in
Table \ref{line_parameters_ebz}. Only four lines appear at the 3$\sigma$ level and are shown in
Fig. \ref{fig_ebz}. The estimated column density is discussed in Sect. \ref{ethynyl_benzene}.

\clearpage

\begin{table*}
\centering
\small
\caption{Observed line parameters of $EBZ$ (ethynyl benzene).}
\label{line_parameters_ebz}
\begin{tabular}{lcccccr}
\hline
Transition       & $\nu_{pred}$\,$^a$ & v$_{LSR}$$^b$            &  $\int$ $T_A^*$ dv $^c$ & $\Delta$v\,$^d$ & $T_A^*$\,$^e$ & N  \\
$(J_{K_{\rm a},K_{\rm c}})_{\rm u}-(J_{K_{\rm a},K_{\rm c}})_{\rm l}$                 & (MHz)              & (kms\,s$^{-1}$)      & (mK\,km\,s$^{-1}$)   & (km\,s$^{-1}$)& (mK)   &  \\
\hline
$11_{2, 9}-10_{2, 8}$&31691.804$\pm$0.001&5.91$\pm$0.09&0.29$\pm$0.08&0.58$\pm$0.16&0.47$\pm$0.15& \\ 
$12_{2,11}-11_{2,10}$&32037.138$\pm$0.001&5.73$\pm$0.19&0.30$\pm$0.10&0.52$\pm$0.30&0.54$\pm$0.15& \\
$13_{1,13}-12_{1,12}$&32505.867$\pm$0.001&             &             &             &$\le$0.54    & \\
$13_{0,13}-12_{0,12}$&32581.787$\pm$0.001&5.93$\pm$0.12&0.37$\pm$0.09&0.85$\pm$0.22&0.41$\pm$0.15& \\
$12_{1,11}-11_{1,10}$&32988.304$\pm$0.001&             &             &             &$\le$0.50    & \\
$12_{2,10}-11_{2, 9}$&34548.672$\pm$0.001&5.63$\pm$0.15&0.18$\pm$0.06&0.47$\pm$0.17&0.37$\pm$0.14& \\ 
$13_{2,12}-12_{2,11}$&34570.607$\pm$0.001&             &             &             &             & A\\ 
$14_{1,14}-13_{1,13}$&34926.745$\pm$0.001&             &             &             &$\le$0.66    & \\
$14_{0,14}-13_{0,13}$&34977.526$\pm$0.001&             &             &             &$\le$0.51    & \\
$13_{1,12}-12_{1,11}$&35391.232$\pm$0.001&             &             &             &$\le$0.48    & \\
$14_{2,13}-13_{2,12}$&37081.142$\pm$0.001&             &             &             &$\le$0.44    & \\
$15_{1,15}-14_{1,14}$&37343.953$\pm$0.001&             &             &             &$\le$0.40    & \\
$13_{2,11}-12_{2,10}$&37344.487$\pm$0.001&             &             &             &$\le$0.40    & \\
$15_{0,15}-14_{0,14}$&37377.435$\pm$0.001&             &             &             &$\le$0.55    & \\
$14_{1,13}-13_{1,12}$&37756.431$\pm$0.001&             &             &             &$\le$0.60    & \\
$15_{2,14}-14_{2,13}$&39570.906$\pm$0.001&             &             &             &$\le$0.54    & \\
$16_{1,16}-15_{1,15}$&39758.590$\pm$0.001&             &             &             &$\le$0.48    & \\
$16_{0,16}-15_{0,15}$&39780.407$\pm$0.001&             &             &             &$\le$0.65    & B\\
$14_{2,12}-13_{2,11}$&40070.665$\pm$0.001&             &             &             &$\le$0.62    & C\\
$15_{1,14}-14_{1,13}$&40103.212$\pm$0.001&             &             &             &$\le$0.56    & D\\
$16_{2,15}-15_{2,14}$&42042.517$\pm$0.001&             &             &             &$\le$0.69    & E\\
$17_{1,17}-16_{1,16}$&42171.467$\pm$0.001&             &             &             &$\le$0.66    & \\%%%
$17_{0,17}-16_{0,16}$&42185.545$\pm$0.001&             &             &             &$\le$0.60    & \\
$16_{1,15}-15_{1,14}$&42446.821$\pm$0.001&             &             &             &$\le$0.60    & \\
$15_{2,13}-14_{2,12}$&42719.693$\pm$0.001&             &             &             &$\le$0.60    & \\
$17_{2,16}-16_{2,15}$&44498.814$\pm$0.001&             &             &             &$\le$0.60    & \\
$18_{1,18}-17_{1,17}$&44583.166$\pm$0.001&             &             &             &$\le$0.60    & \\
$18_{0,18}-17_{0,17}$&44592.174$\pm$0.001&             &             &             &$\le$0.60    & \\
$17_{1,16}-16_{1,15}$&44796.545$\pm$0.001&             &             &             &$\le$0.60    & \\
$16_{2,14}-15_{2,13}$&45286.973$\pm$0.001&             &             &             &$\le$0.60    & \\
$18_{2,17}-17_{2,16}$&46942.629$\pm$0.001&             &             &             &$\le$0.60    & \\
$19_{1,19}-18_{1,18}$&46994.089$\pm$0.001&             &             &             &$\le$0.60    & \\
$19_{0,19}-18_{0,18}$&46999.812$\pm$0.001&             &             &             &$\le$0.60    & \\
$18_{1,17}-17_{1,16}$&47156.405$\pm$0.001&             &             &             &$\le$0.60    & \\
$17_{2,15}-16_{2,14}$&47773.527$\pm$0.001&             &             &             &$\le$0.60    & \\
$19_{2,18}-18_{2,17}$&49376.600$\pm$0.001&             &             &             &$\le$0.60    & \\
$20_{0,20}-19_{0,19}$&49408.126$\pm$0.001&             &             &             &$\le$0.60    & \\
\hline
\end{tabular}
\tablefoot{\\
For the observational parameters we adopted the uncertainty of the Gaussian fit provided by \texttt{GILDAS}.
Upper limits correspond to 3$\sigma$ values.\\
\tablefoottext{a}{Predicted frequency from the rotational and distortion constants
derived from a fit to the lines observed by \citet{Cox1975}, \citet{Dreizler2004}, and \citet{Kisiel2010} 
(see Sect. \ref{ethynyl_benzene}).}\\
\tablefoottext{b}{Observed v$_{LSR}$ in \kms.}\\
\tablefoottext{c}{Integrated line intensity in mK\,km\,s$^{-1}$.}\\
\tablefoottext{d}{Line width at half intensity derived by fitting a Gaussian function to
the observed line profile (in km\,s$^{-1}$).}\\
%\tablefoottext{e}{v$_{LSR}$ assuming that the predicted frequencies as the rest frequencies of the lines (in km\,s$^{-1}$).}\\
\tablefoottext{e}{Antenna temperature in millikelvin.}\\
\tablefoottext{A}{Fully blended with a negative feature produced in the folding of the
 frequency switching data.}\\
\tablefoottext{B}{Partially blended with a unknown feature at 39780.350 MHz. Fit unreliable.}\\
\tablefoottext{C}{Partially blended with a unknown feature at 40070.716 MHz. Fit unreliable.}\\
\tablefoottext{D}{Partially blended with a unknown feature at 40103.150 MHz. Fit unreliable.}\\
\tablefoottext{E}{Partially blended with a unknown feature at 42042.460 MHz. Fit unreliable.}\\
}
\end{table*}

\clearpage

\section{Observed lines of the two isomers of cyano cyclopentadiene (1-$CCP$ and 2-$CCP$)} \label{cyano_cyclopentadiene}

Two isomers of cyano cyclopentadiene, 
1-$CCP$ and  2-$CCP$, have been observed in the laboratory
by different authors \citep{Ford1978,Sakaizumi1987,McCarthy2020}. The structure
of these two isomers is similar to those of 1-$ECP$ and 2-$ECP$ when the CCH group is changed by CN (see Fig. \ref{fig_structure}). 
Dipole moments for both species
were measured by \citet{Ford1978} and \citet{Sakaizumi1987}, who provided values of 
$\mu_a$=4.15$\pm$0.15 and 4.36$\pm$0.25\,D for
1-$CCP$ and 2-$CCP$, respectively. 
Transitions of $b$ type are expected to be much weaker
due to the measured low value of the dipole moment along the $b$ axis of 0.27 and 0.77\,D, respectively \citep{Sakaizumi1987}.

The accuracy of the measurements in \citet{McCarthy2020} is much
higher than those of the early microwave measurements of the two isomers, and their rotational
and distortion constants, improved by additional measurements by \citet{Lee2021}, were used to
predict their frequencies in the domain of our QUIJOTE line survey. 
The two isomers were detected through stacking techniques towards TMC-1 by \citet{McCarthy2021}
and \citet{Lee2021}. However, as mentioned in Sect. \ref{cyano_derivatives} and below, the column densities
they provide are rather uncertain, with a variation for N(1-$CCP$) of a factor of four between both
publications. 
In order to provide a coherent set of column densities for the ethynyl and cyano
derivatives of cyclopentadiene and benzene, we analysed their lines within our survey.

\begin{figure*}
\centering
\includegraphics[scale=0.55,angle=0]{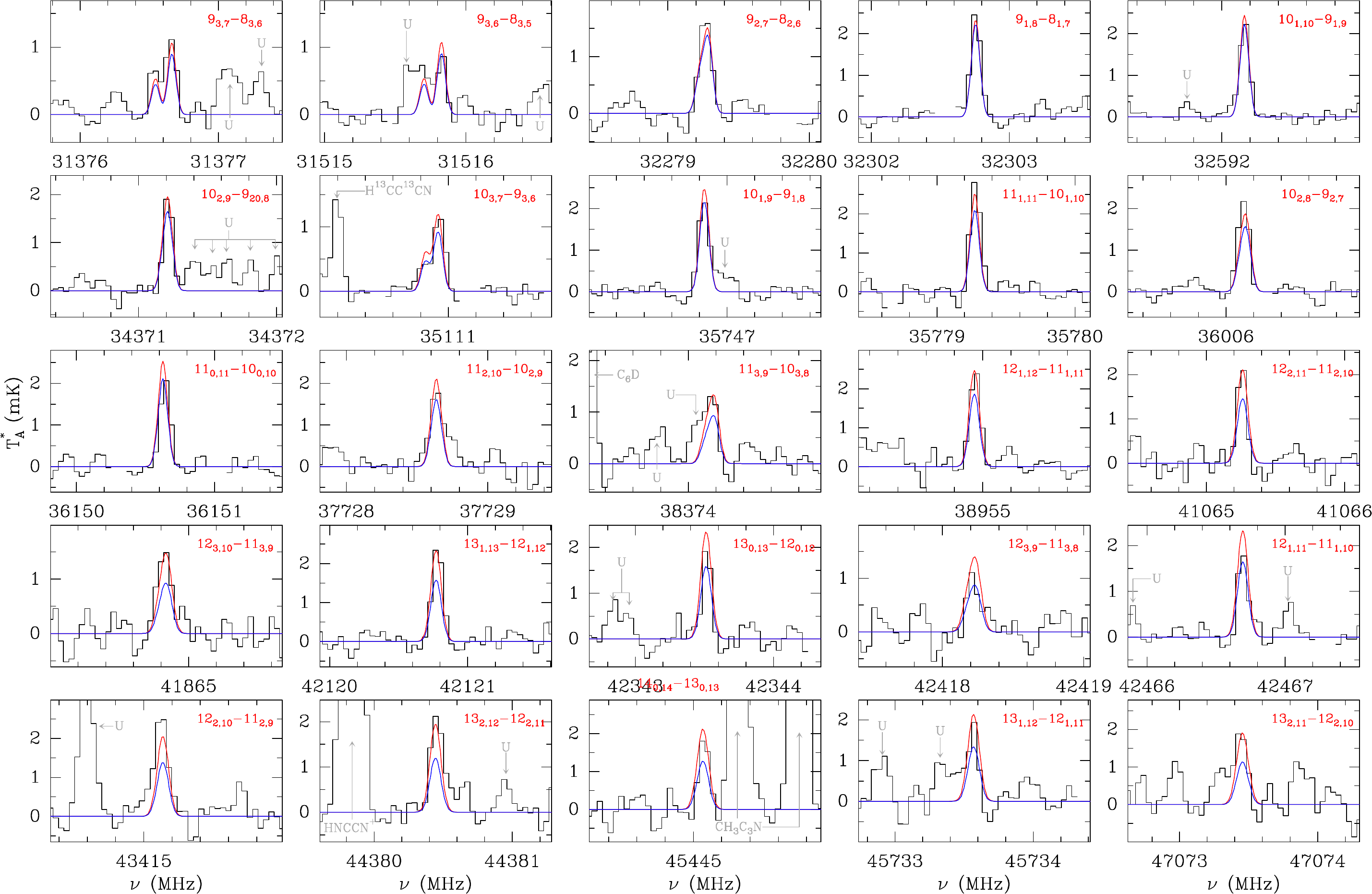}
\caption{Selected lines of 1-$CCP$ in the 31-50 GHz 
frequency range towards TMC-1. 
The abscissa corresponds to the rest frequency assuming a local standard of rest velocity of 5.83
km s$^{-1}$. 
The ordinate is the antenna temperature corrected for atmospheric and telescope losses in mK.
The red line shows the synthetic spectrum obtained from a fit to the observed line profiles,
which provides T$_{rot}$=9.0$\pm$1.0\,K and N(1-$CCP$)=(3.1$\pm$0.3)$\times$\once
(see Appendix \ref{cyano_cyclopentadiene}). The blue line
shows the synthetic spectrum for the same column density and a rotational temperature of 6 K.
The rotational quantum
numbers are indicated in each panel. Blanked channels correspond to negative features produced
in the folding of the frequency switching data.
}
\label{fig_1-CCP}
\end{figure*}

For 1-$CCP$, the published laboratory frequencies of \citet{McCarthy2020} cover the range 7.3-29.9 GHz, with $J_{max}$=9 
and $K_a\le$3. The rotational constants provided by \citet{Lee2021} were derived with unpublished frequencies
up to 36 GHz and provide frequency predictions for the 31-50 GHz domain with uncertainties $\le$15 kHz.
We observed 47 lines of 1-$CCP$ with QUIJOTE, and their
line parameters are given in Table \ref{line_parameters_1-ccp}.
Selected lines for 1-$CCP$ are shown in Fig. \ref{fig_1-CCP}. The red and 
blue lines in Fig. \ref{fig_1-CCP} show
the synthetic spectrum for a column density of 3.1$\times$\once\, and rotational temperatures of 9 and 6\,K, respectively.
It is clear, as it is robustly derived for benzonitrile in Appendix \ref{cyano_benzene}, that the rotational temperature 
has to be close
to 9\,K in order to fit the observed parameters for lines arising from 
levels with $K_a$=3-4. A fit
to the observed line profiles, assuming a source diameter of 80$''$ \citep{Fosse2001}, 
provides T$_{rot}$=9.0$\pm$1.0\,K and N(1-$CCP$)=(3.1$\pm$0.3)$\times$\once. 
The lines shown in Fig. \ref{fig_1-CCP} that involve energy levels below 10\,K can be fitted with a rotational
temperature of 6\,K with practically the same column density, a fact that has been discussed for other
species by \citet{Cernicharo2021e}; it arises from the small dependence of the line opacity with T$_{rot}$
for these transitions as well as from the balance between the term [T$_{rot}$\,$-$\,2.7] and the partition function, which
is proportional to T$_{rot}$ for linear molecules and to T$_{rot}^{3/2}$ for asymmetric species. Only
with observations that provide a reasonable range of upper level energies for the lines is it possible to derive
an accurate rotational temperature and a reliable column density.

\citet{McCarthy2021}
reported for 1-$CCP$ a rotational temperature of 6.0$\pm$0.3\,K and a total column density of 
(1.44$\pm$0.17)$\times$\doce.
However, \citet{Lee2021}
revised these values, also using spectral data stacking, and derive a rotational temperature of 
6.00$\pm$0.03\,K and a total column density 
of (8.27$\pm$0.10)$\times$\once\, for the same isomer (i.e. a factor of three larger than our estimation and a factor of 1.7
lower than in \citealt{McCarthy2021}). 
\citet{Lee2021} argue that they used a larger partition 
function
for 1-$CCP$ than \citet{McCarthy2021}, resulting in a substantially lower column density 
for this isomer.
However, the effect should be the opposite as the column density is proportional
to the partition function and to the rotational temperature, which is the same in
both studies. The difference in the column density between both works
is probably due to the highest data sensitivity in the \citet{Lee2021} 
analysis.
Their results correspond to a fit to four velocity components with different spatial sizes. 
The lack of individual
lines, together with the limited spatial coverage of their data (a single position with all 
information
on spatial sizes arising from the beam size variation with frequency), renders the fit rather 
optimistic,
and   many of the derived parameters may be strongly correlated. The lack of spatial
information also applies to the QUIJOTE data. However, 
a rotational temperature as low as 6\,K cannot explain our 47 observed lines as shown in
Fig. \ref{fig_1-CCP}. Although the stacking procedure is a powerful method for extracting information 
from spectral data below the noise level, the determination of physical parameters from line stacking has to be done 
with extreme caution. The determination of the column density and rotational 
temperature of 1-$CCP$ by \cite{McCarthy2021} and \cite{Lee2021} seems to be 
a bit optimistic given their small quoted uncertainties. The 
differences between the values determined in this work and their studies
could be related to the adopted source diameter and to the beam size
of both telescopes.

\begin{figure*}
\centering
\includegraphics[scale=0.55,angle=0]{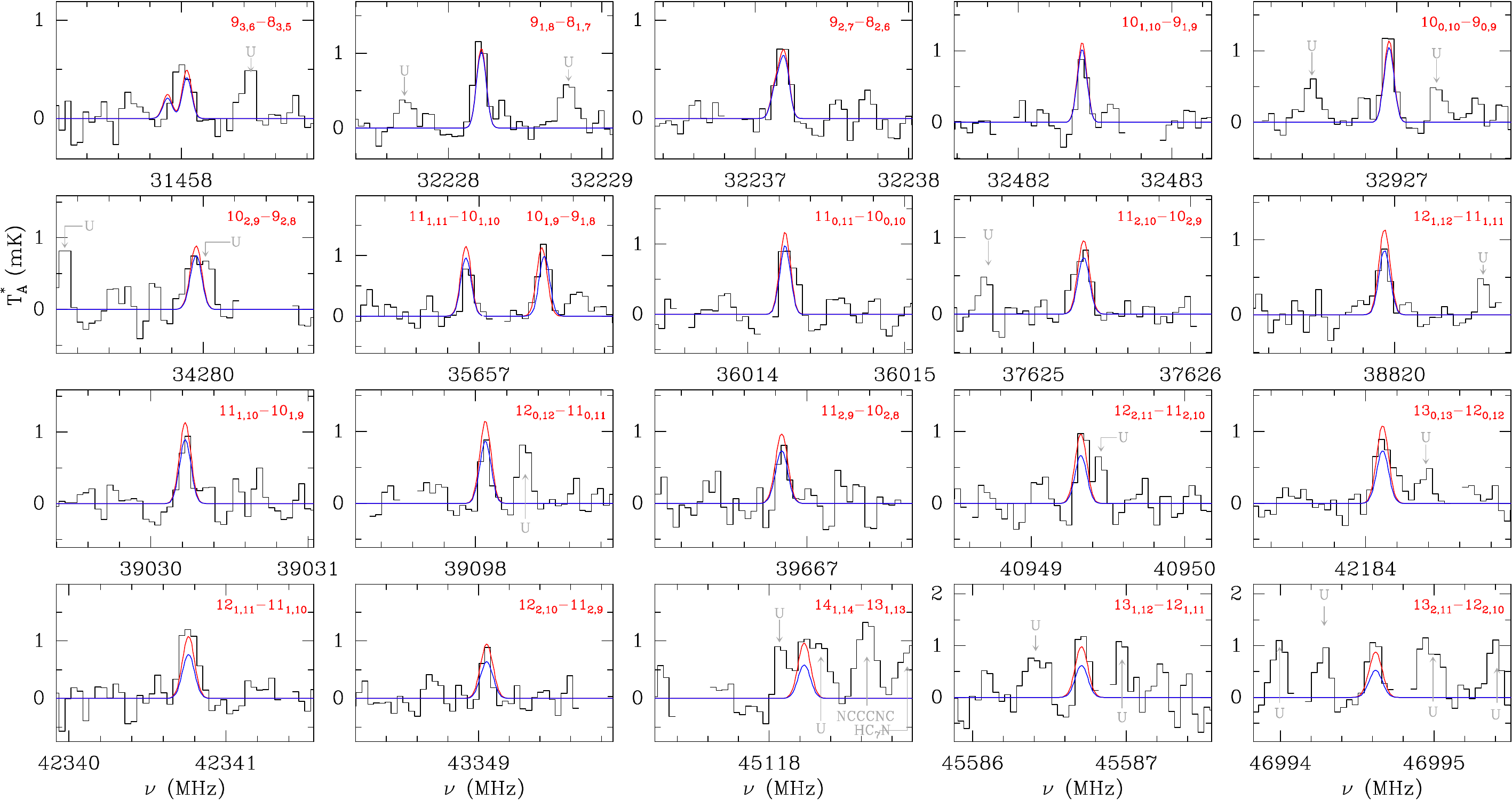}
\caption{Observed lines of 2-$CCP$ in the 31-50 GHz 
frequency range towards TMC-1. 
The abscissa corresponds to the rest frequency assuming a local standard of rest velocity of 5.83
km s$^{-1}$. 
The ordinate is the antenna temperature corrected for atmospheric and telescope losses in mK.
The red line shows the synthetic spectrum obtained from a fit to the observed line profiles,
which provides N(2-$CCP$)=(1.3$\pm$0.2)$\times$\once \ for a rotational temperature of 9\,K.
(see Appendix \ref{cyano_cyclopentadiene}). The blue line
shows the synthetic spectrum for the same column density and a rotational temperature of 6 K.
The rotational quantum
numbers are indicated in each panel. Blanked channels correspond to negative features produced
in the folding of the frequency switching data.
}
\label{fig_2-CCP}
\end{figure*}

For 2-$CCP,$ the laboratory frequencies measured by \citet{McCarthy2020}
cover the range 7.3-18.1 GHz, with $J_{max}$=5 and $K_a\le$2; hence, the
predictions in the frequency domain 31-50 GHz could be more uncertain than those of 1-$CCP$.
The derived line parameters for all the detected lines of 2-$CCP$ are given in Table \ref{line_parameters_2-ccp},
and the lines are shown in Fig. \ref{fig_2-CCP}.
Differences between predicted and observed frequencies reach values as high as 100 kHz. However, 
using the rotational and distortion constants of \citet{Lee2021}, the differences between predicted
and observed frequencies never exceed 20 kHz.
The range of energies
covered by our observations is not large enough to allow a fit to the rotational temperature, and we adopted 
the value derived
for 1-$CCP$ (9\,K). The derived column density for 2-$CCP$ is (1.3$\pm$0.2)$\times$10$^{11}$, a value
1.5 times lower than that of \citet{Lee2021}. Hence, the abundance
ratio between 1-$CCP$ and 2-$CCP$, derived from a significant number of lines for each isomer, 
is $\sim$2.4, a factor of two lower than the ratio derived by \citet{Lee2021}.

We used the observed line frequencies in TMC-1 and those of the laboratory \citep{McCarthy2020} 
to improve the rotational and distortion constants for both isomers (see Appendix \ref{new_constants_ccp}).

\tiny
\begin{table*}
\begin{center}
\small
\caption{Observed lines of 1-$CCP$ towards TMC-1.}
\label{line_parameters_1-ccp}
\begin{tabular}{llrlllr}
\hline       
Transition  & \multicolumn{1}{c}{$\nu_{pred}$\,$^a$} & \multicolumn{1}{c}{$\nu_{obs}$\,$^b$} & 
\multicolumn{1}{c}{$\int T_{\rm A}^* dv$\,$^c$} & \multicolumn{1}{c}{$\Delta$v\,$^d$} & 
\multicolumn{1}{c}{$T_{\rm A}^*$\,$^e$}& Notes\\
$(J_{K_{\rm a},K_{\rm c}})_{\rm u}-(J_{K_{\rm a},K_{\rm c}})_{\rm l}$ & \multicolumn{1}{c}{(MHz)} 
     &   \multicolumn{1}{c}{(MHz)}  & \multicolumn{1}{c}{(mK km s$^{-1}$)} 
     & \multicolumn{1}{c}{(km s$^{-1}$)} & \multicolumn{1}{c}{(mK)} &     \\
\hline
$9_{4,6}-8_{4,5}$               &31348.548$\pm$0.003&                    &             &             & $\le$0.62   &A\\ 
$9_{4,5}-8_{4,4}$               &31353.423$\pm$0.003& 31353.504$\pm$0.010&0.40$\pm$0.09&0.66$\pm$0.13&0.57$\pm$0.15& \\      
$9_{3,7}-8_{3,6}\,\,\,F=9-8$    &31376.544$\pm$0.002& 31376.560$\pm$0.010&0.37$\pm$0.05&0.42$\pm$0.25&0.85$\pm$0.16& \\
$......\,\,\,F=10-9 + F=8-7$    &31376.661$\pm$0.002& 31376.653$\pm$0.010&1.27$\pm$0.06&1.06$\pm$0.12&1.11$\pm$0.16& \\
$9_{3,6}-8_{3,5}\,\,\,F=9-8$    &31515.698$\pm$0.002& 31515.689$\pm$0.020&0.36$\pm$0.10&0.54$\pm$0.30&0.62$\pm$0.16& \\
$......\,\,\, F=10-9 + F=8-7$   &31515.821$\pm$0.002& 31515.833$\pm$0.010&0.52$\pm$0.14&0.57$\pm$0.15&0.86$\pm$0.16& \\
$9_{2,7}-8_{2,6}$               &32279.259$\pm$0.003& 32279.265$\pm$0.010&1.68$\pm$0.12&0.90$\pm$0.07&1.75$\pm$0.16& \\
$9_{1,8}-8_{1,7}$               &32302.755$\pm$0.003& 32302.755$\pm$0.010&2.14$\pm$0.13&0.80$\pm$0.06&2.51$\pm$0.16& \\
$10_{1,10}-9_{1,9}$             &32592.160$\pm$0.003& 32592.161$\pm$0.010&1.92$\pm$0.11&0.79$\pm$0.06&2.30$\pm$0.15& \\
$10_{0,10}-9_{0,9}^a$           &33050.463$\pm$0.003& 33050.408$\pm$0.020&4.35$\pm$0.03&1.45$\pm$0.04&2.80$\pm$0.15&B\\ 
$10_{2,9}-9_{2,8}$              &34371.215$\pm$0.005& 34371.214$\pm$0.010&1.33$\pm$0.08&0.63$\pm$0.04&2.00$\pm$0.14& \\
$10_{4,7}-9_{4,6}\,\,\,F=10-9$  &34856.901$\pm$0.004&                    &             &             &             &C\\
$......\,\,\,F=11-10 + F=9-8$   &34857.055$\pm$0.004& 34857.062$\pm$0.010&0.43$\pm$0.07&0.47$\pm$0.15&0.86$\pm$0.14&D\\ 
$10_{4,6}-9_{4,5}$              &34867.504$\pm$0.006&                    &             &             &$\le$0.54    &A\\ 
$10_{3,8}-9_{3,7}\,\,\,F=10-9$  &34877.024$\pm$0.003& 34877.049$\pm$0.020&0.34$\pm$0.12&0.64$\pm$0.31&0.50$\pm$0.15& \\
$.....\,\,\,F=11-10 + F=9-8$    &34877.109$\pm$0.003& 34877.177$\pm$0.020&1.04$\pm$0.25&1.54$\pm$0.39&0.64$\pm$0.15&E\\ %VERIFICAR
$10_{3,7}-9_{3,6}\,\,\,F=10-9$  &35110.827$\pm$0.003& 35110.822$\pm$0.015&0.36$\pm$0.11&0.84$\pm$0.33&0.40$\pm$0.16& \\
$......\,\,\, F=9-8 + F=11-10$  &35110.919$\pm$0.003& 35110.936$\pm$0.010&1.00$\pm$0.13&0.79$\pm$0.10&1.19$\pm$0.16& \\
$10_{1,9}-9_{1,8}$              &35746.835$\pm$0.004& 35746.834$\pm$0.010&1.67$\pm$0.14&0.73$\pm$0.07&2.16$\pm$0.15& \\
$11_{1,11}-10_{1,10}$           &35779.275$\pm$0.004& 35779.278$\pm$0.010&2.23$\pm$0.12&0.74$\pm$0.05&2.88$\pm$0.17& \\
$10_{2,8}-9_{2,7}$              &36006.125$\pm$0.005& 36006.124$\pm$0.010&1.92$\pm$0.12&0.80$\pm$0.05&2.24$\pm$0.16& \\
$11_{0,11}-10_{0,10}$           &36150.641$\pm$0.004& 36150.642$\pm$0.010&1.60$\pm$0.11&0.71$\pm$0.06&2.13$\pm$0.18& \\
$11_{2,10}-10_{2,9}$            &37728.644$\pm$0.007& 37728.647$\pm$0.010&2.09$\pm$0.18&1.19$\pm$0.15&1.64$\pm$0.17&E\\
$11_{4,8}-10_{4,7}\,\,\,F=11-10$&38372.076$\pm$0.004& 38372.043$\pm$0.030&0.25$\pm$0.11&0.66$\pm$0.24&0.36$\pm$0.16& \\
$......\,\,\, F=12-11 + F=10-9$ &38372.191$\pm$0.004& 38372.194$\pm$0.015&1.16$\pm$0.14&0.82$\pm$0.13&1.33$\pm$0.16& \\
$11_{3,9}-10_{3,8}\,\,\,F=11-10$&38374.124$\pm$0.005& 38374.103$\pm$0.020&0.42$\pm$0.10&0.55$\pm$0.25&0.67$\pm$0.16& \\
$......\,\,\,F=10-9 + F=12-11$  &38374.188$\pm$0.005& 38374.179$\pm$0.010&0.85$\pm$0.14&0.65$\pm$0.25&1.35$\pm$0.16& \\
$11_{4,7}-10_{4,6}\,\,\,F=11-10$&38392.912$\pm$0.004&                    &             &             &$\le$0.55    &A\\ 
$......\,\,\, F=12-11 + F=10-9$ &38393.029$\pm$0.004& 38393.027$\pm$0.015&0.54$\pm$0.10&0.55$\pm$0.13&0.93$\pm$0.18& \\ 
$11_{3,8}-10_{3,7}\,\,\,F=11-10$&38743.577$\pm$0.005& 38743.566$\pm$0.030&0.43$\pm$0.12&0.51$\pm$0.15&0.69$\pm$0.18& \\
$......\,\,\, F=10-9 + F=12-11$ &38743.648$\pm$0.005& 38743.651$\pm$0.030&0.70$\pm$0.13&0.54$\pm$0.11&1.20$\pm$0.18& \\
$12_{1,12}-11_{1,11}$           &38954.943$\pm$0.005& 38954.947$\pm$0.010&1.80$\pm$0.14&0.67$\pm$0.06&2.52$\pm$0.18& \\
$11_{1,10}-10_{1,9}$            &39136.812$\pm$0.005& 39136.808$\pm$0.010&1.43$\pm$0.14&0.62$\pm$0.07&2.16$\pm$0.17& \\
$12_{0,12}-11_{0,11}$           &39246.342$\pm$0.005& 39246.342$\pm$0.010&1.54$\pm$0.40&0.52$\pm$0.17&2.76$\pm$0.60&F\\ 
$11_{2,9}-10_{2,8}$             &39722.554$\pm$0.006& 39722.563$\pm$0.015&0.66$\pm$0.12&0.43$\pm$0.08&1.41$\pm$0.21&D\\
$12_{2,11}-11_{2,10}$           &41065.272$\pm$0.010& 41065.255$\pm$0.010&1.31$\pm$0.17&0.53$\pm$0.08&2.33$\pm$0.22& \\
$12_{3,10}-11_{3,9}$            &41864.830$\pm$0.007& 41864.824$\pm$0.010&1.46$\pm$0.19&0.92$\pm$0.15&1.48$\pm$0.23&E\\
$12_{4,9}-11_{4,8}$             &41893.726$\pm$0.005&                    &             &             & $\le$0.69   &A\\ 
$12_{4,8}-11_{4,7}$             &41932.390$\pm$0.005&                    &             &             & $\le$0.69   &A\\ 
$13_{1,13}-12_{1,12}$           &42120.773$\pm$0.007& 42120.781$\pm$0.010&1.69$\pm$0.13&0.63$\pm$0.06&2.51$\pm$0.18& \\
$13_{0,13}-12_{0,12}$           &42343.514$\pm$0.007& 42343.516$\pm$0.010&1.15$\pm$0.15&0.51$\pm$0.08&2.12$\pm$0.21& \\
$12_{3,9}-11_{3,8}$             &42418.202$\pm$0.007& 42418.209$\pm$0.025&0.67$\pm$0.20&0.55$\pm$0.17&1.14$\pm$0.24& \\
$12_{1,11}-11_{1,10}$           &42466.697$\pm$0.007& 42466.693$\pm$0.010&1.17$\pm$0.11&0.59$\pm$0.07&1.86$\pm$0.17& \\
$12_{2,10}-11_{2,9}$            &43415.138$\pm$0.008& 43415.121$\pm$0.010&2.33$\pm$0.19&0.87$\pm$0.09&2.51$\pm$0.21& \\
$13_{2,12}-12_{2,11}$           &44380.459$\pm$0.013& 44380.458$\pm$0.010&1.82$\pm$0.20&0.82$\pm$0.12&2.10$\pm$0.21& \\
$14_{1,14}-13_{1,13}$           &45278.391$\pm$0.010& 45278.405$\pm$0.015&0.70$\pm$0.14&0.46$\pm$0.09&1.43$\pm$0.27& \\
$13_{3,11}-12_{3,10}$           &45345.911$\pm$0.009&                    &             &             & $\le$0.81   &A\\ 
$13_{4,10}-12_{4,9}$            &45421.042$\pm$0.007&                    &             &             & $\le$0.81   &A\\ 
$14_{0,14}-13_{0,13}$           &45445.085$\pm$0.010& 45445.076$\pm$0.010&1.11$\pm$0.16&0.52$\pm$0.08&1.99$\pm$0.31& \\
$13_{4,9}-12_{4,8}$             &45488.849$\pm$0.007&                    &             &             & $\le$0.94   &A\\ 
$13_{1,12}-12_{1,11}$           &45733.576$\pm$0.010& 45733.564$\pm$0.015&0.85$\pm$0.23&0.44$\pm$0.09&1.79$\pm$0.29& \\
$13_{3,10}-12_{3,9}$            &46135.659$\pm$0.011& 46135.650$\pm$0.020&0.88$\pm$0.22&0.62$\pm$0.18&1.32$\pm$0.36& \\
$13_{2,11}-12_{2,10}$           &47073.451$\pm$0.010& 47073.438$\pm$0.020&1.32$\pm$0.29&0.62$\pm$0.14&2.02$\pm$0.38& \\
$14_{2,13}-13_{2,12}$           &47674.019$\pm$0.016& 47674.018$\pm$0.020&1.51$\pm$0.28&0.98$\pm$0.25&1.45$\pm$0.35& \\
\hline
\end{tabular}
\end{center}
\tablefoot{\\
For the observational parameters we adopted the uncertainty of the Gaussian fit provided by \texttt{GILDAS}.\\
Upper limits correspond to 3$\sigma$ values.\\
\tablefoottext{a}{Predictions based on the rotational constants of \citet{Lee2021} 
(see also \citealt{McCarthy2020}, Sect. \ref{cyano_derivatives}).}\\
\tablefoottext{b}{Observed frequency assuming a v$_{LSR}$ of 5.83 \kms.}\\
\tablefoottext{c}{Integrated line intensity in mK\,km\,s$^{-1}$.}\\
\tablefoottext{d}{Line width at half intensity derived by fitting a Gaussian function to
the observed line profile (in km\,s$^{-1}$).}\\
\tablefoottext{e}{Antenna temperature in millikelvin.}\\
\tablefoottext{A}{Below the detection limit.}
\tablefoottext{B}{Blended with the $J_K=8_2-7_2$ line of CH$_3$C$_3$N.}
\tablefoottext{C}{See next entry.}
\tablefoottext{D}{Affected by a negative feature produced in the folding of the frequency switching data.}
\tablefoottext{E}{Broad line due to the hyperfine structure of the transition.}
\tablefoottext{F}{Noisy zone of the spectrum.}
}
\end{table*}
\normalsize
\clearpage

\begin{table*}
\begin{center}
\small
\caption{Observed lines of 2-$CCP$ towards TMC-1.}
\label{line_parameters_2-ccp}
\begin{tabular}{llrlllr}
\hline\hline       
Transition  & \multicolumn{1}{c}{$\nu_{pred}$\,$^a$} & \multicolumn{1}{c}{$\nu_{obs}$\,$^b$} & 
\multicolumn{1}{c}{$\int T_{\rm A}^* dv$\,$^c$} & \multicolumn{1}{c}{$\Delta$v\,$^d$} & 
\multicolumn{1}{c}{$T_{\rm A}^*$\,$^e$}& Notes\\
$(J_{K_{\rm a},K_{\rm c}})_{\rm u}-(J_{K_{\rm a},K_{\rm c}})_{\rm l}$ & \multicolumn{1}{c}{(MHz)} 
     &   \multicolumn{1}{c}{(MHz)}  & \multicolumn{1}{c}{(mK km s$^{-1}$)} 
     & \multicolumn{1}{c}{(km s$^{-1}$)} & \multicolumn{1}{c}{(mK)} &     \\
\hline
$9_{3,6}-8_{3,5}$      & 31457.995$\pm$0.006& 31458.005$\pm$0.010&0.69$\pm$0.13&1.12$\pm$0.23&0.57$\pm$0.14&E\\
$9_{1,8}-8_{1,7}$      & 32228.215$\pm$0.006& 32228.207$\pm$0.010&1.08$\pm$0.13&0.85$\pm$0.09&1.20$\pm$0.12& \\
$9_{2,7}-8_{2,6}$      & 32237.173$\pm$0.006& 32237.180$\pm$0.010&0.76$\pm$0.10&0.89$\pm$0.13&0.80$\pm$0.12& \\
$10_{1,10}-9_{1,9}$    & 32482.415$\pm$0.009& 32482.420$\pm$0.010&0.62$\pm$0.10&0.60$\pm$0.12&0.97$\pm$0.16& \\
$10_{0,10}-9_{0,9}$    & 32926.951$\pm$0.009& 32926.946$\pm$0.010&0.95$\pm$0.12&0.71$\pm$0.11&1.26$\pm$0.16& \\
$10_{2,9}-9_{2,8}$     & 34279.918$\pm$0.009& 34279.970$\pm$0.020&1.21$\pm$0.19&1.44$\pm$0.23&0.79$\pm$0.19&E\\
$11_{1,11}-10_{1,10}$  & 35656.918$\pm$0.011& 35656.932$\pm$0.010&0.57$\pm$0.08&0.60$\pm$0.11&0.89$\pm$0.13& \\
$10_{1,9}-9_{1,8}$     & 35657.417$\pm$0.008& 35657.412$\pm$0.010&0.95$\pm$0.08&0.73$\pm$0.07&1.20$\pm$0.13& \\
$10_{2,8}-9_{2,7}$     & 35958.391$\pm$0.007&                    &             &             &             &D\\
$11_{0,11}-10_{0,10}$  & 36014.223$\pm$0.012& 36014.255$\pm$0.010&0.98$\pm$0.26&0.85$\pm$0.15&1.00$\pm$0.14& \\
$11_{2,10}-10_{2,9}$   & 37625.343$\pm$0.012& 37625.308$\pm$0.015&0.96$\pm$0.13&0.88$\pm$0.13&0.83$\pm$0.18& \\ 
$12_{1,12}-11_{1,11}$  & 38819.937$\pm$0.014& 38819.928$\pm$0.010&0.68$\pm$0.12&0.75$\pm$0.18&0.85$\pm$0.16& \\ 
$11_{1,10}-10_{1,9}$   & 39030.219$\pm$0.011& 39030.222$\pm$0.010&0.61$\pm$0.11&0.58$\pm$0.14&1.00$\pm$0.18& \\ 
$12_{0,12}-11_{0,11}$  & 39098.064$\pm$0.014& 39098.060$\pm$0.010&0.50$\pm$0.09&0.49$\pm$0.10&0.95$\pm$0.18& \\ 
$11_{2,9}-10_{2,8}$    & 39666.839$\pm$0.009& 39666.848$\pm$0.015&0.49$\pm$0.11&0.54$\pm$0.13&0.86$\pm$0.20& \\ 
$12_{2,11}-11_{2,10}$  & 40949.324$\pm$0.015& 40949.339$\pm$0.015&0.62$\pm$0.11&0.54$\pm$0.12&1.08$\pm$0.18& \\   
$13_{1,13}-12_{1,12}$  & 41973.156$\pm$0.017&                    &             &             &             &D\\ 
$13_{0,13}-12_{0,12}$  & 42184.094$\pm$0.017& 42184.110$\pm$0.030&0.39$\pm$0.10&0.64$\pm$0.18&0.57$\pm$0.17& \\
$12_{1,11}-11_{1,10}$  & 42340.761$\pm$0.015& 42340.761$\pm$0.010&1.29$\pm$0.15&0.86$\pm$0.11&1.32$\pm$0.20& \\ 
$12_{2,10}-11_{2,9}$   & 43349.053$\pm$0.011& 43349.049$\pm$0.015&0.44$\pm$0.11&0.42$\pm$0.13&0.99$\pm$0.22& \\  
$14_{1,14}-13_{1,13}$  & 45118.255$\pm$0.021& 45118.232$\pm$0.030&0.97$\pm$0.26&0.95$\pm$0.38&0.96$\pm$0.24& \\  
$14_{0,14}-13_{0,13}$  & 45274.919$\pm$0.021&                    &             &             &             &D\\ 
$13_{1,12}-12_{1,11}$  & 45586.689$\pm$0.019& 45586.714$\pm$0.020&0.72$\pm$0.17&0.54$\pm$0.12&1.25$\pm$0.32& \\
$13_{2,11}-12_{2,10}$  & 46994.644$\pm$0.014& 46994.606$\pm$0.020&1.07$\pm$0.17&0.70$\pm$0.11&1.44$\pm$0.28& \\
\hline
\end{tabular}
\end{center}
\tablefoot{\\
For the observational parameters we adopted the uncertainty of the Gaussian fit provided by \texttt{GILDAS}.\\
%Predictions based on the rotational constants of \citet{Lee2021} (see also \citealt{McCarthy2020}).\\
%Other notes are the same than for Table \ref{line_parameters_1-ccp}
\tablefoottext{a}{Predictions based on the rotational constants of \citet{Lee2021} 
(see also \citealt{McCarthy2020}, Sect. \ref{cyano_derivatives}).}\\
\tablefoottext{b}{Observed frequency assuming a v$_{LSR}$ of 5.83 \kms.}\\
\tablefoottext{c}{Integrated line intensity in mK\,km\,s$^{-1}$.}\\
\tablefoottext{d}{Line width at half intensity derived by fitting a Gaussian function to
the observed line profile (in km\,s$^{-1}$).}\\
\tablefoottext{e}{Antenna temperature in millikelvin.}\\
%\tablefoottext{A}{Below the detection limit. Upper limits correspond to 3$\sigma$ values.}
%\tablefoottext{B}{Blended with the $J_K=8_2-7_2$ line of CH$_3$C$_3$N.}
%\tablefoottext{C}{See next entry.}
\tablefoottext{D}{Affected by a negative feature produced in the folding of the frequency switching data.}
\tablefoottext{E}{Broad line due to the hyperfine structure of the transition.}
%\tablefoottext{F}{Noisy zone of the spectrum.}
}
\end{table*}
\clearpage

\subsection{Improved rotational constants for 1-$CCP$ and 2-$CCP$} 
\label{new_constants_ccp}
The observed lines of 1-$CCP$ (see Table \ref{line_parameters_1-ccp}) and 2-$CCP$ 
(see Table \ref{line_parameters_2-ccp}) were merged with the
laboratory data \citep{McCarthy2020} to provide a new set of rotational 
and distortion constants. They are given in Table \ref{table_new_constants_ccp} 
and can be used to 
predict the frequencies of the rotational transitions for the two observed isomers
of cyano cyclopentadiene up to 50 GHz with an 
accuracy better than 15 kHz ($K_a\le3$). The calculated and the observed-minus-calculated frequencies
for 1-$CCP$ and 2-$CCP$ are given in Tables \ref{lines_1cpd} and \ref{lines_2cpd}, 
respectively.
\begin{table*}
\centering
\small
\caption{Molecular constants for 1-$CCP$ and 2-$CCP$.}
\label{table_new_constants_ccp}
\begin{tabular}{|lcc|cc|}
\hline
                       & 1-$CCP ^a$       & 1-$CCP ^b$         & 2-$CCP ^a$           & 2-$CCP ^b$ \\
Constant               &   (MHz)          &   (MHz)            &   (MHz)              &   (MHz)   \\
\hline
\hline
$A$                    &   8352.981(10)   &     8352.9758(142) &      8235.592(14)    &      8235.599(37)   \\
$B$                    &   1904.2522(2)   &    1904.251830(231)&       1902.0748(3)   &    1902.07558(49)   \\
$C$                    &   1565.3652(2)   &    1565.365658(197)&       1559.6472(2)   &    1559.64673(38)   \\
$\Delta_J$ 10$^{-3}$   &   0.0743(11)     &       0.07280(110) &      0.0686(11)      &       0.07246(240)  \\
$\Delta_{JK}$ 10$^{-3}$&   2.354(8)       &       2.3493(134)  &      2.287(21)       &         2.248(37)   \\
$\Delta_K$ 10$^{-3}$   &   0.17561$^c$    &                    &      0.32391$^c$     &                     \\
$\delta_J$ 10$^{-5}$   &   1.33(5)        &       1.320(52)    &       1.34(6)        &        1.465(82)   \\
$\delta_K$ 10$^{-3}$   &   1.48(9)        &       1.322(81)    &       1.10(9)        &        1.299(191)  \\
$\chi_{aa}$            &  -4.1810(11)     &      -4.1795  (23) &      -4.2429(6)      &     -4.2316 (55)    \\
$\chi_{bb}$            &   2.3016(14)     &       2.3052  (38) &      2.2475(16)      &      2.2347 (74)    \\
\hline
\hline
N$_{lines}$            & 68               &          175       &          38          &         98        \\
$\nu_{max}$ (GHz)      & 29.9             &           47.6     &         18.1         &         47.0        \\
$(J,K_a)_{max}$        & 9,3              &           14,3     &         5,2          &         13,2        \\
$\sigma$ (kHz)         & 2.2              &           9.1      &         1.9          &         10.3        \\
\hline             
\end{tabular}
\tablefoot{\\
Values between parentheses correspond to the uncertainties of the parameters 
in units of the last significant digits.\\
\tablefoottext{a}{Rotational and distortion constants from \citet{Lee2021}.}\\
\tablefoottext{b}{Merged fit to the laboratory data of \citet{McCarthy2020} and space frequencies 
measured in this work.
The laboratory data used by \citet{Lee2021} are not available.}\\
\tablefoottext{c}{Fixed value.}\\
}
\end{table*}

\clearpage
\onecolumn

\begin{tiny}
\begin{longtable}{cccccccccccc}
\caption[]{Observed and calculated line frequencies for 1-$CCP$.
\label{lines_1cpd}}\\
\hline
\hline
$J'$ & $K'_a$ & $K'_c$ & $J''$ & $K''_a$ & $K''_c$ &  $F'$ & $F''$ & $\nu_{obs}$\,$^a$  & $\nu_{calc}$\,$^b$  & $\nu_{obs}-\nu_{calc}$\,$^c$ & Notes \\
   &   &        &      &        &        &                  &    &   (MHz)      &   (MHz)  &   (MHz) &      \\
\hline
\endfirsthead
\caption{continued.}\\
\hline
\hline
$J'$ & $K'_a$ & $K'_c$ & $J''$ & $K''_a$ & $K''_c$ &  $F'$ & $F''$ & $\nu_{obs}$\,$^a$  & $\nu_{calc}$\,$^b$  & $\nu_{obs}-\nu_{calc}$\,$^c$ & Notes \\
   &   &        &      &        &        &                  &    &   (MHz)      &   (MHz)  &   (MHz) &      \\
\hline
\endhead
\hline
\endfoot
\hline
\endlastfoot
\hline
  2 & 0 & 2 & 2 & 1 & 0 & 1 & 2 &      6924.9737  &   6924.9753   &  -0.0016 &  1 \\
  2 & 0 & 2 & 1 & 1 & 0 & 1 & 0 &      6925.1742  &   6925.1768   &  -0.0026 &  1 \\
  2 & 0 & 2 & 2 & 1 & 0 & 1 & 1 &      6926.2273  &   6926.2290   &  -0.0017 &  1 \\
  2 & 0 & 2 & 3 & 1 & 0 & 1 & 2 &      6926.3145  &   6926.3138   &   0.0008 &  1 \\
  2 & 0 & 2 & 1 & 1 & 0 & 1 & 1 &      6928.3108  &   6928.3114   &  -0.0006 &  1 \\
  2 & 1 & 1 & 2 & 1 & 1 & 0 & 1 &      7277.0535  &   7277.0592   &  -0.0057 &  1 \\
  2 & 1 & 1 & 2 & 1 & 1 & 0 & 2 &      7277.6180  &   7277.6212   &  -0.0032 &  1 \\
  2 & 1 & 1 & 3 & 1 & 1 & 0 & 2 &      7278.3608  &   7278.3621   &  -0.0013 &  1 \\
  3 & 1 & 3 & 3 & 2 & 1 & 2 & 3 &      9891.6689  &   9891.6697   &  -0.0008 &  1 \\
  3 & 1 & 3 & 3 & 2 & 1 & 2 & 2 &      9892.2695  &   9892.2721   &  -0.0026 &  1 \\
  3 & 1 & 3 & 4 & 2 & 1 & 2 & 3 &      9892.6421  &   9892.6412   &   0.0009 &  1 \\
  3 & 1 & 3 & 2 & 2 & 1 & 2 & 2 &      9893.5811  &   9893.5838   &  -0.0027 &  1 \\
  3 & 0 & 3 & 2 & 2 & 0 & 2 & 1 &     10356.7289  &  10356.7293   &  -0.0004 &  1 \\
  3 & 0 & 3 & 3 & 2 & 0 & 2 & 2 &     10356.9498  &  10356.9478   &   0.0020 &  1 \\
  3 & 0 & 3 & 4 & 2 & 0 & 2 & 3 &     10356.9901  &  10356.9899   &   0.0002 &  1 \\
  3 & 2 & 2 & 3 & 2 & 2 & 1 & 2 &     10407.7424  &  10407.7435   &  -0.0011 &  1 \\
  3 & 2 & 2 & 3 & 2 & 2 & 1 & 2 &     10407.7424  &  10407.7435   &  -0.0011 &  1 \\
  3 & 2 & 2 & 4 & 2 & 2 & 1 & 3 &     10409.0876  &  10409.0868   &   0.0008 &  1 \\
  3 & 2 & 1 & 4 & 2 & 2 & 0 & 3 &     10460.9891  &  10460.9913   &  -0.0022 &  1 \\
  3 & 2 & 1 & 2 & 2 & 2 & 0 & 1 &     10461.7333  &  10461.7395   &  -0.0062 &  1 \\
  3 & 1 & 2 & 3 & 2 & 1 & 1 & 3 &     10907.8578  &  10907.8588   &  -0.0010 &  1 \\
  3 & 1 & 2 & 3 & 2 & 1 & 1 & 2 &     10908.5994  &  10908.5997   &  -0.0003 &  1 \\
  3 & 1 & 2 & 2 & 2 & 1 & 1 & 1 &     10908.9552  &  10908.9522   &   0.0030 &  1 \\
  3 & 1 & 2 & 4 & 2 & 1 & 1 & 3 &     10908.9747  &  10908.9738   &   0.0009 &  1 \\
  3 & 1 & 2 & 2 & 2 & 1 & 1 & 2 &     10910.1016  &  10910.1052   &  -0.0036 &  1 \\
  4 & 1 & 4 & 4 & 3 & 1 & 3 & 3 &     13175.4738  &  13175.4710   &   0.0028 &  1 \\
  4 & 1 & 4 & 3 & 3 & 1 & 3 & 2 &     13175.5840  &  13175.5847   &  -0.0007 &  1 \\
  4 & 1 & 4 & 5 & 3 & 1 & 3 & 4 &     13175.6334  &  13175.6333   &   0.0001 &  1 \\
  4 & 0 & 4 & 3 & 3 & 0 & 3 & 2 &     13749.4229  &  13749.4239   &  -0.0010 &  1 \\
  4 & 0 & 4 & 4 & 3 & 0 & 3 & 3 &     13749.5243  &  13749.5253   &  -0.0010 &  1 \\
  4 & 0 & 4 & 5 & 3 & 0 & 3 & 4 &     13749.5483  &  13749.5467   &   0.0016 &  1 \\
  4 & 2 & 3 & 4 & 3 & 2 & 2 & 3 &     13867.8420  &  13867.8411   &   0.0010 &  1 \\
  4 & 2 & 3 & 5 & 3 & 2 & 2 & 4 &     13868.4103  &  13868.4098   &   0.0005 &  1 \\
  4 & 2 & 3 & 3 & 3 & 2 & 2 & 2 &     13868.5578  &  13868.5559   &   0.0019 &  1 \\
  4 & 2 & 2 & 5 & 3 & 2 & 1 & 4 &     13997.3397  &  13997.3399   &  -0.0002 &  1 \\
  4 & 1 & 3 & 4 & 3 & 1 & 2 & 3 &     14529.1170  &  14529.1160   &   0.0010 &  1 \\
  4 & 1 & 3 & 3 & 3 & 1 & 2 & 2 &     14529.2191  &  14529.2201   &  -0.0010 &  1 \\
  4 & 1 & 3 & 5 & 3 & 1 & 2 & 4 &     14529.2828  &  14529.2812   &   0.0016 &  1 \\
  5 & 1 & 5 & 5 & 4 & 1 & 4 & 4 &     16447.0681  &  16447.0680   &   0.0002 &  1 \\
  5 & 1 & 5 & 4 & 4 & 1 & 4 & 3 &     16447.1112  &  16447.1120   &  -0.0008 &  1 \\
  5 & 1 & 5 & 6 & 4 & 1 & 4 & 5 &     16447.1543  &  16447.1549   &  -0.0006 &  1 \\
  5 & 0 & 5 & 4 & 4 & 0 & 4 & 3 &     17093.4579  &  17093.4567   &   0.0012 &  1 \\
  5 & 0 & 5 & 5 & 4 & 0 & 4 & 4 &     17093.5202  &  17093.5204   &  -0.0002 &  1 \\
  5 & 0 & 5 & 6 & 4 & 0 & 4 & 5 &     17093.5326  &  17093.5297   &   0.0029 &  1 \\
  5 & 1 & 4 & 5 & 4 & 1 & 3 & 4 &     18134.9211  &  18134.9206   &   0.0005 &  1 \\
  5 & 1 & 4 & 4 & 4 & 1 & 3 & 3 &     18134.9593  &  18134.9586   &   0.0007 &  1 \\
  5 & 1 & 4 & 6 & 4 & 1 & 3 & 5 &     18135.0103  &  18135.0091   &   0.0012 &  1 \\
  7 & 1 & 7 & 7 & 6 & 1 & 6 & 6 &     22949.1633  &  22949.1652   &  -0.0019 &  1 \\
  7 & 1 & 7 & 6 & 6 & 1 & 6 & 5 &     22949.1747  &  22949.1721   &   0.0026 &  1 \\
  7 & 1 & 7 & 8 & 6 & 1 & 6 & 7 &     22949.1989  &  22949.1986   &   0.0003 &  1 \\
  7 & 0 & 7 & 6 & 6 & 0 & 6 & 5 &     23614.2508  &  23614.2463   &   0.0045 &  1 \\
  7 & 0 & 7 & 8 & 6 & 0 & 6 & 7 &     23614.2823  &  23614.2827   &  -0.0004 &  1 \\
  7 & 0 & 7 & 7 & 6 & 0 & 6 & 6 &     23614.2823  &  23614.2827   &  -0.0004 &  1 \\
  7 & 2 & 6 & 7 & 6 & 2 & 5 & 6 &     24186.1313  &  24186.1327   &  -0.0014 &  1 \\
  7 & 2 & 6 & 6 & 6 & 2 & 5 & 5 &     24186.2294  &  24186.2342   &  -0.0048 &  1 \\
  7 & 2 & 6 & 8 & 6 & 2 & 5 & 7 &     24186.2469  &  24186.2438   &   0.0031 &  1 \\
  7 & 3 & 5 & 8 & 6 & 3 & 4 & 7 &     24376.3271  &  24376.3310   &  -0.0039 &  1 \\
  7 & 3 & 5 & 6 & 6 & 3 & 4 & 5 &     24376.3571  &  24376.3529   &   0.0042 &  1 \\
  7 & 2 & 5 & 7 & 6 & 2 & 4 & 6 &     24858.3337  &  24858.3276   &   0.0061 &  1 \\
  7 & 2 & 5 & 6 & 6 & 2 & 4 & 5 &     24858.4414  &  24858.4448   &  -0.0034 &  1 \\
  7 & 2 & 5 & 8 & 6 & 2 & 4 & 7 &     24858.4555  &  24858.4535   &   0.0020 &  1 \\
  7 & 1 & 6 & 7 & 6 & 1 & 5 & 6 &     25282.4392  &  25282.4374   &   0.0018 &  1 \\
  7 & 1 & 6 & 6 & 6 & 1 & 5 & 5 &     25282.4392  &  25282.4374   &   0.0018 &  1 \\
  7 & 1 & 6 & 8 & 6 & 1 & 5 & 7 &     25282.4669  &  25282.4688   &  -0.0019 &  1 \\
  8 & 0 & 8 & 9 & 7 & 0 & 7 & 8 &     26795.3336  &  26795.3363   &  -0.0027 &  1 \\
  8 & 0 & 8 & 8 & 7 & 0 & 7 & 7 &     26795.3336  &  26795.3363   &  -0.0027 &  1 \\
  9 & 0 & 9 &10 & 8 & 0 & 8 & 9 &     29936.2435  &  29936.2433   &   0.0002 &  1 \\
  9 & 0 & 9 & 9 & 8 & 0 & 8 & 8 &     29936.2435  &  29936.2433   &   0.0002 &  1 \\
  9 & 4 & 5 &10 & 8 & 4 & 4 & 9 &     31353.5040  &  31353.4899   &   0.0141 &  2 \\
  9 & 3 & 7 & 9 & 8 & 3 & 6 & 8 &     31376.5600  &  31376.5447   &   0.0153 &  2 \\
  9 & 3 & 7 & 8 & 8 & 3 & 6 & 7 &     31376.6530  &  31376.6626   &  -0.0096 &  2 \\
  9 & 3 & 7 &10 & 8 & 3 & 6 & 9 &     31376.6530  &  31376.6626   &  -0.0096 &  2 \\
  9 & 3 & 6 & 9 & 8 & 3 & 5 & 8 &     31515.6890  &  31515.7012   &  -0.0122 &  2 \\
  9 & 3 & 6 &10 & 8 & 3 & 5 & 9 &     31515.8330  &  31515.8239   &   0.0091 &  2 \\
  9 & 3 & 6 & 8 & 8 & 3 & 5 & 7 &     31515.8330  &  31515.8239   &   0.0091 &  2 \\
  9 & 2 & 7 & 8 & 8 & 2 & 6 & 7 &     32279.2650  &  32279.2797   &  -0.0147 &  2 \\
  9 & 2 & 7 &10 & 8 & 2 & 6 & 9 &     32279.2650  &  32279.2797   &  -0.0147 &  2 \\
  9 & 1 & 8 & 8 & 8 & 1 & 7 & 7 &     32302.7550  &  32302.7514   &   0.0036 &  2 \\
  9 & 1 & 8 & 9 & 8 & 1 & 7 & 8 &     32302.7550  &  32302.7514   &   0.0036 &  2 \\
  9 & 1 & 8 &10 & 8 & 1 & 7 & 9 &     32302.7550  &  32302.7514   &   0.0036 &  2 \\
 10 & 1 &10 & 9 & 9 & 1 & 9 & 8 &     32592.1610  &  32592.1611   &  -0.0001 &  2 \\
 10 & 1 &10 &10 & 9 & 1 & 9 & 9 &     32592.1610  &  32592.1611   &  -0.0001 &  2 \\
 10 & 1 &10 &11 & 9 & 1 & 9 &10 &     32592.1610  &  32592.1611   &  -0.0001 &  2 \\
 10 & 0 &10 & 9 & 9 & 0 & 9 & 8 &     33050.4080  &  33050.4633   &  -0.0553 &  2 \\
 10 & 0 &10 &11 & 9 & 0 & 9 &10 &     33050.4080  &  33050.4633   &  -0.0553 &  2 \\
 10 & 0 &10 &10 & 9 & 0 & 9 & 9 &     33050.4080  &  33050.4633   &  -0.0553 &  2 \\
 10 & 2 & 9 &10 & 9 & 2 & 8 & 9 &     34371.2140  &  34371.2105   &   0.0035 &  2 \\
 10 & 2 & 9 & 9 & 9 & 2 & 8 & 8 &     34371.2140  &  34371.2105   &   0.0035 &  2 \\
 10 & 2 & 9 &11 & 9 & 2 & 8 &10 &     34371.2140  &  34371.2105   &   0.0035 &  2 \\
 10 & 4 & 7 & 9 & 9 & 4 & 6 & 8 &     34857.0620  &  34857.0623   &  -0.0003 &  2 \\
 10 & 3 & 8 &10 & 9 & 3 & 7 & 9 &     34877.0490  &  34877.0243   &   0.0247 &  2 \\
 10 & 3 & 7 &10 & 9 & 3 & 6 & 9 &     35110.8220  &  35110.8305   &  -0.0085 &  2 \\
 10 & 3 & 7 & 9 & 9 & 3 & 6 & 8 &     35110.9360  &  35110.9222   &   0.0138 &  2 \\
 10 & 3 & 7 &11 & 9 & 3 & 6 &10 &     35110.9360  &  35110.9222   &   0.0138 &  2 \\
 10 & 1 & 9 & 9 & 9 & 1 & 8 & 8 &     35746.8340  &  35746.8311   &   0.0029 &  2 \\
 10 & 1 & 9 &10 & 9 & 1 & 8 & 9 &     35746.8340  &  35746.8311   &   0.0029 &  2 \\
 10 & 1 & 9 &11 & 9 & 1 & 8 &10 &     35746.8340  &  35746.8311   &   0.0029 &  2 \\
 11 & 1 &11 &10 &10 & 1 &10 & 9 &     35779.2780  &  35779.2760   &   0.0020 &  2 \\
 11 & 1 &11 &11 &10 & 1 &10 &10 &     35779.2780  &  35779.2760   &   0.0020 &  2 \\
 11 & 1 &11 &12 &10 & 1 &10 &11 &     35779.2780  &  35779.2760   &   0.0020 &  2 \\
 10 & 2 & 8 &10 & 9 & 2 & 7 & 9 &     36006.1240  &  36006.1266   &  -0.0026 &  2 \\
 10 & 2 & 8 & 9 & 9 & 2 & 7 & 8 &     36006.1240  &  36006.1266   &  -0.0026 &  2 \\
 10 & 2 & 8 &11 & 9 & 2 & 7 &10 &     36006.1240  &  36006.1266   &  -0.0026 &  2 \\
 11 & 0 &11 &10 &10 & 0 &10 & 9 &     36150.6420  &  36150.6422   &  -0.0002 &  2 \\
 11 & 0 &11 &12 &10 & 0 &10 &11 &     36150.6420  &  36150.6422   &  -0.0002 &  2 \\
 11 & 0 &11 &11 &10 & 0 &10 &10 &     36150.6420  &  36150.6422   &  -0.0002 &  2 \\
 11 & 2 &10 &11 &10 & 2 & 9 &10 &     37728.6470  &  37728.6388   &   0.0083 &  2 \\
 11 & 2 &10 &10 &10 & 2 & 9 & 9 &     37728.6470  &  37728.6388   &   0.0083 &  2 \\
 11 & 2 &10 &12 &10 & 2 & 9 &11 &     37728.6470  &  37728.6388   &   0.0083 &  2 \\
 11 & 4 & 8 &11 &10 & 4 & 7 &10 &     38372.0430  &  38372.0795   &  -0.0365 &  2 \\
 11 & 4 & 8 &12 &10 & 4 & 7 &11 &     38372.1940  &  38372.1952   &  -0.0012 &  2 \\
 11 & 4 & 8 &10 &10 & 4 & 7 & 9 &     38372.1940  &  38372.1952   &  -0.0012 &  2 \\
 11 & 3 & 9 &11 &10 & 3 & 8 &10 &     38374.1030  &  38374.1238   &  -0.0208 &  2 \\
 11 & 3 & 9 &10 &10 & 3 & 8 & 9 &     38374.1790  &  38374.1874   &  -0.0084 &  2 \\
 11 & 3 & 9 &12 &10 & 3 & 8 &11 &     38374.1790  &  38374.1874   &  -0.0084 &  2 \\
 11 & 4 & 7 &12 &10 & 4 & 6 &11 &     38393.0270  &  38393.0330   &  -0.0060 &  2 \\
 11 & 4 & 7 &10 &10 & 4 & 6 & 9 &     38393.0270  &  38393.0330   &  -0.0060 &  2 \\
 11 & 3 & 8 &11 &10 & 3 & 7 &10 &     38743.5660  &  38743.5819   &  -0.0159 &  2 \\
 11 & 3 & 8 &10 &10 & 3 & 7 & 9 &     38743.6510  &  38743.6535   &  -0.0025 &  2 \\
 11 & 3 & 8 &12 &10 & 3 & 7 &11 &     38743.6510  &  38743.6535   &  -0.0025 &  2 \\
 12 & 1 &12 &11 &11 & 1 &11 &10 &     38954.9470  &  38954.9446   &   0.0024 &  2 \\
 12 & 1 &12 &12 &11 & 1 &11 &11 &     38954.9470  &  38954.9446   &   0.0024 &  2 \\
 12 & 1 &12 &13 &11 & 1 &11 &12 &     38954.9470  &  38954.9446   &   0.0024 &  2 \\
 11 & 1 &10 &10 &10 & 1 & 9 & 9 &     39136.8080  &  39136.8067   &   0.0013 &  2 \\
 11 & 1 &10 &11 &10 & 1 & 9 &10 &     39136.8080  &  39136.8067   &   0.0013 &  2 \\
 11 & 1 &10 &12 &10 & 1 & 9 &11 &     39136.8080  &  39136.8067   &   0.0013 &  2 \\
 12 & 0 &12 &11 &11 & 0 &11 &10 &     39246.3420  &  39246.3437   &  -0.0017 &  2 \\
 12 & 0 &12 &13 &11 & 0 &11 &12 &     39246.3420  &  39246.3437   &  -0.0017 &  2 \\
 12 & 0 &12 &12 &11 & 0 &11 &11 &     39246.3420  &  39246.3437   &  -0.0017 &  2 \\
 11 & 2 & 9 &11 &10 & 2 & 8 &10 &     39722.5630  &  39722.5548   &   0.0082 &  2 \\
 11 & 2 & 9 &10 &10 & 2 & 8 & 9 &     39722.5630  &  39722.5548   &   0.0082 &  2 \\
 11 & 2 & 9 &12 &10 & 2 & 8 &11 &     39722.5630  &  39722.5548   &   0.0082 &  2 \\
 12 & 2 &11 &12 &11 & 2 &10 &11 &     41065.2550  &  41065.2652   &  -0.0102 &  2 \\
 12 & 2 &11 &11 &11 & 2 &10 &10 &     41065.2550  &  41065.2652   &  -0.0102 &  2 \\
 12 & 2 &11 &13 &11 & 2 &10 &12 &     41065.2550  &  41065.2652   &  -0.0102 &  2 \\
 12 & 3 &10 &12 &11 & 3 & 9 &11 &     41864.8240  &  41864.8275   &  -0.0035 &  2 \\
 12 & 3 &10 &11 &11 & 3 & 9 &10 &     41864.8240  &  41864.8275   &  -0.0035 &  2 \\
 12 & 3 &10 &13 &11 & 3 & 9 &12 &     41864.8240  &  41864.8275   &  -0.0035 &  2 \\
 13 & 1 &13 &12 &12 & 1 &12 &11 &     42120.7810  &  42120.7758   &   0.0052 &  2 \\
 13 & 1 &13 &13 &12 & 1 &12 &12 &     42120.7810  &  42120.7758   &   0.0052 &  2 \\
 13 & 1 &13 &14 &12 & 1 &12 &13 &     42120.7810  &  42120.7758   &   0.0052 &  2 \\
 13 & 0 &13 &12 &12 & 0 &12 &11 &     42343.5160  &  42343.5164   &  -0.0004 &  2 \\
 13 & 0 &13 &14 &12 & 0 &12 &13 &     42343.5160  &  42343.5164   &  -0.0004 &  2 \\
 13 & 0 &13 &13 &12 & 0 &12 &12 &     42343.5160  &  42343.5164   &  -0.0004 &  2 \\
 12 & 3 & 9 &12 &11 & 3 & 8 &11 &     42418.2090  &  42418.2075   &   0.0015 &  2 \\
 12 & 3 & 9 &11 &11 & 3 & 8 &10 &     42418.2090  &  42418.2075   &   0.0015 &  2 \\
 12 & 3 & 9 &13 &11 & 3 & 8 &12 &     42418.2090  &  42418.2075   &   0.0015 &  2 \\
 12 & 1 &11 &11 &11 & 1 &10 &10 &     42466.6930  &  42466.6895   &   0.0035 &  2 \\
 12 & 1 &11 &13 &11 & 1 &10 &12 &     42466.6930  &  42466.6895   &   0.0035 &  2 \\
 12 & 1 &11 &12 &11 & 1 &10 &11 &     42466.6930  &  42466.6895   &   0.0035 &  2 \\
 12 & 2 &10 &12 &11 & 2 & 9 &11 &     43415.1210  &  43415.1388   &  -0.0178 &  2 \\
 12 & 2 &10 &11 &11 & 2 & 9 &10 &     43415.1210  &  43415.1388   &  -0.0178 &  2 \\
 12 & 2 &10 &13 &11 & 2 & 9 &12 &     43415.1210  &  43415.1388   &  -0.0178 &  2 \\
 13 & 2 &12 &13 &12 & 2 &11 &12 &     44380.4580  &  44380.4491   &   0.0089 &  2 \\
 13 & 2 &12 &12 &12 & 2 &11 &11 &     44380.4580  &  44380.4491   &   0.0089 &  2 \\
 13 & 2 &12 &14 &12 & 2 &11 &13 &     44380.4580  &  44380.4491   &   0.0089 &  2 \\
 14 & 1 &14 &13 &13 & 1 &13 &12 &     45278.4050  &  45278.3947   &   0.0103 &  2 \\
 14 & 1 &14 &14 &13 & 1 &13 &13 &     45278.4050  &  45278.3947   &   0.0103 &  2 \\
 14 & 1 &14 &15 &13 & 1 &13 &14 &     45278.4050  &  45278.3947   &   0.0103 &  2 \\
 14 & 0 &14 &13 &13 & 0 &13 &12 &     45445.0760  &  45445.0882   &  -0.0122 &  2 \\
 14 & 0 &14 &15 &13 & 0 &13 &14 &     45445.0760  &  45445.0882   &  -0.0122 &  2 \\
 14 & 0 &14 &14 &13 & 0 &13 &13 &     45445.0760  &  45445.0882   &  -0.0122 &  2 \\
 13 & 1 &12 &12 &12 & 1 &11 &11 &     45733.5640  &  45733.5666   &  -0.0026 &  2 \\
 13 & 1 &12 &14 &12 & 1 &11 &13 &     45733.5640  &  45733.5666   &  -0.0026 &  2 \\
 13 & 1 &12 &13 &12 & 1 &11 &12 &     45733.5640  &  45733.5666   &  -0.0026 &  2 \\
 13 & 3 &10 &13 &12 & 3 & 9 &12 &     46135.6500  &  46135.6655   &  -0.0155 &  2 \\
 13 & 3 &10 &12 &12 & 3 & 9 &11 &     46135.6500  &  46135.6655   &  -0.0155 &  2 \\
 13 & 3 &10 &14 &12 & 3 & 9 &13 &     46135.6500  &  46135.6655   &  -0.0155 &  2 \\
 13 & 2 &11 &13 &12 & 2 &10 &12 &     47073.4380  &  47073.4511   &  -0.0131 &  2 \\
 13 & 2 &11 &12 &12 & 2 &10 &11 &     47073.4380  &  47073.4511   &  -0.0131 &  2 \\
 13 & 2 &11 &14 &12 & 2 &10 &13 &     47073.4380  &  47073.4511   &  -0.0131 &  2 \\
 14 & 2 &13 &14 &13 & 2 &12 &13 &     47674.0180  &  47674.0074   &   0.0106 &  2 \\
 14 & 2 &13 &13 &13 & 2 &12 &12 &     47674.0180  &  47674.0074   &   0.0106 &  2 \\
 14 & 2 &13 &15 &13 & 2 &12 &14 &     47674.0180  &  47674.0074   &   0.0106 &  2 \\
\hline
\hline
\end{longtable}
\tablefoot{\\
\tablefoottext{a}{Observed line frequencies and assigned uncertainties.}\\
\tablefoottext{b}{Calculated frequencies and uncertainties.}\\
\tablefoottext{c}{Observed-minus-calculated frequencies.}\\
\tablefoottext{1}{Laboratory data from \citet{McCarthy2020}.}\\
\tablefoottext{2}{Frequencies observed in TMC-1 assuming a v$_{LSR}$ of 5.83 km\,s$^{-1}$.}}\\
\end{tiny}
\clearpage

\begin{tiny}
\begin{longtable}{cccccccccccc}
\caption[]{Observed and calculated line frequencies for 2-$CCP$.
\label{lines_2cpd}}\\
\hline
\hline
$J'$ & $K'_a$ & $K'_c$ & $J''$ & $K''_a$ & $K''_c$ &  $F'$ & $F''$ & $\nu_{obs}$\,$^a$  & $\nu_{calc}$\,$^b$  & $\nu_{obs}-\nu_{calc}$\,$^c$ & Notes \\
   &   &        &      &        &        &                  &    &   (MHz)      &   (MHz)  &   (MHz) &      \\
\hline
\endfirsthead
\caption{continued.}\\
\hline
\hline
$J'$ & $K'_a$ & $K'_c$ & $J''$ & $K''_a$ & $K''_c$ &  $F'$ & $F''$ & $\nu_{obs}$\,$^a$  & $\nu_{calc}$\,$^b$  & $\nu_{obs}-\nu_{calc}$\,$^c$ & Notes \\
   &   &        &      &        &        &                  &    &   (MHz)      &   (MHz)  &   (MHz) &      \\
\hline
\endhead
\hline
\endfoot
\hline
\endlastfoot
\hline
  2 &  1 &  1 &  3 &  1 &  1 &  0 &  2 &   7266.1158  &  7266.1160  &  -0.0002  &    1 \\
  3 &  1 &  3 &  3 &  2 &  1 &  2 &  3 &   9862.3228  &  9862.3225  &   0.0003  &    1 \\
  3 &  1 &  3 &  3 &  2 &  1 &  2 &  2 &   9862.9633  &  9862.9643  &  -0.0010  &    1 \\
  3 &  1 &  3 &  4 &  2 &  1 &  2 &  3 &   9863.3416  &  9863.3396  &   0.0020  &    1 \\
  3 &  1 &  3 &  2 &  2 &  1 &  2 &  2 &   9864.3366  &  9864.3377  &  -0.0011  &    1 \\
  3 &  0 &  3 &  2 &  2 &  0 &  2 &  1 &  10331.0354  & 10331.0363  &  -0.0009  &    1 \\
  3 &  0 &  3 &  3 &  2 &  0 &  2 &  2 &  10331.2534  & 10331.2546  &  -0.0012  &    1 \\
  3 &  0 &  3 &  4 &  2 &  0 &  2 &  3 &  10331.2998  & 10331.2996  &   0.0002  &    1 \\
  3 &  2 &  2 &  3 &  2 &  2 &  1 &  2 &  10384.0499  & 10384.0474  &   0.0025  &    1 \\
  3 &  2 &  2 &  4 &  2 &  2 &  1 &  3 &  10385.4117  & 10385.4074  &   0.0043  &    1 \\
  3 &  1 &  2 &  3 &  2 &  1 &  1 &  2 &  10889.8976  & 10889.8986  &  -0.0010  &    1 \\
  3 &  1 &  2 &  2 &  2 &  1 &  1 &  1 &  10890.2582  & 10890.2612  &  -0.0030  &    1 \\
  3 &  1 &  2 &  4 &  2 &  1 &  1 &  3 &  10890.2772  & 10890.2767  &   0.0005  &    1 \\
  4 &  1 &  4 &  3 &  3 &  1 &  3 &  2 &  13135.9674  & 13135.9701  &  -0.0027  &    1 \\
  4 &  1 &  4 &  5 &  3 &  1 &  3 &  4 &  13136.0237  & 13136.0221  &   0.0016  &    1 \\
  4 &  1 &  4 &  3 &  3 &  1 &  3 &  3 &  13137.3441  & 13137.3436  &   0.0005  &    1 \\
  4 &  0 &  4 &  3 &  3 &  0 &  3 &  2 &  13712.9093  & 13712.9081  &   0.0012  &    1 \\
  4 &  0 &  4 &  4 &  3 &  0 &  3 &  3 &  13713.0091  & 13713.0071  &   0.0020  &    1 \\
  4 &  0 &  4 &  5 &  3 &  0 &  3 &  4 &  13713.0319  & 13713.0319  &   0.0000  &    1 \\
  4 &  2 &  3 &  4 &  3 &  2 &  2 &  3 &  13835.8635  & 13835.8654  &  -0.0019  &    1 \\
  4 &  2 &  3 &  5 &  3 &  2 &  2 &  4 &  13836.4421  & 13836.4416  &   0.0005  &    1 \\
  4 &  2 &  3 &  3 &  3 &  2 &  2 &  2 &  13836.5882  & 13836.5897  &  -0.0015  &    1 \\
  4 &  2 &  2 &  4 &  3 &  2 &  1 &  3 &  13969.7215  & 13969.7222  &  -0.0007  &    1 \\
  4 &  2 &  2 &  5 &  3 &  2 &  1 &  4 &  13970.3043  & 13970.3056  &  -0.0013  &    1 \\
  4 &  2 &  2 &  3 &  3 &  2 &  1 &  2 &  13970.4531  & 13970.4547  &  -0.0016  &    1 \\
  4 &  1 &  3 &  4 &  3 &  1 &  2 &  3 &  14503.5437  & 14503.5444  &  -0.0007  &    1 \\
  4 &  1 &  3 &  3 &  3 &  1 &  2 &  2 &  14503.6554  & 14503.6529  &   0.0025  &    1 \\
  4 &  1 &  3 &  5 &  3 &  1 &  2 &  4 &  14503.7120  & 14503.7117  &   0.0003  &    1 \\
  5 &  1 &  5 &  5 &  4 &  1 &  4 &  4 &  16396.7422  & 16396.7420  &   0.0002  &    1 \\
  5 &  1 &  5 &  4 &  4 &  1 &  4 &  3 &  16396.7864  & 16396.7864  &   0.0000  &    1 \\
  5 &  1 &  5 &  6 &  4 &  1 &  4 &  5 &  16396.8316  & 16396.8315  &   0.0002  &    1 \\
  5 &  1 &  5 &  4 &  4 &  1 &  4 &  4 &  16398.2745  & 16398.2737  &   0.0008  &    1 \\
  5 &  0 &  5 &  4 &  4 &  0 &  4 &  3 &  17044.4234  & 17044.4268  &  -0.0034  &    1 \\
  5 &  0 &  5 &  5 &  4 &  0 &  4 &  4 &  17044.4853  & 17044.4869  &  -0.0016  &    1 \\
  5 &  0 &  5 &  6 &  4 &  0 &  4 &  5 &  17044.5023  & 17044.5003  &   0.0020  &    1 \\
  5 &  1 &  4 &  5 &  4 &  1 &  3 &  4 &  18101.8650  & 18101.8636  &   0.0014  &    1 \\
  5 &  1 &  4 &  4 &  4 &  1 &  3 &  3 &  18101.9048  & 18101.9045  &   0.0003  &    1 \\
  5 &  1 &  4 &  6 &  4 &  1 &  3 &  5 &  18101.9544  & 18101.9537  &   0.0007  &    1 \\
  9 &  3 &  6 &  9 &  8 &  3 &  5 &  8 &  31458.0050  & 31457.9951  &   0.0099  &    2 \\
  9 &  3 &  6 & 10 &  8 &  3 &  5 &  9 &  31458.0050  & 31457.9951  &   0.0099  &    2 \\
  9 &  3 &  6 &  8 &  8 &  3 &  5 &  7 &  31458.0050  & 31457.9951  &   0.0099  &    2 \\
  9 &  1 &  8 &  8 &  8 &  1 &  7 &  7 &  32228.2070  & 32228.2150  &  -0.0080  &    2 \\
  9 &  1 &  8 &  9 &  8 &  1 &  7 &  8 &  32228.2070  & 32228.2150  &  -0.0080  &    2 \\
  9 &  1 &  8 & 10 &  8 &  1 &  7 &  9 &  32228.2070  & 32228.2150  &  -0.0080  &    2 \\
  9 &  2 &  7 &  8 &  8 &  2 &  6 &  7 &  32237.1800  & 32237.1870  &  -0.0070  &    2 \\
  9 &  2 &  7 & 10 &  8 &  2 &  6 &  9 &  32237.1800  & 32237.1870  &  -0.0070  &    2 \\
 10 &  1 & 10 &  9 &  9 &  1 &  9 &  8 &  32482.4200  & 32482.4148  &   0.0052  &    2 \\
 10 &  1 & 10 & 10 &  9 &  1 &  9 &  9 &  32482.4200  & 32482.4148  &   0.0052  &    2 \\
 10 &  1 & 10 & 11 &  9 &  1 &  9 & 10 &  32482.4200  & 32482.4148  &   0.0052  &    2 \\
 10 &  0 & 10 &  9 &  9 &  0 &  9 &  8 &  32926.9460  & 32926.9508  &  -0.0048  &    2 \\
 10 &  0 & 10 & 11 &  9 &  0 &  9 & 10 &  32926.9460  & 32926.9508  &  -0.0048  &    2 \\
 10 &  0 & 10 & 10 &  9 &  0 &  9 &  9 &  32926.9460  & 32926.9508  &  -0.0048  &    2 \\
 10 &  2 &  9 & 10 &  9 &  2 &  8 &  9 &  34279.9700  & 34279.9239  &   0.0461  &    2 \\
 10 &  2 &  9 &  9 &  9 &  2 &  8 &  8 &  34279.9700  & 34279.9239  &   0.0461  &    2 \\
 10 &  2 &  9 & 11 &  9 &  2 &  8 & 10 &  34279.9700  & 34279.9239  &   0.0461  &    2 \\
 11 &  1 & 11 & 10 & 10 &  1 & 10 &  9 &  35656.9320  & 35656.9174  &   0.0146  &    2 \\
 11 &  1 & 11 & 11 & 10 &  1 & 10 & 10 &  35656.9320  & 35656.9174  &   0.0146  &    2 \\
 11 &  1 & 11 & 12 & 10 &  1 & 10 & 11 &  35656.9320  & 35656.9174  &   0.0146  &    2 \\
 10 &  1 &  9 &  9 &  9 &  1 &  8 &  8 &  35657.4120  & 35657.4171  &  -0.0051  &    2 \\
 10 &  1 &  9 & 10 &  9 &  1 &  8 &  9 &  35657.4120  & 35657.4171  &  -0.0051  &    2 \\
 10 &  1 &  9 & 11 &  9 &  1 &  8 & 10 &  35657.4120  & 35657.4171  &  -0.0051  &    2 \\
 11 &  0 & 11 & 11 & 10 &  0 & 10 & 10 &  36014.2550  & 36014.2229  &   0.0321  &    2 \\
 11 &  2 & 10 & 11 & 10 &  2 &  9 & 10 &  37625.3080  & 37625.3501  &  -0.0421  &    2 \\
 11 &  2 & 10 & 10 & 10 &  2 &  9 &  9 &  37625.3080  & 37625.3501  &  -0.0421  &    2 \\
 11 &  2 & 10 & 12 & 10 &  2 &  9 & 11 &  37625.3080  & 37625.3501  &  -0.0421  &    2 \\
 12 &  1 & 12 & 11 & 11 &  1 & 11 & 10 &  38819.9280  & 38819.9366  &  -0.0086  &    2 \\
 12 &  1 & 12 & 12 & 11 &  1 & 11 & 11 &  38819.9280  & 38819.9366  &  -0.0086  &    2 \\
 12 &  1 & 12 & 13 & 11 &  1 & 11 & 12 &  38819.9280  & 38819.9366  &  -0.0086  &    2 \\
 11 &  1 & 10 & 10 & 10 &  1 &  9 &  9 &  39030.2220  & 39030.2201  &   0.0019  &    2 \\
 11 &  1 & 10 & 11 & 10 &  1 &  9 & 10 &  39030.2220  & 39030.2201  &   0.0019  &    2 \\
 11 &  1 & 10 & 12 & 10 &  1 &  9 & 11 &  39030.2220  & 39030.2201  &   0.0019  &    2 \\
 12 &  0 & 12 & 11 & 11 &  0 & 11 & 10 &  39098.0600  & 39098.0636  &  -0.0036  &    2 \\
 12 &  0 & 12 & 13 & 11 &  0 & 11 & 12 &  39098.0600  & 39098.0636  &  -0.0036  &    2 \\
 12 &  0 & 12 & 12 & 11 &  0 & 11 & 11 &  39098.0600  & 39098.0636  &  -0.0036  &    2 \\
 11 &  2 &  9 & 11 & 10 &  2 &  8 & 10 &  39666.8480  & 39666.8294  &   0.0187  &    2 \\
 11 &  2 &  9 & 10 & 10 &  2 &  8 &  9 &  39666.8480  & 39666.8294  &   0.0187  &    2 \\
 11 &  2 &  9 & 12 & 10 &  2 &  8 & 11 &  39666.8480  & 39666.8294  &   0.0187  &    2 \\
 12 &  2 & 11 & 12 & 11 &  2 & 10 & 11 &  40949.3390  & 40949.3329  &   0.0061  &    2 \\
 12 &  2 & 11 & 11 & 11 &  2 & 10 & 10 &  40949.3390  & 40949.3329  &   0.0061  &    2 \\
 12 &  2 & 11 & 13 & 11 &  2 & 10 & 12 &  40949.3390  & 40949.3329  &   0.0061  &    2 \\
 13 &  0 & 13 & 12 & 12 &  0 & 12 & 11 &  42184.1100  & 42184.0934  &   0.0166  &    2 \\
 13 &  0 & 13 & 14 & 12 &  0 & 12 & 13 &  42184.1100  & 42184.0934  &   0.0166  &    2 \\
 13 &  0 & 13 & 13 & 12 &  0 & 12 & 12 &  42184.1100  & 42184.0934  &   0.0166  &    2 \\
 12 &  1 & 11 & 11 & 11 &  1 & 10 & 10 &  42340.7610  & 42340.7647  &  -0.0037  &    2 \\
 12 &  1 & 11 & 13 & 11 &  1 & 10 & 12 &  42340.7610  & 42340.7647  &  -0.0037  &    2 \\
 12 &  1 & 11 & 12 & 11 &  1 & 10 & 11 &  42340.7610  & 42340.7647  &  -0.0037  &    2 \\
 12 &  2 & 10 & 12 & 11 &  2 &  9 & 11 &  43349.0490  & 43349.0406  &   0.0084  &    2 \\
 12 &  2 & 10 & 11 & 11 &  2 &  9 & 10 &  43349.0490  & 43349.0406  &   0.0084  &    2 \\
 12 &  2 & 10 & 13 & 11 &  2 &  9 & 12 &  43349.0490  & 43349.0406  &   0.0084  &    2 \\
 14 &  1 & 14 & 15 & 13 &  1 & 13 & 14 &  45118.2320  & 45118.2539  &  -0.0219  &    2 \\
 13 &  1 & 12 & 12 & 12 &  1 & 11 & 11 &  45586.7140  & 45586.6964  &   0.0176  &    2 \\
 13 &  1 & 12 & 14 & 12 &  1 & 11 & 13 &  45586.7140  & 45586.6964  &   0.0176  &    2 \\
 13 &  1 & 12 & 13 & 12 &  1 & 11 & 12 &  45586.7140  & 45586.6964  &   0.0176  &    2 \\
 13 &  2 & 11 & 13 & 12 &  2 & 10 & 12 &  46994.6060  & 46994.6288  &  -0.0228  &    2 \\
 13 &  2 & 11 & 12 & 12 &  2 & 10 & 11 &  46994.6060  & 46994.6288  &  -0.0228  &    2 \\
 13 &  2 & 11 & 14 & 12 &  2 & 10 & 13 &  46994.6060  & 46994.6288  &  -0.0228  &    2 \\
\hline
\hline
\end{longtable}
\tablefoot{\\
\tablefoottext{a}{Observed line frequencies and assigned uncertainties.}\\
\tablefoottext{b}{Calculated frequencies and uncertainties.}\\
\tablefoottext{c}{Observed-minus-calculated frequencies.}\\
\tablefoottext{1}{Laboratory data from \citet{McCarthy2020}.}\\
\tablefoottext{2}{Frequencies observed in TMC-1 assuming a v$_{LSR}$ of 5.83 km\,s$^{-1}$.}}\\
\end{tiny}

\clearpage

\section{Benzonitrile, C$_6$H$_5$CN} \label{cyano_benzene}

\begin{figure*}
\centering
\includegraphics[scale=0.54,angle=0]{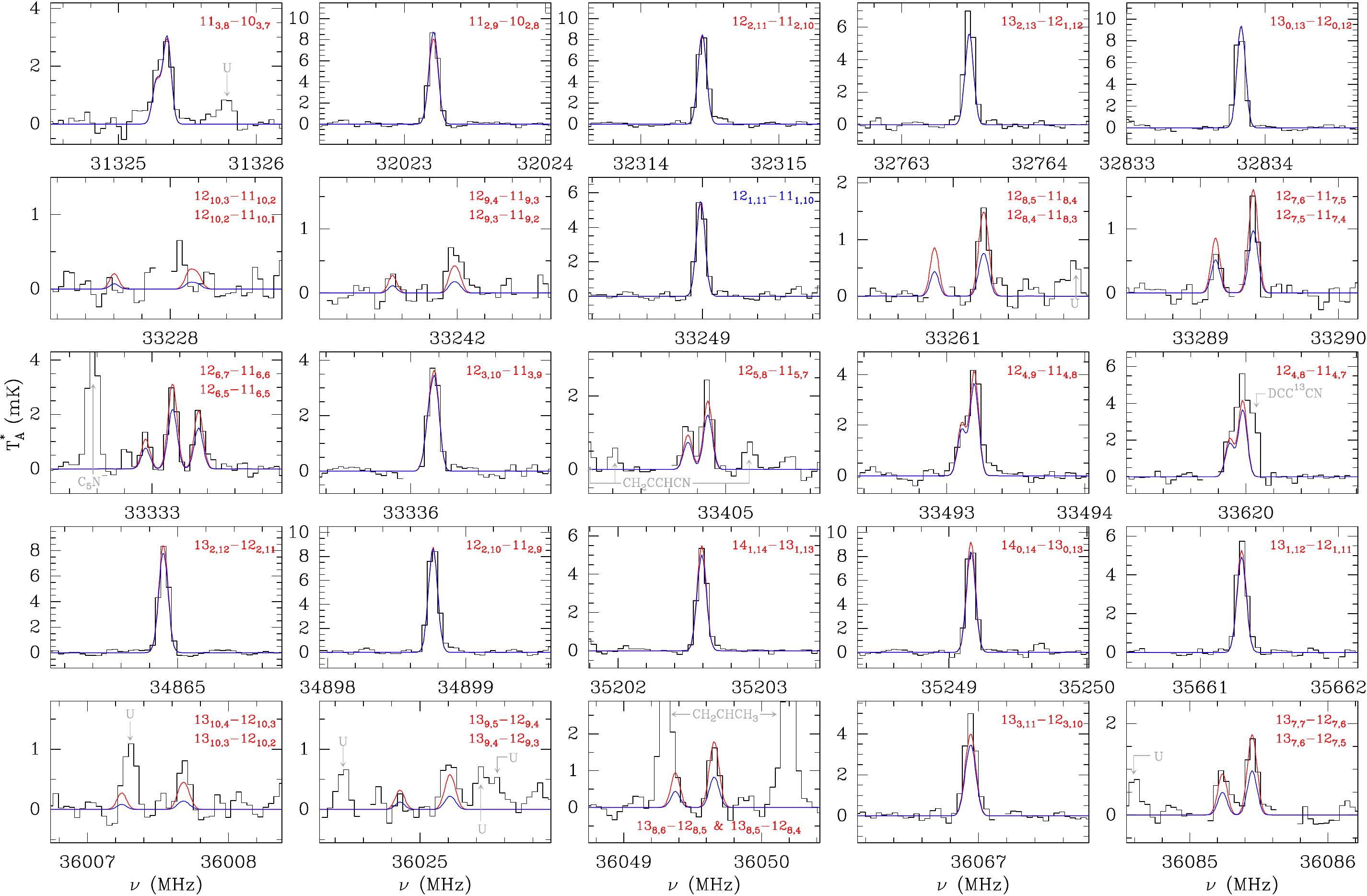}
\caption{Observed lines of C$_6$H$_5$CN  in the 31-50 GHz 
frequency range towards TMC-1. 
The abscissa corresponds to the rest frequency assuming a local standard of rest velocity of 5.83
km s$^{-1}$. 
The ordinate is the antenna temperature corrected for atmospheric and telescope losses in mK.
The red line shows the synthetic spectrum obtained from a fit to the observed line profiles,
which provides T$_r$=9$\pm$0.5\,K and N(C$_6$H$_5$CN)=(1.2$\pm$0.1)$\times$\doce. The blue line
shows the synthetic spectrum for a rotational temperature of 6 K and a column density of
1.5$\times$\doce. The rotational quantum
numbers are indicated in each panel. Blanked channels correspond to negative features produced
in the folding of the frequency switching data. While the blue line produces a very good
agreement with the observed intensities for $K_a\le$4, for higher values of $K_a$ it
underestimates the observed intensities by a factor of two.
}
\label{fig_bn1}
\end{figure*}

\begin{figure*}
\centering
\includegraphics[scale=0.54,angle=0]{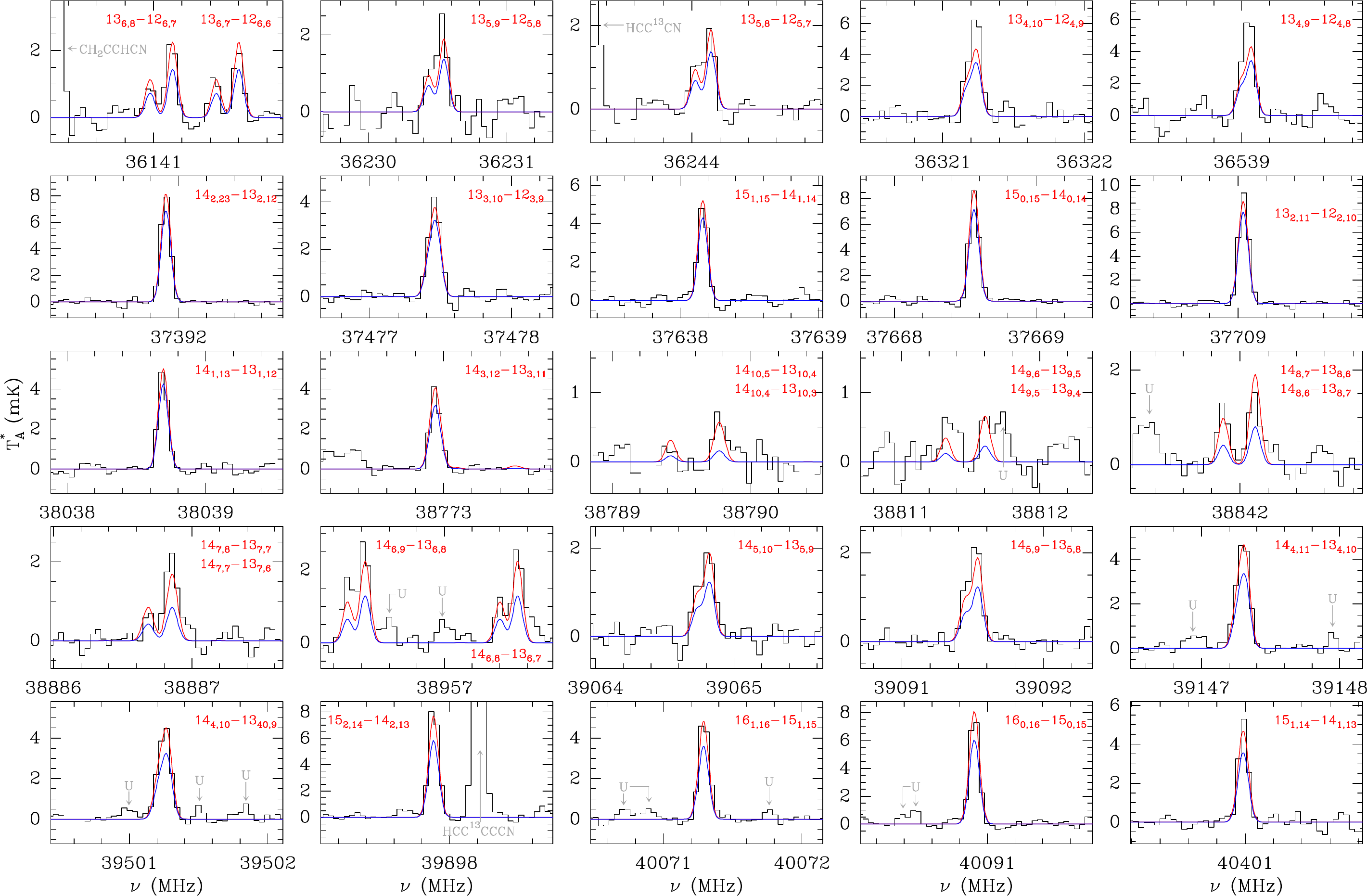}
\caption{Same as Fig. \ref{fig_bn1}.
}
\label{fig_bn2}
\end{figure*}

\begin{figure*}
\centering
\includegraphics[scale=0.54,angle=0]{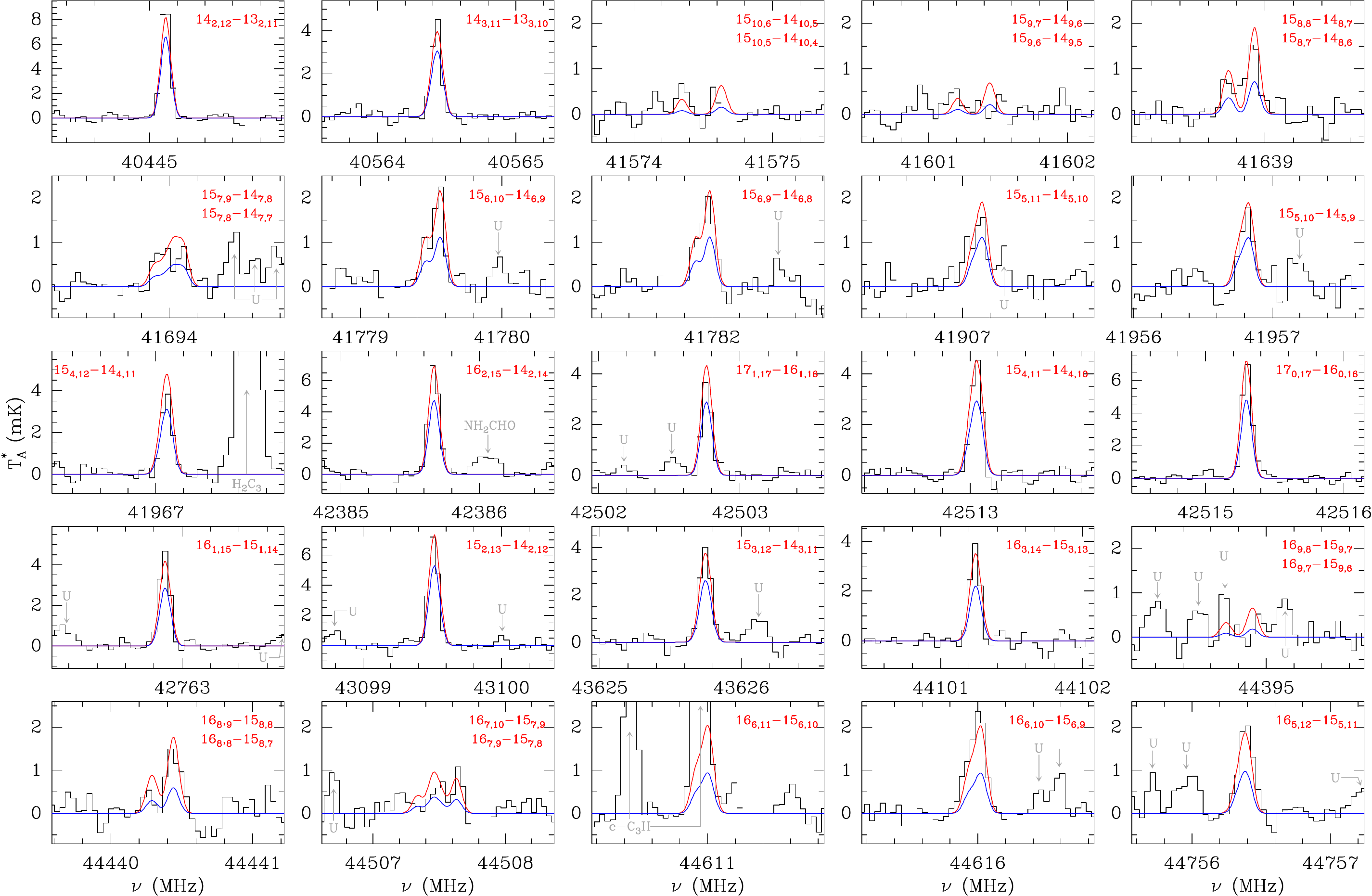}
\caption{Same as Fig. \ref{fig_bn1}.
}
\label{fig_bn3}
\end{figure*}

\begin{figure*}
\centering
\includegraphics[scale=0.54,angle=0]{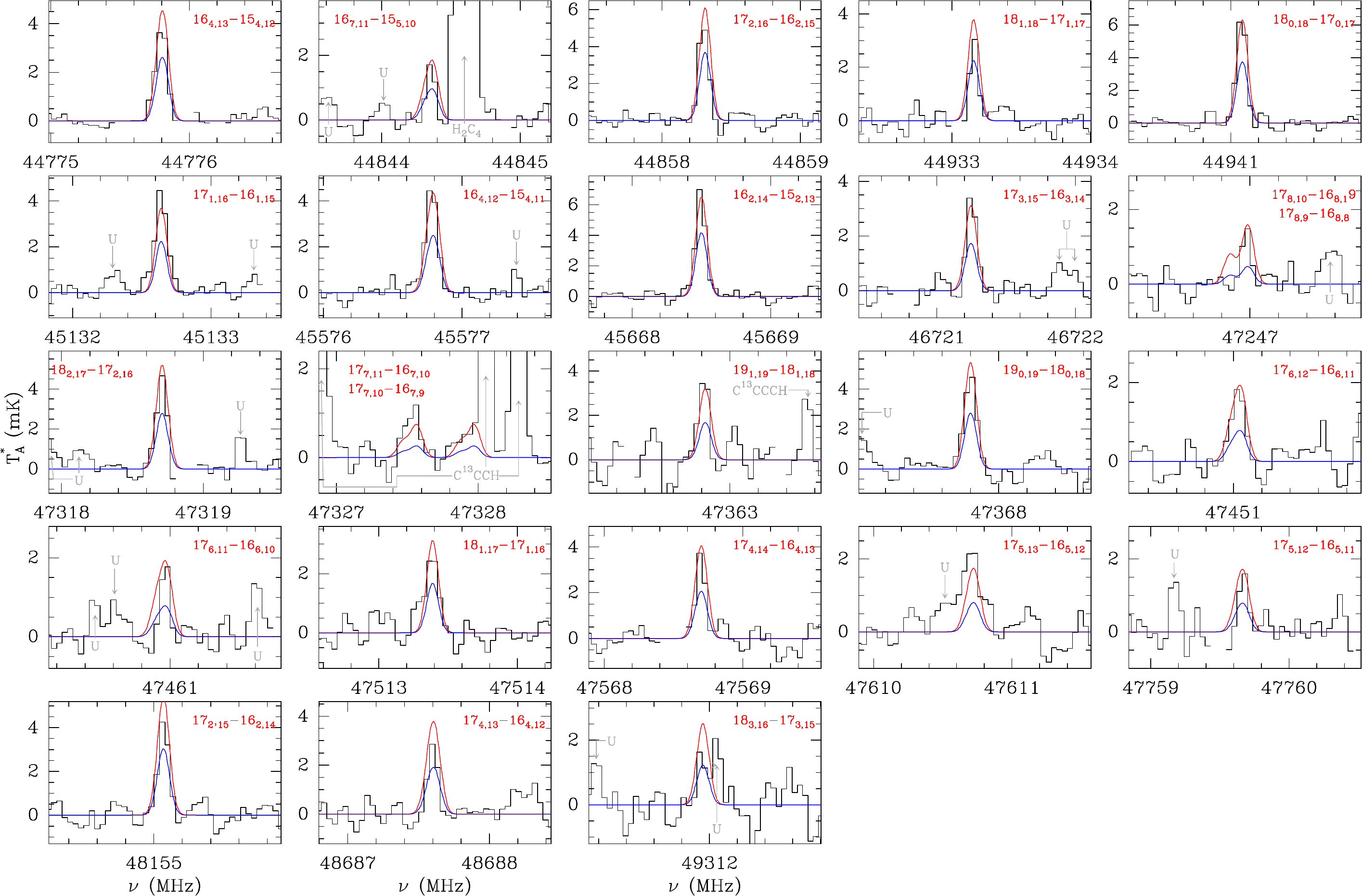}
\caption{Same as Fig. \ref{fig_bn1}.
}
\label{fig_bn4}
\end{figure*}

Benzonitrile, C$_6$H$_5$CN, was detected in TMC-1 by \citet{McGuire2018}. 
Assuming a rotational temperature of 7\,K, these authors derived a column density of 4$\times$\once. 
In a more recent work by \citet{Burkhardt2021}, the total derived column density is 1.6$\times$\doce\ (i.e.
a factor of four higher than previously reported), and the derived rotational temperature is 6.1$\pm$0.3\,K.
We explored all lines of benzonitrile in our QUIJOTE line survey. A hundred lines were detected,
some of which show hyperfine splitting. They are shown in Figs. \ref{fig_bn1}, \ref{fig_bn2}, \ref{fig_bn3}, 
and \ref{fig_bn4}. A fit to the observed lines, assuming the same source parameters as for $ECP$ and $CCP$,
provides a rotational temperature of 9.0$\pm$0.5 K and a column density of (1.2$\pm$0.1)$\times$\doce. The
corresponding synthetic spectrum is shown by the red line in these figures.
Adopting a lower rotational temperature of 6\,K, and maintaining the same column density, we could
reproduce the lines reasonably well with $K_a\le$3 (blue line in the figures). However, the synthetic
line profiles for T$_{rot}$=6\,K systematically fail to reproduce the lines involving $K_a\ge$4. Definitively, the
rotational temperature has to be close to 9\,K (i.e. near the kinetic temperature of the cloud).
This result is consistent with the rotational temperatures derived for indene \citep{Cernicharo2021a} 
and for the ethynyl and cyano derivatives of cyclopentadiene, as discussed in 
Sects. 3.1 and 3.3.
The small discrepancy between the column densities for benzonitrile derived from GOTHAM (1.6$\times$
10$^{12}$ cm$^{-2}$) and QUIJOTE (1.2$\times$10$^{12}$ cm$^{-2}$) data
is  probably related to the difference in the beam size, the assumed and/or fitted source size, 
and the rotational temperature used in each set of data for this species.

\clearpage

\section{Chemical scheme} \label{app:chem_scheme}
\begin{figure*}
\centering
\includegraphics[scale=0.7,angle=0]{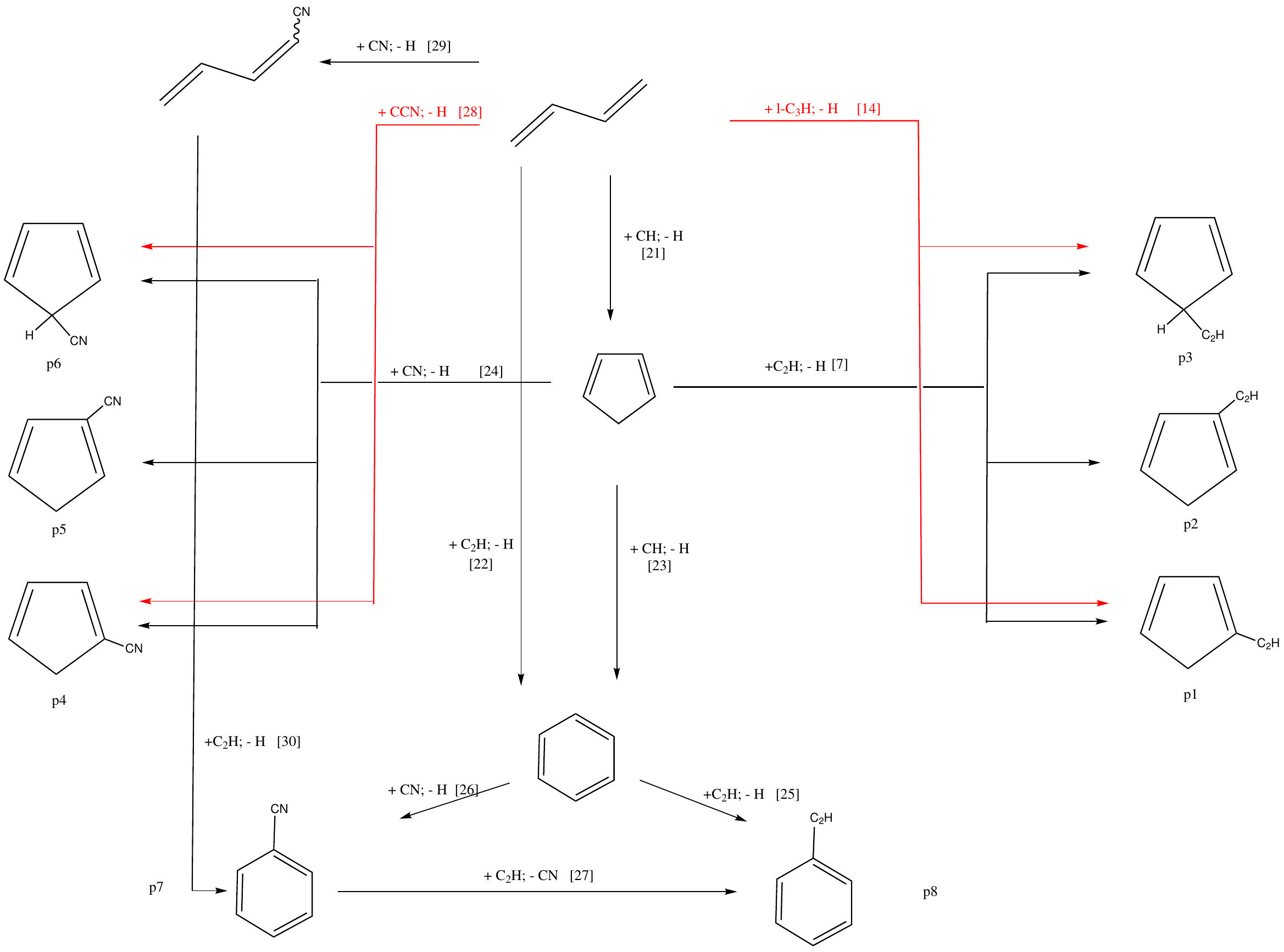}
\caption{Chemical scheme of formation of $c$-C$_5$H$_6$ and C$_6$H$_6$ and their CCH and 
CN derivatives. 
Numbers correspond to the reactions discussed in Appendix 
\ref{app:chem_scheme}. 
}
\label{fig:chem_scheme}
\end{figure*}

TMC-1 represents a prototype of a cold molecular 
cloud, that is, an interstellar environment characterized by low average gas temperatures 
of about 10 K and number densities of molecular hydrogen ranging from 10$^4$ to 10$^6$ 
cm$^3$\,s$^{-1}$.  In the gas phase, these low temperatures and pressures require exoergic 
bimolecular reactions of the type A + B $\rightarrow$ C + D to proceed without 
entrance barriers, such as rapid neutral-neutral reactions \citep{Kaiser2021}.  
With respect to the newly detected 1-$ECP$ and 2-$ECP$ 
(labelled as products $p1$ and $p2$ in Fig. \ref{fig:chem_scheme}), 
which resemble the product C, and atomic hydrogen as the light 
counter fragment D, this involves indirect reactions accessing the C$_7$H$_7$ 
potential energy surface. This C$_7$ hydrocarbon can be formed from 
the A plus B reactants via bimolecular collisions of C$_1$-C$_6$, C$_2$-C$_5$, and 
C$_3$-C$_4$ hydrocarbon species at various degrees of hydrogenation, as detailed 
via reactions [1]-[5], [6]-[12], and [13]-[20] (see Fig. \ref{fig:chem_scheme}): \newline

\noindent
[1]  C + C$_6$H$_7$              \newline
[2]  CH + C$_6$H$_6$             \newline
[3]  CH$_2$ +C$_6$H$_5$          \newline
[4]  CH$_3$ +C$_6$H$_4$          \newline
[5]  CH$_4$ + C$_6$H$_3$         \newline  
                                 \newline  
[6] C$_2$ + C$_5$H$_7$           \newline
[7] C$_2$H + C$_5$H$_6$          \newline
[8] C$_2$H$_2$ + C$_5$H$_5$      \newline
[9] C$_2$H$_3$ + C$_5$H$_4$      \newline
[10] C$_2$H$_4$ + C$_5$H$_3$     \newline
[11] C$_2$H$_5$ + C$_5$H$_2$     \newline   
[12] C$_2$H$_6$ + C$_5$H         \newline   
                                 \newline
[13] C$_3$ + C$_4$H$_7$          \newline
[14] C$_3$H + C$_4$H$_6$         \newline
[15] C$_3$H$_2$ + C$_4$H$_5$     \newline
[16] C$_3$H$_3$ + C$_4$H$_4$     \newline
[17] C$_3$H$_4$ + C$_4$H$_3$     \newline
[18] C$_3$H$_5$ + C$_4$H$_2$     \newline
[19] C$_3$H$_6$ + C$_4$H         \newline
[20] C$_3$H$_7$ + C$_4$.          \newline

Among these reactions, multiple bimolecular reactions potentially leading to 
C$_7$H$_7$ isomers plus atomic hydrogen are either direct or follow indirect 
dynamics with entrance barriers; these processes are therefore closed under 
the physical conditions in TMC-1. In detail, reaction [5] does not lead to 
any C$_7$H$_7$ isomer since any doublet C$_6$H$_3$ radical abstracts a 
hydrogen atom from the closed shell methane (CH$_4$) reactant, forming the 
methyl radical (CH$_3$) plus C$_6$H$_4$ isomers through transition states 
located between 5 and 30 kJ\,mol$^{-1}$ above the separated reactants 
(\citealt{Kaiser2011}, and therein references). 
The direct 
nature and exclusive hydrogen abstraction also prohibit reaction [12] (i.e. C$_5$H radicals 
forming ethyl (C$_2$H$_5$) radicals plus C$_5$H$_2$ isomer),  
once again through barriers from 5 and 30 kJ\,mol$^{-1}$. The remaining reactions 
are indirect, via C$_7$H$_7$ complex formation. Among them, reactions [8]–[10] 
as well as [16]-[18] involve entrance barriers of addition of the doublet 
radical to the carbon-carbon double and triple bonds of 90 - 144 kJ\,mol$^{-1}$ 
(reaction [8]; \citealt{daSilva2010}),  8 - 40 kJ\,mol$^{-1}$ (reactions 
[9] and [10]), 44 - 130 kJ\,mol$^{-1}$ for the propargyl radical 
(C$_3$H$_3$) reaction with vinylacetylene (C$_4$H$_4$) 
(reaction [16]; \citealt{daSilva2017}, and references therein),  8 - 40 kJ\,mol$^{-1}$ (reaction [17]), and 
29 - 104 kJ\,mol$^{-1}$ 
for reaction [18] of the allyl radical (C$_3$H$_5$) with diacetylene (C$_4$H$_2$) 
(\citealt{Bodi2015}, and therein references). 
The remaining reactions have no entrance barriers.

Here, reaction [2] was explored under single-collision conditions in 
crossed molecular beams as well as computationally \citep{He2020b}. The results reveal a 
strong energy and hence temperature dependence of the branching ratios with 
acetylene (C$_2$H$_2$) and cyclopentadienyl (C$_5$H$_5$) formed almost exclusively 
at temperatures of 10 K. Both reactions [1] and [3] access the same surface 
through distinct barrier-less entrance channels of carbon atom addition to the 
cyclic C$_6$H$_7$ radical (reaction [1]) followed by ring opening and/or hydrogen 
migration of the collision complex and carbene (CH$_2$)–phenyl radical 
(C$_6$H$_5$) recombination that leads to the benzyl radical (C$_6$H$_5$CH$_2$). 
Although reaction [4] has never been explored experimentally or computationally, 
a barrier-less methyl (CH$_3$)–$o$-benzyne (C$_6$H$_4$) reaction leads to the 
o-tolyl radical (C$_6$H$_4$CH$_3$), which isomerizes through hydrogen shift to 
the benzyl radical (C$_6$H$_5$CH$_2$). These reaction intermediates are coupled 
to the intermediates accessed through reaction [2]. Consequently, reactions 
[1]–[4] are not expected to lead to 1-$ECP$ and 2-$ECP$
isomers (C$_5$H$_5$CCH).

Reactions of the carbon clusters C$_2$, C$_3$, and C$_4$ (reactions [6], [13], and [20]) 
have not been explored computationally or experimentally. Likewise, the 
co-reactants C$_5$H$_7$, C$_4$H$_7$, and C$_3$H$_7$ have not been included 
in any astrochemical model. Therefore, the actual effect on the production 
of 1-$ECP$ and 2-$ECP$ isomers is unknown. Although reaction 
[19] is barrier-less and both the propylene (C$_3$H$_6$) and butadiynyl 
reactants (C$_4$H) have been observed in TMC-1, the reaction is not 
expected to lead to 1-$ECP$ and 2-$ECP$. Recent 
crossed molecular beam and computational studies of the ethynyl (CCH) reaction 
with propylene (C$_3$H$_6$) revealed reaction dynamics dictated by ethynyl 
addition – hydrogen loss mechanisms, but not to the thermodynamically most 
stable cyclopentadiene (C$_5$H$_6$) isomer (Kaiser and Mebel, in preparation). Since butadiynyl (C$_4$H) is 
isolobal to ethynyl (CCH) and hence can be considered as an ethynyl-substituted 
ethynyl radical, the dynamics of reaction [19] are not expected to form 1-$ECP$ and 
2-$ECP$, but rather butadiynyl-substituted propylene 
isomers. Further, the reactivities of the interstellar $c$-C$_3$H$_2$ isomer 
cyclopropenylidene and vinylidene carbene (H$_2$CCC) with any 
doublet C$_4$H$_5$ radical are unknown as well. However, considering the molecular 
structures of the cyclic and carbene-type reactants, formation of 1-$ECP$ and 
2-$ECP$ is unlikely. 

Overall, these considerations leave us with reactions [7] and [14] as the most 
likely pathways to forming 1-$ECP$ and 2-$ECP$ in TMC-1. Here, 
reaction [7] represents an addition–hydrogen elimination process that leads to 
three distinct C$_5$H$_5$CCH isomers, $p1$ to $p3$ (see Fig. \ref{fig:chem_scheme}). 
Crossed molecular 
beams merged with electronic structure calculations revealed that reactions of 
ethynyl radicals with unsaturated hydrocarbons are barrier-less, exoergic, and 
lead via ethynyl addition to the carbon-carbon triple or double bond followed 
by hydrogen atom loss to ethynyl-substituted hydrocarbons via molecular mass 
growth from the bottom up (\citealt{Jones2011}, and references therein).  In the case of vinylacetylene (C$_4$H$_4$)   
and 1,3-butadiene (C$_4$H$_6$),  the initial addition intermediates isomerize 
via hydrogen migration and cyclization, leading eventually to $o$-benzyne (C$_6$H$_4$) 
and benzene (C$_6$H$_6$), respectively (Fig. \ref{fig:chem_scheme}; 
\citealt{Zhang2011,Jones2011}). Consequently, the addition 
of ethynyl to the chemically non-equivalent C$_1$ and C$_2$ carbon atoms of 
cyclopentadiene followed by hydrogen elimination is expected to lead to three 
distinct C$_5$H$_5$CCH isomers, among them the astronomically 
observed 1-$ECP$ and 2-$ECP$ ($p1$ and $p2$ in Fig. \ref{fig:chem_scheme}). 

Finally, reaction [14] is worth exploring. Once again, this reaction has not been 
studied computationally or experimentally. The linear propynylidyne radical 
(C$_3$H) is ubiquitous in TMC-1. It can be considered as an ethynyl-substituted 
methylidyne (CH) radical. Considering that at 10 K methylidyne (CH) reacts with 
1,3-butadiene (C$_4$H$_6$) to cyclopentadiene (C$_5$H$_6$),  ethynyl-substituted 
methylidyne radicals might react to ethynyl-substituted cyclopentadiene isomers 
$p1$ and $p3$ in barrier-less, overall exoergic bimolecular neutral-neutral reactions 
(see Fig. \ref{fig:chem_scheme}).

Overall, reactions [7] and [14] are plausible pathways to forming the ethynyl-substituted 
cyclopentadiene isomers 
$p1$ to $p3$  under the conditions present in TMC-1 (Fig. 
\ref{fig:chem_scheme}).

\begin{table*}
\small
\caption{Reactions included in the chemical scheme of formation of 
$c$-C$_5$H$_6$ and C$_6$H$_6$ and their CCH and CN derivatives.}
\label{table:reactions}
\centering
\begin{tabular}{llcl}
\hline \hline
& \multicolumn{1}{l}{Reaction} & \multicolumn{1}{c}{$k$ (cm$^3$\,s$^{-1}$)} & \multicolumn{1}{l}{Comment} \\
\hline
21  & CH + CH$_2$CHCHCH$_2$ $\rightarrow$ $c$-C$_5$H$_6$ + H    & $4.0 \times 10^{-10}$ &  \\
22  & C$_2$H + CH$_2$CHCHCH$_2$ $\rightarrow$ C$_6$H$_6$ + H    & $4.0 \times 10^{-10}$ & \\
23  & CH + $c$-C$_5$H$_6$ $\rightarrow$ C$_6$H$_6$ + H    & $4.0 \times 10^{-10}$ & \\
7   & C$_2$H + $c$-C$_5$H$_6$ $\rightarrow$ $c$-C$_5$H$_5$CCH (1-$ECP$) + H    & $1.0 \times 10^{-10}$ & \\
    & ~~~~~~~~~~~~~~~~~~~~~~~~~~~~~ $c$-C$_5$H$_5$CCH (2-$ECP$) + H    & $1.0 \times 10^{-10}$ & \\
24  & CN + $c$-C$_5$H$_6$ $\rightarrow$ $c$-C$_5$H$_5$CN (1-$CCP$) + H    & $2.0 \times 10^{-10}$ & \\
    & ~~~~~~~~~~~~~~~~~~~~~~~~~~~~~ $c$-C$_5$H$_5$CN (2-$CCP$) + H    & $2.0 \times 10^{-10}$ & \\
25  & C$_2$H + C$_6$H$_6$ $\rightarrow$ C$_6$H$_5$CCH + H    & $4.2 \times 10^{-10}$ & Value at 105 K \citep{Goulay2006}. \\
26  & CN + C$_6$H$_6$ $\rightarrow$ C$_6$H$_5$CN + H    & $5.4 \times 10^{-10}$ & Value at 15 K \citep{Cooke2020}. \\
27  & C$_2$H + C$_6$H$_5$CN $\rightarrow$ C$_6$H$_5$CCH + CN    & $2.0 \times 10^{-11}$ & \\
14  & $l$-C$_3$H + CH$_2$CHCHCH$_2$ $\rightarrow$ $c$-C$_5$H$_5$CCH (1-$ECP$) + H    & $4.0 \times 10^{-10}$ & \\
28  & CCN + CH$_2$CHCHCH$_2$ $\rightarrow$ $c$-C$_5$H$_5$CN (1-$CCP$) + H    & $4.0 \times 10^{-10}$ & \\
29  & CN  + CH$_2$CHCHCH$_2$ $\rightarrow$ CH$_2$CHCHCHCN + H & $4.0 \times 10^{-10}$ \\
30  & CCH + CH$_2$CHCHCHCN  $\rightarrow$ C$_6$H$_5$CN + H    & $4.0 \times 10^{-10}$ \\
\hline
\end{tabular}
\end{table*}

\end{appendix}

\end{document}